 \RequirePackage[]{lineno}
\documentclass[prd,amsfonts,nofootinbib,amssymb,amsmath,graphicx,epsfig]{revtex4-1}
\usepackage{epsfig}
\usepackage{xcolor}
\usepackage{ulem}
\usepackage{appendix}
\begin{document}

\title{Quantum Reference Frames in terms of Tetrads}

\author{Jeremy Butterfield}\email{jb56@cam.ac.uk}
\affiliation{Trinity College, University of Cambridge, Cambridge U.K.}

\author{Daniel A.\ Turolla Vanzella}\email{vanzella@ifsc.usp.br}
\affiliation{S\~ao Carlos Institute of Physics,  University of S\~ao Paulo,  IFSC-USP,
13566-590, S\~ao Carlos, SP, Brazil }

\medskip

\date{10 October 2026}

\begin{abstract}

We adapt to the context of quantum theory the treatment, familiar in relativity theory, of a reference frame as a tetrad. Since a tetrad is a set of four orthonormal vectors at a spacetime point, it encodes a timelike direction, three orthogonal spatial directions, and units; and so it gives a configuration in spacetime (a spatial orientation and state of motion) of an idealized infinitesimal apparatus (``rulers and a clock"). So a natural ``first guess" is that the state of a quantum reference frame should be an assignment of complex amplitudes to a set of tetrads. We develop this idea, in terms of bundles over spacetime of bases of the tangent spaces. 

This paper is a companion to two other papers. It emphasises the main conceptual aspects of our proposal, while the other two   papers give physical details. It also explains the main contrast between those two papers. Namely: (i) the first considers a single spacetime geometry, so that a quantum reference frame, on our proposal, gives a superposition of perspectives on that geometry; while (ii)  the second considers the spacetime manifold with  no geometry given {\it a priori}, which will imply that a quantum reference frame, on our proposal, can describe a superposition of perspectives on a superposition of geometries---not ``just" a superposition of perspectives on a single geometry. The main mathematical idea for representing this contrast will be choosing between (i) the orthonormal frame bundle, for a given spacetime geometry; (ii) the frame bundle in the usual sense of containing all bases of all spacetime points' tangent spaces.

\end{abstract}

\maketitle

\tableofcontents

\section{Introduction}
\label{sec:intro}

In relativity theory, it is customary to treat a reference frame at a spacetime point as a {\it tetrad}, i.e.\ a set of four orthonormal vectors at the point that thus encode a timelike direction, three orthogonal spatial directions, and units.\footnote{We will always work with Lorentzian not Euclidean signature; so for brevity we will say `orthonormal’, not the more cumbersome `pseudo-orthonormal’.} This gives a configuration in spacetime, i.e.\ a state of motion and a spatial orientation, of an idealized infinitesimal apparatus (``a clock and three rulers") at the spacetime point. For it enables a compact and precise description of the results of measuring time intervals and spatial lengths (a reading on a clock or ruler): namely, in terms of the coupling of the tetrad with vectors representing e.g.\ the motion of a test-particle. 

In this paper, it is this use of reference frames to measure 
temporal and spatial intervals, that we will be mostly concerned with. 
 But we note two considerations that give tetrads as reference frames a wider physical significance. First: The result of  all  measurements (‘pointer-readings’) on   classical systems, e.g.\ the electromagnetic field, can be expressed by the coupling between a tetrad and the system, in a manner wholly parallel to our expressing temporal and spatial intervals by coupling a tetrad to e.g.\ a vector representing the motion of a test-particle. Second: a good case can be made for {\it all} measurements being ultimately reducible to measurements of temporal and spatial intervals: cf.\ for example Ref.~\cite{MPSV23}.  

Our aim is to adapt this compact and successful treatment of classical reference frames to the context of quantum theory. In this endeavour, a natural  “first guess” is that the state of a {\it quantum reference frame} should be taken as an assignment (of course, subject to various conditions) of complex amplitudes to tetrads. For since a tetrad represents  an idealized infinitesimal apparatus’  configuration in spacetime (i.e.\ spatial orientation and state of motion),  the state of a quantum reference frame should surely be a complex ``wave-function” defined on the set of these configurations.\footnote{\label{sptconfig}{Our `surely' is of course in homage to Schr\"{o}dinger's invention of wave mechanics. But whereas wave mechanics uses complex functions of purely spatial configurations, the configuration of a tetrad has a timelike component. Unsurprisingly, this difference will call for some careful treatment in what follows. 
%{\jtod Decided it is best to have a footnote explanation-plus-warning, once for all!}
}} We will show that this first guess is indeed viable. That is: our proposal is, in short, that states of quantum reference frames are complex functions on tetrads.

But this proposal will be spread out over three papers. For as we will see, this proposal is also motivated by the fact that it renders in geometric (especially coordinate-free) language many aspects of some recent proposals about quantum reference frames which were formulated using coordinates. This includes proposals that describe the gravitational field of a massive quantum system in a superposition of two spatial positions, i.e.\ describe a superposition of geometries. As a result of these connections, it will be clearest to split our exposition in three papers. This paper will emphasise the conceptual aspects of our proposal, describing how our quantum reference frames couple to classical systems, and so measure a classical quantity on them.   The other two papers will give more physical details. The first of them, which we call $\bf {[I]}$, discusses measuring quantum quantities, and transformations between quantum reference frames. The second, which we call $\bf {[II]}$, gives the details about our proposal's description of a superposition of geometries. (The ``coordinate-based'' papers that were the original motivation for our geometrically-formulated proposal are mainly Refs.\ \cite{GCB19}, \cite{CGBB20}, \cite{FB22} and \cite{HKC23}. And the main ideas about how our proposal relates to these papers are presented, in draft form, in Ref.\ \cite{VB24}, especially its Sections I, V and VI. So Ref.\ \cite{VB24} is in effect a predecessor of this paper and $\bf {[I]}$ and  $\bf {[II]}$.)

\bigskip

The  contrast between $\bf {[I]}$ and  $\bf {[II]}$ will be fully explained in Section \ref{ssec:bundle}. But in short, it is as follows. \\
\indent \indent \indent (i): $\bf {[I]}$ considers a given spacetime geometry, so that a quantum reference frame, on our proposal, gives a superposition of perspectives on that geometry; while \\
\indent \indent \indent (ii): $\bf {[II]}$  considers different spacetime geometries, which will imply that a quantum reference frame, on our proposal, can describe a superposition of perspectives on a superposition of geometries---not ``just" a superposition of perspectives on a single geometry.

Accordingly,  in $\bf {[I]}$ and in this paper's discussion of its ideas,  we will talk of a {\it restricted} quantum reference frame; which we will abbreviate as `r-QRF'. Similarly, we will talk of the {\it restricted framework}. And in $\bf {[II]}$ and in this paper's discussion of its ideas, we will talk of a {\it general} quantum reference frame, which we abbreviate as `g-QRF'; and of the {\it general framework}.

In all three papers, we will adopt the language of fibre bundles. In all three, the broad idea will be to take spacetime as the base-manifold, and a point in the fibre above a spacetime point $p$ will be given by some basis of the tangent space at $p$. So it will be to the points in such a fibre that the state of a quantum reference frame assigns  a (possibly zero) complex amplitude. 

Here, the distinction between a tetrad, i.e.\ orthonormal basis, and a generic (i.e.\ maybe not  orthonormal) basis will be crucial. For it provides the explanation  (already in this paper)  of this main contrast between $\bf {[I]}$ and  $\bf {[II]}$---and of how our proposal can be adapted to both sides of this contrast. More details below. But in short, the answer will be as follows.\\
\indent \indent \indent   In Sections \ref{ssec:bundle} and \ref{sec:perspsing}  (and in  $\bf {[I]}$): the relevant bundle will take the  base-manifold to be, not just a manifold, but a {\it spacetime}: that is, a Lorentzian manifold, i.e.\ a manifold $\cal M$ equipped with a Lorentzian metric $g$. (We will of course also sometimes write $g_{ab}$.) And the fibre above a spacetime point $p \in {\cal M}$ will consist of all those bases of the tangent space $T_p{\cal M}$ at $p$ that are orthonormal according to the  metric $g$. That is: bases that are tetrads for $g$. We write this bundle as ${\cal F}_{\rm o}[g]$ (or ${\cal F}_{\rm o}[g_{ab}]$); with `o' standing for `orthonormal', and $[g]$ signalling the dependence on the given metric. And we will call ${\cal F}_{\rm o}[g]$ {\it the orthonormal frame bundle}.\\
\indent \indent \indent  In Section \ref{ssec:bundle} and \ref{sec:basicsing}  (and in $\bf {[II]}$): the relevant bundle will take the base-manifold to be {\it just} a manifold $\cal M$, not endowed with a metric; and the fibre above a  point $p \in {\cal M}$ will consist of {\it all} bases of the tangent space $T_p{\cal M}$ at $p$. This is usually called {\it the frame bundle}. We will write it as $\cal {F(M)}$; (so the absence of $g$  signals that no metric is assumed).  The ``extra freedom" of considering all bases allows one to represent any spacetime (i.e.\ any Lorentzian manifold) built from $\cal M$.  The idea is a standard one in differential  geometry; (and is also of course standard for Riemannian metrics). Namely: one specifies a metric on $\cal M$ by (a) giving a section of this bundle, and then (b) stipulating that the points (i.e.\ bases) in this section are to be orthonormal in the metric one is hereby defining.\footnote{\label{sadjargon}The usual jargon is that $\cal {F(M)}$ is called `the frame bundle'. But for our purposes this  jargon is confusing. For in physics, the word `frame' connotes orthonormality, while $\cal {F(M)}$ does not assume that a metric (or any other geometric structure)  is given. So a better name for it would be `the basis bundle': which would signal that its points, i.e.\ arbitrary bases of tangent spaces, are not {\it given} as being orthonormal---though, as just explained, they can be stipulated to be. But  we will abide by the usual jargon. (Accordingly, our calling ${\cal F}_{\rm o}[g]$ `the orthonormal frame bundle', i.e.\ our using both the words `orthonormal' and  `frame', is not really repetitive. For we have gone along, for $\cal {F(M)}$, with the usual jargon by saying  `the frame bundle'  for what would be better called `the basis bundle'.)}  

Here, we should clarify our phrase above `represent any spacetime (i.e.\ any Lorentzian manifold) built from $\cal M$'. For we must recall that  (i) a manifold admits a Lorentzian metric iff it admits a  continuous, nowhere-vanishing vector field; and (ii) not every  manifold admits a global, continuous basis field, i.e.\ it does so iff the tangent bundle is trivial (such a manifold being called  `parallelizable').  Notice that admitting a Lorentzian metric---which is the sole mandatory requirement to build a spacetime on top of ${\cal M}$---does not guarantee existence of a global,  continuous basis field. This motivates us to  (i') {\it assume} that  the manifold $\cal M$ with which we work admits a nowhere-vanishing vector field, and  (ii') ``work locally", confining all our constructions to  parallelizable regions ${\cal O} \subset {\cal M}$, in such a way that the metric on ${\cal M}$ can be determined/defined ``patch by patch''.  Indeed, one can hope that for any physical application of our quantum reference frames, the region $\cal O$ could be chosen small enough to guarantee the features (i') and (ii'). 

 In $\bf {[I]}$ and  $\bf {[II]}$, we will present the physical consequences and applications of formulating quantum reference frames in terms of these proposed bundles.  In $\bf {[I]}$, the main innovation compared with this paper will be: (a) to formulate transformations between our proposed quantum reference frames (including transformations that are superpositions of classical reference frame transformations); and (b) to express quantum observables, i.e.\ physical quantities, in our formalism. Doing (a) and (b) will imply that our proposal can represent a much-discussed feature of the quantum reference frame literature, viz.\ the ``transfer" of a superposition from the object-system to a reference frame.  In $\bf {[II]}$, the main innovation (compared with $\bf {[I]}$ and this paper) will be showing that, once we use the frame bundle (albeit misnamed: cf.\ footnote \ref{sadjargon}),  our proposal  can be adapted to represent superpositions of different  geometries---and so can represent  the gravitational field sourced by a massive quantum system in a spatial superposition.

 \bigskip
 
 So much by way of briefly stating the positive agenda of the three papers. We should also state some limitations of their scope.  These  limitations have a common theme, viz.\ that we will treat our quantum reference frames more simply than does (so far as we know) much of the quantum reference frame literature.  Although this paper will not make detailed comparisons with other work, we note two main differences. 
 
 (1):  While all proposals (including ours) take a quantum reference frame to have a Hilbert space, some proposals treat the frame’s preferred quantity (its ``pointer-observable'') very generally---while we will use only the tetrad itself; though, as we note at the end of (2), this difference is one of generality rather than of kind.
  
 (2): So far as we know, those proposals that, like us, {\it do} invoke tetrads---namely, Refs.\ \cite{JG24} and \cite{FG25}---do so in a more general and abstract way than we do. More precisely: while we couple a tetrad to a field (e.g.\ the electromagnetic field) of the measured system, they have a map (a POVM) sending appropriate (Borel) subsets of the frame bundle to positive operators on the Hilbert space of the quantum reference frame.  Cf.\ e.g.\ Ref.\ \cite{JG24}, Section 5.1 and 5.2, especially Eq.\ 58 and  Ref.\ \cite{FG25} Definition 2.2.  But the two proposals are closely related: our restricted framework's quantum reference frames, and  expectation values with respect to them, are recovered in the operational approach of Refs.\ \cite{JG24} and \cite{FG25}  by choosing a particular sharp frame observable and a pure frame state \cite{GK26}.\footnote{\label{operational}The relation between the proposals is discussed in Ref.\ \cite{GK26}; cf.\ also Ref.~\cite{FIM26}. Let us summarize, with some ``looking ahead’’ to our own later notation. Section 11.3 of Ref.\ \cite{GK26}  considers a trivial principal bundle ${\cal M}\times G$, with the frame observable given by the canonical projection-valued measure (multiplication by characteristic functions) on $L^2({\cal M}\times G,\nu\otimes m)$, where $\nu$ is a measure on ${\cal M}$ and $m$ a Haar measure on $G$. Restricting a relativized observable to the pure frame state given by a unit vector $\psi$ then yields an average of the system's quantity weighted by $|\psi|^2\,\nu\otimes m$. For $G=O(3,1)$, with ${\cal F}_{\rm o}[g]$ trivialized by a fiducial tetrad field, $\nu=\epsilon$ and $m=\mu_L$, this is our expectation value for a region (Eq.\ \ref{EvalueQRFnewspt} below), with a classical state for the measured system; and our normalisation condition, Eq.\ \ref{norm20Oct}, is just the normalisation of $\psi$. So our using ``only the tetrad itself'' amounts, in these authors' operational approach, to choosing this canonical sharp frame observable together with a pure frame state; while our coupling of the tetrad to the measured system's field corresponds to their relativization map. 
 
 The operational approach is also general in allowing unsharp frame observables, mixed frame states and general operator algebras. But note that in Ref.\ \cite{GK26} (Section 11.3), the group action defining covariance and relativization is given in a chosen trivialization; (a gauge formulation is postponed to future work). On the other hand, our approach's expectation values are explicitly independent of the trivialization, i.e.\ of the fiducial tetrad field (Section \ref{ssec:ps:normn}). And our approach extends to the general framework of Section \ref{sec:basicsing}, with its volume element varying along the fibres of ${\cal F}({\cal M})$, so that the natural measure there is not of the form $\nu\otimes m$ with $m$ a Haar measure (cf.\ Eq.\ \ref{eq:muO}). 
 
 Note also that the integration theory of Ref.\ \cite{GK26} concerns {\it bounded} operator fields. So for Ref.\ \cite{GK26}, the relativized operator corresponding to our coupling of the tetrad to the measured system exists only when this coupling is a bounded function on the bundle (e.g.\ for the probability that a reading lies in a given interval). Our expectation values, by contrast, involve integrands that are unbounded along the non-compact fibres---e.g.\ $F_{ab}(\Lambda{\bf e})^a_j(\Lambda{\bf e})^b_0$ grows like $\cosh\eta$ under a boost of rapidity $\eta$---for which no such operator is constructed in Ref.\ \cite{GK26}. However, for frame states with sufficiently fast decay along the fibres, the scalar averages (cf. Eq.\ (32) in Section 11.1 of Ref.\ \cite{GK26}) still make sense, and coincide with our approach's expectation values.
 }

 In short, there will be four  limitations. We state them in increasing order of specificity, which will mean that each item adds detail to its predecessors; as follows. 
 
  \bigskip
 
\indent \indent \indent \indent \indent (A):  {\it Reference frames as perspectival}:--- The first limitation begins with a comment about our use (e.g.\ in the indented (i) and (ii) above)  of the word `perspective'. The comment  is that since in classical relativity theory, measurement results are determined by how a tetrad (representing an infinitesimal apparatus) couples to quantities for the measured system, it is natural to say that in classical relativity, a tetrad gives a {\it perspective} on the measured system. We will adopt this metaphor, and extend it to quantum reference frames. So we will also say that a restricted QRF (an r-QRF) gives a superposition of perspectives on a given geometry. And as in the indented (ii) above:  we will also say that a general QRF (a g-QRF) can give a superposition of perspectives on a superposition of geometries---where, in this paper, the latter phrase is meant in a {\it pointwise} sense: the metric {\it at the (uncertain) location of the apparatus} is itself uncertain (cf.\ Section \ref{ssec:bs:definition}).  But the fact that our quantum reference frames are perspectival will amount to a limitation of scope. 

For it will imply that we do {\it not} address the detailed physics of a superposition of geometries: in particular, what constraints should be put on such a superposition, and how to understand it as the time-evolution of superposed 3-geometries. As we will see, this limitation is ``hidden in plain sight”, in a way that could be misleading. The reason is that the mathematical object we will use (in Section \ref{sec:basicsing} and in detail in $\bf {[II]}$) to specify a spacetime geometry, viz.\ a section of the frame bundle (though cf.\ footnote \ref{sadjargon}), also specifies a tetrad field on that same geometry. This fact 
\newpage 
\noindent makes it tempting (a) to think that a general QRF, a g-QRF, will assign a complex amplitude to the section, and then (b) to interpret this as assigning a complex amplitude to a  spacetime geometry---as well as to a ``perspective” on a geometry. Agreed, this interpretation would surely lead in to the detailed physics of superpositions of geometries. But in fact (and as we will see in (B) just below): our g-QRFs will {\it not} assign complex amplitudes to sections. And accordingly,  we will avoid that detailed physics: sufficient unto the day is the work thereof.\footnote{\label{setasidebasic}In the predecessor to these three papers (Ref.\ \cite{VB24}), we suggested a label for the effort to describe this detailed physics in terms of our  idea of complex amplitudes for bases of tangent spaces of  spacetime points---an effort which, there also, we set aside (\cite{VB24}, Section IV). We called it `the basic conception’ of a quantum reference frame, as against the `perspectival conception’ that we do adopt. But note that the basic conception is really a conception, not of a quantum reference frame, but of how to do quantum gravity. More precisely, since one is describing gravitation with the traditional metric variables: it is a conception of how to do quantum geometrodynamics.} 

 \bigskip

\indent \indent \indent  \indent \indent(B): {\it A single apparatus}:--- For the second limitation, we begin by recalling that  in classical relativity theory, a reference frame is often defined to be: not (i) a single tetrad at spacetime point (as so far discussed), but (ii) a {\it field} of tetrads, i.e.\ an assignment to every spacetime point---maybe not throughout $\cal M$, but throughout some region ${\cal O} \subset {\cal M}$---of a tetrad at it. Thus one imagines an infinitesimal apparatus at every point, in a certain classical configuration, where `configuration' is again taken in a spacetime sense, as including a state of motion as well as a spatial orientation. 

This single-vs.-field contrast carries over to our proposed quantum reference frames. Thus, one can think of either:\\
\indent \indent \indent (i): A single infinitesimal apparatus: which is to be treated as a quantum system, so that it has  non-zero amplitudes ``across spacetime and along the fibre”---amplitudes that encode quantum uncertainty about its spacetime location and configuration. Or: \\
\indent \indent \indent (ii): A field of many infinitesimal apparatuses spread across spacetime. Classically, each such apparatus would be modelled as a tetrad: but again, one  takes them to be subject to quantum uncertainty. Note that in this option (ii), the word ‘many’ is ambiguous between some finite number of tetrads and infinitely many. But the word `field' and the classical definition  suggest continuously many apparatuses,  filling spacetime or a region ${\cal O}$ of it.\footnote{Such a continuous infinity of apparatuses would of course be treated as some kind of continuum limit of increasingly but finitely many, ever smaller, apparatuses.}  \\ 
\indent We will call these: (i)  the {\it single conception} and (ii) the {\it field conception}, respectively.

Our second  limitation is now easily stated. Namely: for both our restricted and our general frameworks (here, and in $\bf {[I]}$ and $\bf {[II]}$), we will focus exclusively on the single conception---which obviously will be simpler. 

 \bigskip
 
\indent \indent \indent \indent \indent (C): {\it Without `hats'}:--- Both the third and fourth limitations of all three papers are  about our setting aside, to a large extent, the dynamics of our quantum reference frames.  In this regard, the first thing to say is that---although our proposal is ``quantum”  in that it assigns complex amplitudes to classical configurations of an infinitesimal apparatus---we will not ``add hats” to, nor in any other sense quantize, variables describing these bases. This limitation applies to both our restricted and our general frameworks.  In other words: although we agree that the quantum reference frame is indeed a physical system, and so has degrees of freedom that in principle should be described quantum-theoretically (either with operator dynamical variables (``hats”) or perhaps with path integrals), in neither of our frameworks will we undertake such a description. But note that this limitation will not prevent our treating transformations, both classical and quantum, between our QRFs (in both $\bf {[I]}$ and $\bf {[II]}$); and thereby ``transferring" a superposition from the object-system to the QRF.\footnote{\label{twoclarifctns} Note also that this limitation will be somewhat clearer for the restricted framework. For in the general framework, we use a field of bases to define a geometry, by stipulating that the bases are to be orthonormal in the geometry thereby defined. So the bases do not ``merely" encode or reflect a geometry, as one might say of the tetrads in the restricted framework with its metric given {\it ab initio}. But obviously this difference does {\it not} amount to the general framework treating variables describing the bases as quantum. 
%\\
% 20July2026: surely we can now delete the following, as we  DO discuss the ``double interpretation"...\\
% {\jtod (left here as aide-memoire for drafting: delete in final version): I am proud that I managed to say (2) without referring to technicalia of sections, let alone the (for-us-now[due to (B) above]-less-important) field conception. But NB we surely must say later, even on the single conception,  that the general framework involves a ``double interpretation", in that we will interpret  a `tetrad-by-stipulation' field as {\it also} representing an idealized infinitesimal apparatus whose readings, through its representing the metric, register features of the stipulated geometry.} \\
}

\bigskip

\indent \indent \indent \indent \indent (D): {\it Setting aside dynamics}:---  As our fourth limitation, let us be more specific  about largely setting aside  the dynamics of quantum reference frames.

We can best introduce this by recalling a familiar spectrum of broad conceptions of a reference frame. This spectrum is general: it is wholly independent of the ideas of tetrads and fibre bundles, and it applies equally to classical and quantum physics. We can articulate it as three alternatives: with successive alternatives being more realistic. Thus one can distinguish:\\
\indent \indent \indent (i): Treating a reference frame as ``having no physics of its own”. We treat it as without degrees of freedom, and {\it a fortiori} without dynamics, and without  effects on other systems. This means it is similar to a coordinate system, at least in spacetime theories, and as regards measurements of spatial and temporal intervals. (We only say `similar to' because if, like us, one takes a reference frame as a tetrad or tetrad field, then of course: it carries information about the metric, which a coordinate system does not.)\\
\indent \indent \indent (ii): Treating a reference frame as ``having physics of its own”. It has degrees of freedom, and maybe also dynamics (equations of motion) for these degrees of freedom. But its dynamics is not coupled to other systems or fields (including the gravitational field). In particular, in spacetime theories with a dynamical (i.e.\ causally influenced) geometry: there is no back-reaction on the geometry.\\
\indent \indent \indent (iii): Treating a reference frame as having a dynamics that {\it is} coupled to other systems and fields. In particular, in spacetime theories with a dynamical  geometry: there is back-reaction on the geometry.

Obviously, these three alternatives are successively more realistic, and so harder to implement in one’s theory. And as our discussion above makes clear: our treatment of quantum reference frames will be an unambitious version of (ii). That is: we will attribute to the reference frame degrees of freedom. It will have, firstly,  {\it locational} degrees of freedom, about where it is in spacetime: about ``on which fibre of the bundle”. And it will have  {\it orientational} (in a spacetime sense of `orientation'), or {\it internal}, degrees of freedom, about ``where on a given fibre”. But we will not give any dynamics for these degrees of freedom (though we will  briefly mention some possibilities for future work). So in particular, we will say nothing about back-reaction {\it a la} (iii).\footnote{\label{Carlo}We have treated (B), (C) and (D) separately and briefly. But their topics are of course related, and they have considerable and overlapping literatures. For example, general relativity’s diffeomorphism invariance and the legacy of Einstein’s hole argument prompt one to define spacetime points in terms of the values of appropriate spacetime-filling material fields, rather like (B)’s field conception; though one then faces the question whether, and how, to treat the dynamics of these fields, i.e.\ the topic of (D).  We will return to these issues in Section \ref{ssec:conceptual}, especially its Part (II). Suffice it to say here that Ref.\ \cite{R91a}, especially Sections 1-3, and Ref.\ \cite{R91b}, especially Section 2, are excellent discussions of these topics, also for the quantum aspects.}

\bigskip

But despite our limiting the  ``quantumness” of our reference frames to ``just” assigning complex amplitudes to tetrads, considered as (spacetime) configurations of an infinitesimal apparatus (implementing the basic idea of Schr\"{o}dinger's wave mechanics):  it will turn out  that  there is plenty to say. As we mentioned, we will leave to $\bf {[I]}$ and  $\bf {[II]}$ the physical consequences and applications of our proposals. But in this paper, our task will be to  formulate quantum reference frames in terms of the proposed bundles. We will also be able, given some care over issues about normalisation, to define expectation values of classical observables, using an elementary treatment in terms of tetrads' coupling to the measured system; ($\bf {[I]}$ will give a fuller treatment of quantum observables).

The plan for the paper is thus as follows. Section \ref{sec:notations}  gives notations and other preliminaries. Then Section \ref{sec:perspsing}  develops the restricted framework, and  Section \ref{sec:basicsing} develops the general framework. Section \ref{sec:conclusion} concludes.  Some calculations justifying our proposals, for example as suitably independent of a fiducial choice, are placed in the Appendix, Section \ref{sec:appdx}.

 \section{Notations for tetrads,  measurement results, and bundles}\label{sec:notations}
 
 We first state some notations and abbreviations that will be used throughout the three papers. 
\begin{enumerate}
\item We use units in which $c = 1$ and the abstract-index notation for tensorial quantities (see, e.g.,
Ref.~\cite{Wald84}). According to this notation, Greek letters are used 
for ``concrete'' indices---i.e., those which assume numerical values---from 
$0$ to $3$. Latin letters $i,j,...$ are used for ``concrete'' indices 
from $1$ to $3$. Finally, Latin letters from the beginning of the alphabet, 
$a,b,...$ (up to $h$), are used for ``abstract'' indices---i.e.\ indices which
do {\it not} assume numerical values but instead indicate, by their position and number,
the type of tensor which the indexed object is. As usual, we will usually suppress indices when they are clear from the context. For example, for a spacetime $({\cal M}, g_{ab})$, with 
$g_{ab}$ the metric tensor  on a  manifold ${\cal M}$, we will  write $({\cal M}, g)$.  

\item Given a spacetime $({\cal M}, g)$, we take the classical  notion of a reference frame at a point  $p \in {\cal M}$ to be a tetrad $\{{\bf e}_\mu^a(p)\}_{\mu = 0,1,2,3}$. 
Here, using abstract-index notation: $a$ simply says that each element of 
$\{{\bf e}_\mu^a(p)\}_{\mu = 0,1,2,3}$ is a tensor of type $(1,0)$---i.e.\ a 4-vector---and $\mu$ labels the four elements of this set of 
4-vectors.\footnote{This concrete index $\mu$  in ${\bf e}_\mu^a$ 
must not be confused with the concrete indices which label components of 
tensors. Some authors avoid this confusion by using Latin capital letters
in the former case, writing, e.g., ${\bf e}_A^a$. On the other hand, Latin capital
letters are also often used to represent algebra-valued quantities. So, here, 
we stick to the simpler index notation which only distinguishes between
concrete and abstract indices.} So, the fact that
$\{{\bf e}_\mu^a(p)\}_{\mu = 0,1,2,3}$ is a tetrad means that  
$g_{ab}{\bf e}_\mu^a(p){\bf e}_\nu^b(p)
= \eta_{\mu\nu } :={\rm diag}(-1,1,1,1)$. At
$p$, 
${\bf e}_0^a(p)$ sets {\it both} direction {\it and} 
unit for what is to be considered as ``time''---i.e.\ ${\bf e}_0^a(p)$ {\it is}
the 4-velocity of the observer/apparatus which the tetrad is supposed to describe at
$p$---while
$\{{\bf e}_j^a(p)\}_{j= 1,2,3}$ then gives the space directions and 
units in such a way that $c =1$ is ensured. 

\item As we said in Section \ref{sec:intro}, we will write ${\cal F}_{\rm o}[g]$ for  the bundle  
of orthonormal frames of $({\cal M},g)$. We also use the notation ${\bf {P_e}} := (p ; \{{\bf e}_\mu^a(p)\})$ for a point of this bundle ``above" the point $p \in {\cal M}$.

\item Similarly to the last two items, as regards the frame bundle $\cal {F(M)}$, i.e.\ the bundle of all bases at tangent spaces to $\cal M$. That is: we will write  $\{{\bf b}_\mu^a(p)\}_{\mu = 0,1,2,3}$ for a  (not necessarily orthonormal) basis of $T_p{\cal M}$ (`b' for `basis').  We also use the notation ${\bf {P_b}} := (p ; \{{\bf b}_\mu^a(p)\})$ for a point of this bundle ``above" the point $p \in {\cal M}$.

\end{enumerate}

We also need to give more details about how in classical relativity, one uses a tetrad to represent measurement results at a spacetime point. So these details,  in items 1 and 2 below, assume a spacetime $({\cal M}, g)$ is given; and should be understood in the context of its bundle ${\cal F}_{\rm o}[g]$.

\begin{enumerate}
\item We can represent measurements of spatial and temporal intervals (more precisely, infinitesimal intervals) by inner products, as defined by the metric, of the tetrad at the point $p$ with appropriate vectors. 

Given two infinitesimally close points  $p, q$---i.e.\ joined by a curve with tangent vector $\xi^a$ with an infinitesimal parameter difference $d\lambda$---the time separation and spatial separations (along each direction) of $p$ and $q$, according to the apparatus at $p$ described by the tetrad  ${{\bf e}_\mu^a(p)}$ (whose dual basis is ${{\bf E}^\mu_a(p)}$) is:
\begin{equation}\label{basicinnerprod}
d\tau = -d\lambda\, g_{ab} {\bf e}_0^a(p) \xi^b = d\lambda\, {\bf E}^0_a(p) \xi^a  \; \; ; \; \;
dl_j = d\lambda \, g_{ab} {\bf e}_j^a(p) \xi^b = d\lambda\, {\bf E}^j_a(p) \xi^a.
\end{equation}
These values are of course independent of the parameterization of the curve joining $p$ and $q$, since the tangent vector $\xi^a$ is $(d/d\lambda)^a$. In flat spacetime, $d\lambda \,\xi^a$ is the separation vector between $p$ and $q$.  But allowing for the space to be curved, these expressions have to be infinitesimal. One can use the identity $g_{ab} {\bf e}^b_\mu = \eta_{\mu \nu} {\bf E}^\nu_a$ to relate these expressions.\footnote{Note that the expressions in terms of the dual basis ${{\bf E}^\mu_a}$ are more easily generalized to the case where we superpose metrics, as in $\bf {[II]}$. Thus writing $\bf b$ for a generic basis, as in item 4 just above, and so $\bf B$ for its dual basis: ${{\bf E}^\mu_a} \to {{\bf B}^\mu_a}$, the dual basis w.r.t.\ ${{\bf b}_\mu^a}$.}\\

\item The main conceptual point about Eq.\ \ref{basicinnerprod} is the {\it coupling} of the tetrad, representing the metric, to $\xi^a$. For this is a template for how a tetrad's coupling can represent measuring many other quantities: namely, by coupling to the quantity concerned.  For instance, an apparatus described by the tetrad $\{{\bf e}_\mu^a(p)\}_{\mu = 0,1,2,3}$, which measures the classical electric
field at $p$ along the spatial direction ${\bf e}_j^a$, is modelled by the coupling  $F_{ab}(p){\bf e}^a_j(p) {\bf e}^b_0(p)$, where $F_{ab}$ is the Faraday tensor  that encodes all the information about the electromagnetic field of the system
being measured. Were we measuring the energy density, then the measuring procedure would be modelled by the coupling $T_{ab}(p){\bf e}^a_0(p) {\bf e}^b_0(p)$, with $T_{ab}(p)$ being the system’s stress-energy-momentum tensor at $p$. Furthermore, it can be argued that this modelling of measurement by a tetrad’s coupling to the system is adequate to encode whatever the observer’s apparatus happens to be, because all measurements reduce, at bottom, to measurements of lengths and time-intervals; cf.\ for example Ref.~\cite{MPSV23}. \end{enumerate}

\subsection{Frames in terms of fibre bundles}\label{ssec:bundle}
In this Subsection, we will first use  fibre bundle ideas to formulate the notion of a classical reference frame; which will return us to the single-vs.-field contrast in (B) at the end of Section \ref{sec:intro}. Then we discuss  our quantum reference frames, first restricted and then general. But for these quantum reference frames, this Subsection's emphasis will be on just the frames' classical configuration spaces. Details about complex amplitudes, and so normalisation etc.\ will be in later Sections.

In classical relativity theory, the usual notion of a reference frame is---not a single tetrad as we have so far mostly  considered, but---a {\it field} of tetrads spread across space or spacetime. Here, the phrase `space or spacetime'  signals  the need to avoid confusing spatial and spacetime notions (`3+1' and `4'). But in line with our setting aside the dynamics of frames, as explained in (C) and (D) at the end of Section \ref{sec:intro}, we will mostly think of spacetime. 

For the bundle ${\cal F}_{\rm o}[g]$  of orthonormal frames of $({\cal M},g)$, we will often write a  point in the bundle above $p \in {\cal M}$ as ${\bf {P_e}} := (p ; \{{\bf e}_\mu^a(p)\})$. We note that the structure group of ${\cal F}_{\rm o}[g]$ is the Lorentz group. Since for a  4-dimensional spacetime $({\cal M}, g)$, the Lorentz group is 6-dimensional (three for rotations and three for boosts), ${\cal F}_{\rm o}[g]$ has dimension 4+6 = 10: (4 for the base-manifold,  6 for a fibre).

So given a spacetime $({\cal M}, g)$, a classical  reference frame on a region ${\cal O} \subset {\cal M}$ is a smooth assignment to each $p \in {\cal O}$ of a tetrad there. That is: the frame is a smooth {\it local section} of the bundle ${\cal F}_{\rm o}[g]$ 
of orthonormal frames of $({\cal M},g)$. Such a section is a smooth mapping $\sigma:{\cal O}\to {\cal F}_{\rm o}[g]$, such that $\pi_{\rm o}\circ\sigma = {\rm id}_{\cal O}$ ($\pi_{\rm o}:{\cal F}_{\rm o}\to {\cal M}$ being the canonical projection of the bundle
and ${\rm id}_{\cal O}$ the identity map on ${\cal O}$). 

 We will write the set of these local sections over $\cal O$ as $Sec_{\cal O}{\cal F}_{\rm o}[g]$.  We note that $Sec_{\cal O}{\cal F}_{\rm o}[g]$ is an infinite-dimensional  space.  We will mostly think of a section, not as a smooth mapping, but as the mapping's range; i.e.\ as the embedded (4-dimensional) manifold, a subset of the bundle ``above" the base-manifold.

The notion of a section, and the notation $Sec$, will have two main roles in this paper, and in $\bf [I]$ and  $\bf [II]$. The first is minor, and we state it mainly to avoid confusion; the second will be  important. \\
\indent \indent \indent \indent \indent (i): {\it If} we were to think of a (local) classical reference frame on $({\cal M}, g)$ as a continuous field on $\cal O$ of infinitesimal apparatuses, then $Sec_{\cal O}{\cal F}_{\rm o}[g]$ could be called the frame's (i.e.\ the continuous field's) configuration space.  (We say `{\it could} be called' because `configuration space' usually means the set of possible {\it instantaneous} classical configurations. But as discussed, we will work (almost entirely) with spacetime (``4D-thinking"), not with space (``(3+1)-thinking); so we will  abjure this connotation of `configuration space'.\footnote{\label{config}{One might suggest we should replace the jargon `configuration space' with e.g.\ `history space'. But we will resist this (even for a classical reference frame) since `history space' suggests: (i) that each point of the space represents a history, which is not so; and-or (ii) that the theory gives details about dynamics---which as we admitted in (C) and (D) at the end of Sec \ref{sec:intro}, we do not give.}}) But in (B) at the end of Sec \ref{sec:intro}, we limited ourselves (here, and in $\bf [I]$ and  $\bf [II]$) to what we called the single conception, not the field conception: i.e.\ to a single infinitesimal apparatus. So we will not have much use for this interpretation of a set of sections; in particular, of $Sec_{\cal O}{\cal F}_{\rm o}[g]$  as a configuration space of a field of apparatuses.\\
\indent \indent \indent \indent \indent (ii): As we briefly explained at the start of Section \ref{sec:intro}, and will see in more detail in a moment: In our {\it general} framework, which uses $\cal {F(M)}$ not ${\cal F}_{\rm o}[g]$, we use a section  to also specify a geometry, i.e.\ a metric on $\cal M$. 

\bigskip

We now introduce the two frameworks in more detail. The general framework will need a  more extended introduction.

\bigskip

\noindent (A): {\it The restricted framework}:--- \\
 In order to work locally, we begin by assuming that the single infinitesimal (in its space and time extensions) apparatus  probing the spacetime $({\cal M}, g)$ is confined to---and possibly,   due to quantum uncertainty in its location, ``spread'' over---the region  ${\cal O} \subset {\cal M}$. Then the classical configuration space for the apparatus is the set $\pi^{-1}_{\rm o}[{\cal O}]$. That is: the ``column" subset of the bundle ${\cal F}_{\rm o}[g]$ that stands above the region  ${\cal O}$ of the base-manifold ${\cal M}$.\footnote{\label{config2}{We note again that our phrase `configuration space' is a little strained. Agreed: the elements of the space are instantaneous configurations in the sense that they are ``very short-lived", i.e.\ at a spacetime point. But  a usual connotation of `instantaneous configuration space' is lacking. For  not  every curve through this space represents a kinematically possible history; (since countless curves will project down by $\pi_{\rm o}$ to  curves in $\cal M$ that are spacelike). Compare the preceding footnote.}}  So we say:
\begin{quote} 
On {\bf the restricted framework} (again: for a single apparatus) that we will expound in Sec.\ \ref{sec:perspsing}: \\
A quantum reference frame (a r-QRF) on $\cal O$ will be an assignment of a complex amplitude to every element of this column, $\pi^{-1}_{\rm o}[{\cal O}]$. Writing the r-QRF as $\Psi$ and ${\bf {P_e}} := (p ; \{{\bf e}_\mu^a(p)\})$ for a point of this column, the complex amplitude will be: $\Psi({\bf {P_e}}) \equiv \Psi(p ; \{{\bf e}_\mu^a(p)\})$.\footnote{\label{avoidstate1}{Agreed: a more careful jargon would be that {\it the state of} the quantum reference frame is such an assignment, with the quantum reference frame being {\it the system} whose state-space is the set of such assignments. But since this paper sets aside the dynamics of quantum reference frames (cf.\ the end of  Sec.\ \ref{sec:intro}), it will be briefer, and just as clear, to call such an assignment `a quantum reference frame', rather than `a state of a quantum reference frame'. 

Note incidentally that a similar abbreviation happens in the classical context. For a local section of the bundle ${\cal F}_{\rm o}[g]$ really represents {\it the state} (motion and spatial orientation) of a system comprising continuously many infinitesimal apparatuses. But for brevity, the literature calls the section `a frame' not `a state of a frame'.}}
\end{quote}

\bigskip

\noindent (B): {\it The general framework}:--- \\
In order to describe different spacetime geometries, and thereby superpositions of them and superposition of perspectives on them, the key point we need to recall from differential geometry is as follows. Given a manifold $\cal M$, one can define a geometry i.e.\ a Lorentzian metric $g$ on $\cal M$, by specifying a field of bases at the tangent spaces. That is, one says that at each spacetime point, the basis one has specified there is stipulated to be orthonormal in the geometry one is thereby defining. 

Here, we should again note the qualification we stated in Section \ref{sec:intro} (see (i) and (ii) and (i') and (ii'), before (A)). Namely:  since (i) a  manifold admits a Lorentzian metric iff it admits a continuous nowhere-vanishing vector field, and (ii) a manifold admits a global basis field iff its tangent bundle is trivial (called being {\it parallelizable}), our strategy for defining a Lorentzian metric depends on (i') the manifold having feature (i) and on (ii') our ``working  locally", confining all our constructions to  regions ${\cal O} \subset {\cal M}$ which for any physical applications can be chosen small enough to guarantee feature (ii) for $\pi^{-1}[{\cal O}]\subset {\cal F}({\cal M})$ (where $\pi:{\cal F}({\cal M})\to{\cal M}$ is the canonical projection of the bundle ${\cal F}({\cal M})$): this is only a mild qualification.

So let us express this strategy for defining a Lorentzian metric, in terms of  the  frame bundle. For clarity and simplicity, we will cast the discussion in terms of the bundle $\cal {F(M)}$. But if $\cal M$ is not parallelizable, then as just mentioned, we simply confine our constructions to $\cal {F(O)}$ for an appropriate region ${\cal O} \subset {\cal M}$. 
In $\cal {F(M)}$ the fibre above $p \in {\cal M}$ is the set of all {\it bases} of $T_p{\cal M}$. So the structure group of ${\cal F}({\cal M})$ is the general linear group, and the bundle makes no assumption that there is a metric. (Despite its name! Cf.\ footnote \ref{sadjargon}.)   Since the general linear group for a 4-dimensional spacetime is 16-dimensional ($4 \times 4$), this fibre bundle has 16-dimensional fibres, and its total dimension is 4+16 = 20.  So we {\it define} a metric, i.e.\ a spacetime $({\cal M}, g)$, by specifying a {\it section} of ${\cal F}({\cal M})$. To be precise: we specify a ``covering by compatible sections'': i.e.\
a set of sections $\{\sigma_\alpha\}_{\alpha\in{\cal I}}$ defined for a
covering $\{{\cal O}_\alpha\}_{\alpha\in {\cal I}}$ of ${\cal M}$, where each $\sigma_\alpha$ is a section of 
$\pi^{-1}[{\cal O}_\alpha]$, such that 
on any non-empty intersection $\pi^{-1}[{\cal O}_\alpha]\cap\pi^{-1}[{\cal O}_\beta]$,  the sections $\sigma_\alpha$ and $\sigma_\beta$ are fibre-wise related by a Lorentz transformation. (This is what we meant,  before (A) in Section I,  by ``defining a metric on ${\cal M}$ `patch by patch' \!''.)

A section of $\cal {F(M)}$, i.e.\ a smooth choice of one element from each fibre, thus represents two items, taken together. Namely: (i) a spacetime, in the usual physicists’ sense of a manifold equipped with a Lorentzian metric, taken together with: (ii) a tetrad field, i.e.\ a smooth assignment to each $p \in {\cal M}$ of a basis that is orthonormal according to the metric in (i). So (ii) represents a (classical) reference frame on the  spacetime (i); and (ii)
is the field of bases that we used to define the geometry (i), by stipulating them to be orthonormal.

It is helpful to express the ideas of the last paragraph a bit more formally. For we will thereby see the relationship to the restricted framework's bundle ${\cal F}_{\rm o}[g]$, with its given metric $g$.  In the set $Sec({\cal F(M)})$  of sections of $\cal F(M)$ (with sections understood as embedded sub-manifolds of $\cal F(M)$), the relation `is a pointwise Lorentz transform of’ is of course an equivalence relation. (The set $Sec({\cal F(M)})$  of sections is of course infinite-dimensional.) Any two sections  in the same equivalence class generate the same metric $g$, and so the same spacetime (Lorentzian manifold), i.e.\ $({\cal M}, g)$.\footnote{\label{HintOfHole}But note that this is {\it not} the only way that different sections of ${\cal F}({\cal M})$ generate the same {\it geometry}---in the (widespread) sense of `geometry' according to which any two isometric Lorentzian manifolds using a common base-manifold $\cal M$ encode the same geometry, though they differ about which manifold point has which metrical properties. That is: sections related by a push-forward map induced by a diffeomorphism of ${\cal M}$ also represent the same geometry. This feature is a consequence of diffeomorphism invariance. We will discuss it further at the end of this Section, at the start of Section \ref{sec:basicsing}, and in more detail in  [{\bf II}].} Indeed, again taking sections as embedded sub-manifolds of ${\cal F(M)}$: the union of these sections (considered as a subset of $\cal F(M))$ is precisely ${\cal F}_{\rm o}[g]$, for the metric $g$ that is defined in the above manner by any of the sections. So in terms of items (i) and (ii) in the preceding paragraph: (i) is an equivalence class, and  by ``looking inside” it, we see its elements, each of which is a tetrad field according to the geometry (i).

Thus we see that $Sec({\cal F(M)})$ (taken as the set of embedded sub-manifolds of $\cal F(M)$) is partitioned into continuously many equivalence classes. Each class is labelled by a Lorentzian metric  $g$ on $\cal M$. And each class defines (just by the set-theoretic union of its elements) the orthonormal frame bundle ${\cal F}_{\rm o}[g]$ for that metric.

 As we mentioned, we have for clarity and simplicity cast  this discussion  in terms of {\it global} sections. But of course, parallel remarks apply to local sections. So from now on (and in later Sections) we will almost always work locally, with a region  ${\cal O} \subset {\cal M}$ that is assumed to meet the requirements above, i.e.\ to be parallelizable and so admit a nowhere-vanishing vector field.   So we now state the basic idea of a general quantum reference frame, a g-QRF, as much as possible in parallel to the discussion above of a restricted quantum reference frame (r-QRF); as follows.
 
We begin by assuming that the single (spatially- and temporally-)infinitesimal apparatus  is confined to the region  ${\cal O} \subset {\cal M}$---again due to spacetime location uncertainty. And now that we allow for different geometries, we define the apparatus' classical configuration space using $\cal {F(M)}$, not any specific ${\cal F}_{\rm o}[g]$: or rather, we use  $\cal {F(O)}$. So the configuration space is the set ${\cal {F(O)}} = \pi^{-1}[{\cal O}] \subset {\cal {F(M)}}$. In words, it is the ``column" subset of the bundle $\cal F(M)$ that stands above the region  ${\cal O}$. So we say:
\begin{quote} 
On {\bf the general framework} (again: for a single apparatus) that we will expound in Sec.\ \ref{sec:basicsing}:\\
A  quantum reference frame (a g-QRF) on $\cal O$ will be an assignment of a complex amplitude to every element of the column, $\pi^{-1}[{\cal O}]\subset {\cal {F(M)}}$. Writing the g-QRF as $\Psi$ and ${\bf {P_b}} := (p ; \{{\bf b}_\mu^a(p)\})$ for a point of this column (again: `b' for `basis'), the complex amplitude will be: $\Psi({\bf {P_b}}) \equiv \Psi(p ; \{{\bf b}_\mu^a(p)\})$.\footnote{Of course, the comments in footnotes \ref{config}, \ref{config2} and \ref{avoidstate1}, about our using the phrase `configuration space' and our not using the phrase `state of quantum reference frame', apply again here in the general framework.} 
\end{quote}

We can now make two interpretative comments about the general framework, which we will expand on at the start of Section \ref{sec:basicsing}. The first is physical, the second philosophical.

\bigskip

First,  the amplitude assigned by a g-QRF to a point  ${\bf {P_b}} \equiv (p ; \{{\bf b}_\mu^a(p)\})$ is of course the amplitude for the apparatus to be (i) at the point $p$ and (ii) in the configuration $\{{\bf b}_\mu^a(p)\}$. But (ii) deserves a comment. For (a) we now work with $\cal {F(O)}$, even while (b) the apparatus is to be a tetrad, i.e.\ to determine three orthogonal spatial directions, all orthogonal to a timelike direction. So by (ii) we mean  `the configuration $\{{\bf b}_\mu^a(p)\}$  understood as orthonormal according to some or other {\it single} metric on $\cal O$'.

\bigskip

Second,  the above discussion makes it clear that our formalism seems to be what philosophers call {\it haecceitistic} in how it treats a spacetime point $p$  existing in two spacetimes.  We will say more about this at the start of Section \ref{sec:basicsing}. But there is an obvious concern about such a treatment that we should note already here. For it implies that  one and the same point $p$ can exist in two spacetimes, while having very different metrical properties and relations in them. This allowance is the doctrine that philosophers call {\it haecceitism}. (In this comment, it will be clearer to revert to the usual idea of entire spacetimes, and so the notation $\cal M$, not our local region  $\cal O$; but the discussion is unaffected.)

Indeed, this allowance can happen in either of two ways:\\
\indent \indent \indent  (i): the two spacetimes using the common base-manifold $\cal M$ are isometric; they ``merely" assign their common geometry's metrical properties and relations differently across the points of $\cal M$ (cf.\ footnote \ref{HintOfHole});\\
\indent \indent \indent  (ii): the two spacetimes using the common base-manifold $\cal M$ are not isometric.\\
In case (i), the allowance  is widely regarded as wrong. For  in view of Einstein's famous hole argument of 1913, it surely makes general relativity indeterministic---in a way that is surely spurious. For the indeterminism's different (but isometric!) possibilities differ only as regards which spacetime point has which metrical properties and relations: which is surely unphysical. And in case (ii), the allowance  is widely regarded as problematic. For  when considering two non-isometric spacetimes, it is not at all clear how one should answer questions like ``Which spacetime point in this spacetime is (or in some good sense: corresponds to) this point in this other spacetime?". So our formalism's treating a spacetime point $p$ as  existing in two, indeed many, such spacetimes---as ``sitting under" its fibre,  the selfsame point enjoying diverse incompatible metrical properties and relations in the countless non-isometric spacetimes, $({\cal M}, g^1)$, $({\cal M}, g^2),...,$ that are specified by countlessly many sections of $\cal {F(M)}$---needs to be   justified.

To this, we reply: Fair comment.  But we will see at the start of Section \ref{sec:basicsing} that our haecceitism is only apparent. It is a formal convenience, to simplify our formalism. Besides, we will see that our formalism is flexible---for both case (i) and case (ii)---about how one resolves questions, when comparing two spacetimes, about ``which spacetime point in this spacetime is (or corresponds to) which point in this other spacetime?".

\section{The restricted framework}\label{sec:perspsing}

In this Section, we will: (i) formally define a quantum reference frame, on the restricted framework, and make some comments about it (Sec \ref{ssec:ps:definition}); (ii) discuss normalisation, and the (partial) analogy with a classical probability distribution  on tetrads (Sec \ref{ssec:ps:normn}); (iii) write down the ensuing expectation values of measurement results  (Sec \ref{ssec:ps:expnvals}).\\

\subsection{The definition}\label{ssec:ps:definition}
We adopt the restricted framework, and consider a single apparatus (as in (B) of Section \ref{sec:intro}). So we assume a spacetime with a given metric $({\cal M},g)$, and its bundle of orthonormal frames ${\cal F}_{\rm o}[g]$. We model a single quantum apparatus as an assignment $\Psi$ of complex amplitudes to tetrads at points $p$ in a region ${\cal O} \subset {\cal M}$.  We shall refer to this
notion of quantum reference frame (QRF) as a  `r-QRF'.  So  we {\bf define}: 
\begin{quote}
A {\bf r-quantum reference frame} (r-QRF) $\Psi$ on a region ${\cal O} \subset {\cal M}$ is an assignment of a complex amplitude (possibly zero) to every element of the column, $\pi^{-1}_{\rm o}[{\cal O}]$, of the orthonormal frame bundle above $\cal O$. Writing ${\bf {P_e}} := (p ; \{{\bf e}_\mu^a(p)\})$ for a point of this column, the complex amplitude is: $\Psi({\bf {P_e}}) \equiv \Psi(p ; \{{\bf e}_\mu^a(p)\})$. This assignment is also required to obey the normalisation condition Eq.\ \ref{norm20Oct} in Section \ref{ssec:ps:normn} below; and its fibre integrals are required to be finite, Eq.\ \ref{normpjfiducialfnoinfy}.\footnote{For our reason for calling such an assignment a `frame', rather than a `state of a frame', cf.\ footnote \ref{avoidstate1}.}
\end{quote}
To put this in the language of bundles, we recall  that $\pi_{\rm o}:{\cal F}_{\rm o}[g]\to {\cal M}$ is the canonical projection of the bundle ${\cal F}_{\rm o}[g]$. So a r-QRF $\Psi$ on $\cal O$ is an assignment of  amplitudes to all the elements of all the fibres above $\cal O$. That is: $\Psi:{\cal F}_{\rm o}[g] \supset \pi^{-1}_{\rm o}[{\cal O}] \to {\mathbb C}$.

Thus for a global reference frame, we can write more simply---by identifying ${\cal F}_{\rm o}[g] $ with $\pi^{-1}_{\rm o}[{\cal M}]$---that a r-QRF $\Psi$ on $\cal M$ is an assignment of  amplitudes to all  elements of the bundle. That is:  $\Psi:{\cal F}_{\rm o}[g] \to {\mathbb C}$. Notice that globally-defined r-QRFs do not
depend on the existence of globally-defined continuous sections of ${\cal F}_{\rm o}[g]$---i.e., on ${\cal F}_{\rm o}[g]$ being a trivial bundle (equivalently, on ${\cal M}$ being parallelizable).

Since two tetrads at the point $p$ are Lorentz transforms of each other, we can write the amplitude for a Lorentz transform of the tetrad $\{{\bf e}_\mu^a(p)\}_{\mu = 0,1,2,3}$ as: $ \Psi({\bf P}_{\Lambda{\bf e}}):= \Psi(p;\{\Lambda^\nu_\mu{\bf e}^a_\nu(p)\})$. Here, $\Lambda\in O(3,1)$, the Lorentz group; i.e.\ the entries $\Lambda^\nu_\mu$ of matrix $\Lambda$ satisfy
$\Lambda^\alpha_\mu\Lambda^\beta_\nu \eta_{\alpha\beta} = \eta_{\mu\nu}$. So with elementary wave mechanics in mind, we intuitively expect that  $|\Psi({\bf P}_{\Lambda{\bf e}})|^2$ will be a Born-rule probability density for the apparatus to be at $p$, and with its internal configuration (spatial orientation and state of motion) being the tetrad $\{\Lambda^\nu_\mu{\bf e}^a_\nu(p)\}$. This will indeed be so, {\it modulo} some comments in Sec \ref{ssec:ps:normn} below about our needing care over normalisation and the apparatus’ possible dynamics.

Before those comments, this definition prompts two immediate remarks.

(A): We stress again that the concept of an r-QRF introduces quantum uncertainties on measurements of spatial distances and temporal intervals; but without making the metric
tensor uncertain. So it expresses the idea of superpositions of ``perspectives" or ``points of view''---
due to the possible configurations of the apparatus at each point---on a
well-defined spacetime geometry. 

(B): Furthermore, one can consider linear combinations of r-QRFs 
in the obvious way, i.e.\ point-wise in  ${\cal F}_{\rm o}[g]$.  Thus for a given region ${\cal O} \subset {\cal M}$, the r-QRFs on $\cal O$  are the normalised elements of a complex vector space of ``wave-functions" (a linear combination being re-normalised, as usual in quantum mechanics), each of them defined on the union of the fibres above $\cal O$, i.e.\ on $\pi_{\rm o}^{-1}[{\cal O}]$.

\subsection{Normalisation and classical analogy}\label{ssec:ps:normn}

As we said in the last Subsection: we intuitively expect  $|\Psi({\bf P}_{{\bf e}})|^2$ to be the Born-rule probability density for the apparatus to be at $p$, and with its configuration (i.e.\ spatial orientation and state of motion) being the tetrad  $\{{\bf e}^a_\mu(p)\}$. Obviously, this interpretation can only make sense if $\Psi$ is normalized,  not only in what one might
call ``space'', but also in ``time'' {\it and} in the apparatus's
possible ``orientations'' and ``states of motion'' (i.e.,  elements of the fibres). 
This contrasts with the usual idea that a quantum system, in particular an apparatus: (i) can have arbitrary extension in the time direction, which implies that there is no rationale for  imposing normalization of its state (say, its wave-function) {\it along} its existence; though it (ii) should be normalized over space, to reflect the idea that ``at any time, it must be somewhere". However, our notion of r-QRF as a function on 
$\pi_{\rm o}^{-1}[{\cal O}]\subset {\cal F}_{\rm o}[g]$ takes the apparatus to be infinitesimal both in
spatial and  in time extensions---a ``single-instant,  point-like apparatus''---with the support of $\Psi$ projected on ${\cal M}$,  $\pi_{\rm o}[{\rm supp}(\Psi)]\subseteq {\cal O}$, representing {\it not} the classical  extension of a spacetime-extended apparatus, but the  {\it uncertainty} of this infinitesimal apparatus's  spacetime location. Thus  an apparatus with a time-like extension (though spatially infinitesimal) could presumably be modelled by a one-parameter family of our r-QRFs. The family might be written as $\Psi_\tau$, where the parameter  $\tau$ is an appropriate internal degree of freedom, such as the apparatus's proper time. In this model, a r-QRF would represent the uncertainty of  the apparatus at a given value of this internal parameter. Since at each value of this parameter it must be ``somewhere/sometime", the spacetime normalization would be justified for each r-QRF.

So we now state how an r-QRF will be normalised. We first state the condition, and then make  four comments. We require that an r-QRF $\Psi$ on $\cal O$ be normalised in the following sense:
\begin{eqnarray}
\fbox{ $
 \|\Psi\|^2:=\int_{\cal O} \epsilon(p) \int_{O(3,1)}d\mu_L(\Lambda) \left|\Psi(p;\{\Lambda^\nu_\mu{\bf e}^a_\nu(p)\})\right|^2 = 1 $ }
\label{norm20Oct}
\end{eqnarray} 
for   {\it any fixed} (fiducial) tetrad field $\{{\bf e}_\mu^a\}_{\mu = 0,1,2,3}$ on $\cal O$;
where (i) $\epsilon$  is  the positive preferred 4-volume
element selected by the metric $g_{ab}$, i.e.\ $\epsilon = \sqrt{|{\rm det}(g_{\mu\nu})|} dx^0 dx^1 dx^2 dx^3$ in an arbitrary  coordinate system $p\mapsto x^\mu(p)$; (ii) $\Lambda\in O(3,1)$;
and (iii) $\mu_L$ is the (unique up to 
a multiplicative factor) invariant Haar 
measure defined on $O(3,1)$.  In other words, $\Psi\in L^2({\cal F}_{\rm o}[g])$ (with $\epsilon$ and $\mu_L$ understood in the $L^2$ notation); and we require that $\Psi$ has norm 1.\footnote{\label{footdelta}A more general treatment could enlarge the space of QRFs to {\it distributions} on ${\cal F}_{\rm o}[g]$ (e.g.\ via a rigged Hilbert space), so as to include idealised, sharply peaked QRFs---analogous to the ``position eigenstates'' $|x\rangle$ of quantum mechanics. Note, however, that quantities quadratic in $\Psi$, such as the normalisation Eq.~\ref{norm20Oct} and the expectation values below, are not defined for such distributions (e.g.\ $|\delta|^2$ makes no sense). A classical observer at $(p;\{{\bf e}_\mu^a(p)\})$ is better represented as a limit of normalised $\Psi_n\in L^2({\cal F}_{\rm o}[g])$ whose probability measures $|\Psi_n|^2\,\epsilon\, d\mu_L$ converge (weakly) to the Dirac measure at that point; the expectation values below then converge to the corresponding classical values. That is: the $\delta$ belongs to the probability measure, not to the amplitude. We will work with $L^2({\cal F}_{\rm o}[g])$ throughout. It will be clear in Section \ref{sec:basicsing} that this comment applies {\it mutatis mutandis} to the general framework.}

\bigskip

There are four comments to make about this normalisation condition. The first two are mathematical and the third and fourth are about probability. \\
 
(1): The first mathematical topic to address is the  positive volume element $\epsilon$.   Since $\epsilon$ is determined by the metric $g$, and all elements of the fibre are tetrads of this same metric, $\epsilon(p)$ is independent of the point on the fibre (consistently with $|{\rm det} \Lambda|=1$ for $\Lambda \in O(3,1)$) and so we take it out of the rightmost integral.\footnote{We note that since $\cal O$ is parallelizable, it is orientable and {\it does} admit a volume element.}\\

(2): The second mathematical topic is the
 invariant Haar measure $\mu_L$ on $O(3,1)$. We said `the': but it is only unique up to 
a multiplicative factor. So from a purely mathematical viewpoint, the fibre integral in Eq.\ \ref{norm20Oct} could  use different measures at different $p \in {\cal O}$, differing from each other by a multiplicative factor. 
However, the fixed fiducial tetrad field $\{{\bf e}_\mu^a\}$---which is a section of $\pi_{\rm o}^{-1}[{\cal O}]$---used to make sense of the fibre integration in Eq.~\ref{norm20Oct},  provides a corresponding identification among measures in different fibres. This is so because
the Lorentz group $O(3,1)$ acts freely and transitively on 
each fiber $\pi^{-1}_{\rm o}(p)$, which means that each fibre is a ``copy'' of  $O(3,1)$ with the identity ``rubbed out''.\footnote{In technical terms, each fibre is a {\it torsor}.  It will be clear in Section~\ref{sec:basicsing} that this comment applies to the general framework, just substituting $GL(4)$ for  $O(3,1)$. }
The role of the fiducial tetrad field is exactly to select, at each $p\in {\cal O}$, the fibre element
$(p;\{{\bf e}_\mu^a(p)\})$ which is to be identified with the group identity.  Once this is done,  all fibers become
uniquely identified among themselves and to $O(3,1)$: $(p;\{\Lambda^\nu_\mu{\bf e}_\nu^a(p)\})$ is identified to $(q;\{\Lambda^\nu_\mu{\bf e}_\nu^a(q)\})$ and to $\Lambda$,  for all $p,q\in{\cal O}$, $\Lambda \in O(3,1)$.
Hence,  fixing one invariant Haar measure on $O(3,1)$ fixes the integration measure on all fibres of $\pi_{\rm o}^{-1}[{\cal O}]$.  Note that this is already built into Eq.~\ref{norm20Oct} where the fibre integration is
implemented by the integration over the Lorentz group.

But the discussion above prompts a natural question: what if a different fiducial tetrad field,  say 
$\{\bar{\bf e}_\mu^a\}$,  were used in Eq.~\ref{norm20Oct}? To what extent is our 
formalism dependent on such an arbitrary choice? We show,  in what follows, that Eq.~\ref{norm20Oct}
is actually  {\it invariant} under change of fiducial tetrad field---as one wants, for
a physical quantity under a change of an arbitrary convention.

 We begin by relating the two fiducial tetrad fields.  Let $\bar{\Lambda}(p)\in O(3,1)$ be defined by
$\bar{\bf e}_\mu^a(p) = \bar{\Lambda}_\mu^\nu(p){\bf e}_\nu^a(p)$. Hence,  
$\Lambda_\mu^\nu\bar{\bf e}_\nu^a(p) =\Lambda_\mu^\nu \bar{\Lambda}_\nu^\alpha(p){\bf e}_\alpha^a(p)
=[\bar{\Lambda}(p)\Lambda]^\alpha_\mu {\bf e}_\alpha^a(p)$. So, considering the innermost integration
in Eq.~\ref{norm20Oct},  but with $\{{\bf e}_\alpha^a\}$ substituted by $\{\bar{\bf e}_\alpha^a\}$, we have:
\begin{eqnarray}
\int_{O(3,1)}d\mu_{L}(\Lambda) \left|\Psi(p;\{\Lambda^\nu_\mu\bar{\bf e}^a_\nu(p)\})\right|^2 &=&\int_{O(3,1)}d\mu_{L}(\Lambda) \left|\Psi(p;\{[\bar{\Lambda}(p)\Lambda]^\alpha_\mu {\bf e}_\alpha^a(p)\})\right|^2\nonumber \\
&=&\int_{O(3,1)}d\mu_{L}(\bar{\Lambda}^{-1}(p)\Lambda') \left|\Psi(p;\{\Lambda'^\alpha_\mu {\bf e}_\alpha^a(p)\})\right|^2
\nonumber \\
&=&\int_{O(3,1)}d\mu_{L}(\Lambda') \left|\Psi(p;\{\Lambda'^\alpha_\mu {\bf e}_\alpha^a(p)\})\right|^2,
\end{eqnarray}
where  in passing from the first to the second line we have defined $\Lambda':=\bar{\Lambda}(p)\Lambda$
and  in passing from the second to the third  line we used that  $\mu_L$ is an invariant measure on $O(3,1)$---meaning that $\mu_L({\Lambda}G) = \mu_L(G)$
for any ${\Lambda}\in O(3,1)$ and $G$ a measurable subset of $O(3,1)$.\footnote{Recall that the Lorentz group is unimodular, meaning that its invariant measures are both left- and right-invariant.} This means that the
quantity\footnote{Note that Eq.~\ref{normpjfiducialfnoinfy}  
{\it defines} the symbol $\|\Psi(p)\|^2$, {\it not} an object $\Psi(p)$.}

\begin{eqnarray}
\fbox{ $
 \|\Psi(p)\|^2:=\int_{O(3,1)}d\mu_{L}(\Lambda) \left|\Psi(p;\{\Lambda^\nu_\mu{\bf e}^a_\nu(p)\})\right|^2 
 $ }
\label{normpjfiducialfnoinfy}
\end{eqnarray}
is independent of the fiducial tetrad field $\{{\bf e}_\mu^a\}$. It can be seen as a marginal probability density
over ${\cal O}$; and we will require it to be finite, in order that we can divide by it, when in Section \ref{ssec:ps:expnvals} below we define expectation values at a point $p$.

  We conclude that it is therefore legitimate to impose our normalisation condition Eq.~\ref{norm20Oct}. \\

(3): Returning to  probability, there are two straightforward comments to make. The first  is about quantum theory in general. Recall that for a single quantity, a probability distribution over its possible values that is given by the Born rule applied to a quantum state, is of course ``wholly classical”. It is only by considering a quantity that does not commute with the given one, that one can see the quantum aspects of probability. So also here. A probability distribution over tetrad configurations given by the Born rule is wholly classical; and so is a lot like postulating a classical probability distribution over tetrads, without any reference to complex amplitudes, a quantum state of a reference frame etc. It will only be in {\bf  [I]},  when we consider quantities in general, and so non-commutation,  that the quantum aspects of our probability distributions will be visible. \\

(4): The second, and less important, comment about probability is just that also for a postulated classical probability distribution, one can of course consider dynamics: some sort of transport equation, like a Fokker-Planck equation. But as explained in (C) and (D) at the end of Sec \ref{sec:intro},  we anyway will set dynamics aside.

\subsection{Expectation values}\label{ssec:ps:expnvals}

We will now combine the Born-rule probability interpretation we have given to  Eq.\ \ref{norm20Oct} with Section \ref{sec:notations}’s discussion of the measurement of various quantities, so as to write down expectation values for measurement results. We begin, in (1) below, with  the measurement of spatial and temporal intervals at a point $p$, as in Section \ref{sec:notations}’s  Eq.\ \ref{basicinnerprod}.  Then in (2) we similarly adapt the classical expressions for the quantities taken as examples in item 2 at the start of of Section \ref{sec:notations}:  the electric field, and the energy density. 
In (2) as in (1), the expectation values will be for measurement results at a given point $p \in {\cal O}$.  We stress that this will mean `the expectation value, conditional on the apparatus firing at that point': which implies that we will  divide by the squared norm of the fibre integral, as defined in Eq.\ \ref{normpjfiducialfnoinfy} (which, recall, we required to be finite).  Then in (3), we write down expectation values for (a measurement occurring somewhere in) a whole region  ${\cal O}$. Finally, in (4), we write down expectation values for a given time, i.e.\ a spatial slice across the region  ${\cal O}$.     \\

We should note an important conceptual point that applies equally to all of (1) to (4). All their expectation values (and similarly, probabilities of individual results), whether for a point, a region or a time, will obviously be sensitive to where in the frame bundle the support of $\Psi$ happens to be---however the state of the measured system, e.g.\ a classical electromagnetic field, happens to vary across spacetime. For the weight $\left|\Psi(p;\{\Lambda^\nu_\mu{\bf e}^a_\nu(p)\})\right|^2$, which in general varies along a fibre and across spacetime,  will be a factor in the integrand of all of (1) to (4)'s  formulas  for expectation values. Of course, that the expectation values are thus sensitive is no surprise. For it reflects how our proposal for QRFs attributes to the frame some physical degrees of freedom\footnote{Albeit without our giving details about the dynamics of a QRF; cf.\ (C) and (D) at the end of Section \ref{sec:intro}.} that couple to the measured system's degrees of freedom, and which therefore contribute to determining expectation values.\\

\bigskip
 
\noindent (1):  {\it The expectation value, at a point $p$, of time and spatial separations}:---

Accordingly, we propose that for two infinitesimally close points  $p$ and $q$, joined by a curve with tangent vector $\xi^a$, the expectation value---according to the r-QRF $\Psi$ that we assume has support at (the fibre above) the point $p$---of their time separation and spatial separations (along each direction) is as follows:
\begin{equation}\label{ps:expntime}
\langle d\tau\rangle= - \|\Psi(p)\|^{-2}
\int_{O(3,1)}d\mu_L(\Lambda) 
g_{ab}(p){\bf e}_\alpha^a \Lambda_0^\alpha d\lambda\, {\xi}^b
\left|\Psi({\bf P}_{\Lambda{\bf e}})\right|^2 	
= 
\|\Psi(p)\|^{-2}
\int_{O(3,1)}d\mu_L(\Lambda) 
(\Lambda^{-1})^0_\alpha{\bf E}^\alpha_a  d\lambda\, {\xi}^a
\left|\Psi({\bf P}_{\Lambda{\bf e}})\right|^2
\end{equation}
and
\begin{equation}\label{ps:expnlength}
\langle dl_j\rangle =  \|\Psi(p)\|^{-2}
\int_{O(3,1)}d\mu_L(\Lambda) 
g_{ab}(p){\bf e}_\alpha^a \Lambda_j^\alpha d\lambda \,{\xi}^b
\left|\Psi({\bf P}_{\Lambda{\bf e}})\right|^2 	
= \|\Psi(p)\|^{-2}
\int_{O(3,1)}d\mu_L(\Lambda) 
 (\Lambda^{-1})^j_\alpha{\bf E}^\alpha_a d\lambda \,{\xi}^a
\left|\Psi({\bf P}_{\Lambda{\bf e}})\right|^2 .
\end{equation}
(Note that $j=1,2,3$ in the latter equation is just a concrete index labeling the different equations, 
as is $0$ in the former one; they do not need to match positions on both sides of the equation.) 
In passing from the first to the second equalities in each one of the expressions above, 
we used the identity 
$\eta_{\alpha\beta} \Lambda^\beta_\gamma= \eta_{\mu\beta}\Lambda^\beta_\gamma \Lambda^\mu_\nu (\Lambda^{-1})^\nu_\alpha =\eta_{\gamma\nu} (\Lambda^{-1})^\nu_\alpha $,  which together with the fact that $\{{\bf E}^\mu_a\}$
is the dual basis to the tetrad  $\{{\bf e}_\mu^a\}$ imply
 $ g_{ab} {\bf e}^b_\alpha\Lambda^\alpha_\mu= \eta_{\alpha\beta} {\bf E}^\beta_a\Lambda^\alpha_\mu=\eta_{\alpha\mu}(\Lambda^{-1})^\alpha_\beta{\bf E}^\beta_a\equiv
\eta_{\mu\alpha}(\Lambda^{-1}{\bf E})^\alpha_a$---hence,  
$ g_{ab} {\bf e}^a_\alpha\Lambda^\alpha_0=
-(\Lambda^{-1}{\bf E})^0_a$ and  $ g_{ab} {\bf e}^a_\alpha\Lambda^\alpha_j=
(\Lambda^{-1}{\bf E})^j_a$.
This just confirms that one can express the scalar integrands equally well in terms of the tetrad $\{{\bf e}_\mu^a\}$ or of its dual basis, $\{{\bf E}^\mu_a\}$,  as long as one recalls that if 
${\bf e}^a_\mu \mapsto (\Lambda {\bf e})^a_\mu$,  then the duality relation demands 
${\bf E}_a^\mu \mapsto (\Lambda^{-1}{\bf E})^\mu_a$.  

We stress that here `the expectation value, at a point' means `the expectation value, conditional on the apparatus firing at that point'. Hence equations \ref{ps:expntime} and \ref{ps:expnlength} divide by the squared norm of the fibre integral, as defined in Eq.\ \ref{normpjfiducialfnoinfy} (which, recall, we required to be finite). 

Note also that, since the Lorentz group is non-compact, the integrands of these expectation values are unbounded along the fibre (e.g.\ $(\Lambda^{-1}{\bf E})^0_a\xi^a$ grows like $\cosh\eta$ under a boost of rapidity $\eta$). So the expectation values are finite only if, besides being normalised, $\Psi$ decays sufficiently fast along the fibre---just as, in elementary quantum mechanics, $\langle x\rangle$ requires the wave-function to be in the domain of $\hat{x}$. The same holds for all the expectation values discussed below--and in the general framework, as well as the restricted one.

Note also that here we do {\it not} transform the given vector $d\lambda \,\xi^a$. For in the restricted framework, we are ``only" superposing points of view, i.e.\ apparatuses represented by tetrads. So the given manifold points $p,q$ which determine $d\lambda\, \xi^a$---representing the quantity to be measured---are fixed, and the ``only" question  is how the different apparatus configurations (spatial orientations and states of motion) in the superposition attribute time separation and spatial separation (along the different spatial directions) to this fixed pair of points.\\

\bigskip
 
\noindent (2):  {\it The expectation value, at a point $p$, of a classical  field}:---\\
Let us consider the electric field.  We assume that the electromagnetic field is completely described, at $p$, by the Faraday tensor $F_{ab}$. Then as in item 2 near the start of Section \ref{sec:notations}: for a {\it classical} apparatus passing through $p$ with 4-velocity $u^a$, the electric field measured along the spatial direction given by the unit vector $n^a$ (with $g_{ab}u^a n^b = 0$) is $E_n=F_{ab}n^au^b$.

To adapt this expression to  our framework, we recall that for us,  `expectation value, at a point' means `expectation value, conditional on the apparatus firing at that point'. So as in (1), we will divide by the squared norm of the fibre integral in Eq.\ \ref{normpjfiducialfnoinfy}. Thus for a r-QRF $\Psi$ with support at $p$, we define the expectation value at $p$ for the $j$ component of the electric field by:
\begin{eqnarray}
\langle E_j\rangle|_{p} :=\|\Psi(p)\|^{-2} 
\int_{O(3,1)}d\mu_L(\Lambda) \; F_{ab}{\bf e}_\alpha^a{\bf e}_\beta^b
\Lambda_j^\alpha \Lambda_0^\beta
\left|\Psi({\bf P}_{\Lambda{\bf e}})\right|^2 ,
    \label{EvalueQRFnew}
\end{eqnarray}
where $\{{\bf e}_\mu^a\}$ is any (fixed)  tetrad at $p$; and recall that  ${\bf P}_{\Lambda{\bf e}}$ stands for 
  $ (p;\{\Lambda^\nu_\mu {\bf e}_\nu^a\})$.\footnote{\label{meaningofj}
%{\jtod I suggest we add a clarifying/reassuring sentence,  as a footnote, saying: [DV: OK]
Note that although the apparatus can have various possible spatial orientations,  our phrase  `{\it the} $j$ component' has its meaning fixed by the {\it label} $j$. In effect, we can think of one of the three spatial rods as having `$j$' painted on it.}

 Similarly of course for any classical quantity, $A$ say, whose measurement results can be written in terms of  the coupling between the system and the tetrad characterizing the apparatus.  That is: we would similarly define  the expectation value $\langle A\rangle|_{p}$ at 
$p$ for a r-QRF $\Psi$ with support at $p$, by (i) averaging over the  tetrads  $\{{\bf e}\}$ with the probability density $\left|\Psi(p;\{{\bf e}\})\right|^2$, and (ii) dividing by the squared norm of the fibre integral. 

For example, let us return to energy density $\rho$, as the ``00-component" of the stress-energy-momentum: which was the other example in  item 2 near the start of Section \ref{sec:notations}. For this quantity, a classical apparatus at $(p;\{{\bf e}\})$---i.e., at $p$ with state of motion and orientation given by $\{{\bf e}_\mu^a\}$---registers the result  $T_{ab}(p){\bf e}^a_0(p) {\bf e}^b_0(p)$, where $T_{ab}(p)$ is the system’s stress-energy-momentum tensor at $p$.  So in our framework, for an r-QRF $\Psi$ with support at $p$, we define the expectation value at $p$  of the energy density  by:
\begin{eqnarray}
\langle \rho \rangle|_{p} :=\|\Psi(p)\|^{-2} 
\int_{O(3,1)}d\mu_L(\Lambda) \; T_{ab}{\bf e}_\alpha^a{\bf e}_\beta^b
\Lambda_0^\alpha \Lambda_0^\beta
\left|\Psi({\bf P}_{\Lambda{\bf e}})\right|^2 .
    \label{EvalueQRFnewclassicalrho}
\end{eqnarray}

\bigskip

\noindent (3):  {\it The expectation value, for a region $\cal O$, of a classical field}:---\\
So much for the expectation value of a classical quantity at the point $p$, i.e.\ conditional on the apparatus being, i.e.\ firing, at $p$. But in the restricted framework, we can  also adapt an expression such as Eq.\ \ref{EvalueQRFnew} for the electric field, by integrating over spacetime, i.e.\ by averaging over the spacetime location of our quantum-uncertain but single apparatus. 

We thus return to our assumption at the start of Section \ref{ssec:ps:normn}, that the apparatus is infinitesimal in temporal and spatial extension but the uncertainty in
its location is spread over a spacetime region ${\cal O}$. Mathematically, we impose the normalisation condition Eq.\ \ref{norm20Oct}, and its probability interpretation.

So, again taking the example of the electric field: given a r-QRF $\Psi$ on $\cal O$, we define the expectation value for $\cal O$ of the $j$ component of the electric field, by integrating over $\cal O$ taking the absolute value of the volume element.  That is, we define:
\begin{eqnarray}
\langle E_j\rangle|_{\cal O} := \int_{\cal O}  \epsilon(p)   \; 
\int_{O(3,1)}d\mu_L(\Lambda) \; 
F_{ab}{\bf e}_\alpha^a{\bf e}_\beta^b
\Lambda_j^\alpha \Lambda_0^\beta
\left|\Psi({\bf P}_{\Lambda{\bf e}})\right|^2 ,
    \label{EvalueQRFnewspt}
\end{eqnarray}
where $\{{\bf e}_\mu^a\}$ is any (fixed)  tetrad  field.

 Similarly of course for any classical quantity whose measurement results can be written in terms of  the coupling between the system and the tetrad characterizing the apparatus: for example, the energy density $\rho$  as in Eq.\ \ref{EvalueQRFnewclassicalrho}, in item (2) above.

\bigskip

\noindent (4):  {\it The expectation value, at a time, of a classical  field}:---\\
In physics, one usually  considers probabilities, and so expectation values, at a time. Thus on the ``instrumentalist’’ understanding of quantum probabilities which we adopt, one considers the probability of a result, were one to measure the system at a given time, i.e.\ on a given spatial slice. (One assumes that the apparatus is somewhere on the slice, and guaranteed to ``fire”.) But for an apparatus in a relativistic setting, with quantum uncertainty about its spacetime location (and its spatial configuration and state of motion), how should we understand `at a given time’? 

 The most direct way to make this precise, staying with r-QRFs on $\cal O$ normalised by Eq.\ \ref{norm20Oct}, is to {\it condition} on the time of firing---just as, in (1) and (2) above, we conditioned on the point $p$. So let $t$ be a time function on $\cal O$, with level sets (spacelike slices) $\Sigma_t$ and lapse $N$, so that $\epsilon = N\,dt\wedge\epsilon|_{\Sigma_t}$. Then the probability measure $|\Psi|^2\,\epsilon\, d\mu_L$ decomposes into: the marginal probability density for the apparatus to fire at time $t$, viz.
\begin{equation}
P(t) := \int_{\Sigma_t} N(p)\, \epsilon|_{\Sigma_t}(p)\, \|\Psi(p)\|^2
\label{marginaltime}
\end{equation}
(so that $\int dt\, P(t) = 1$, by Eq.\ \ref{norm20Oct}); and conditional probabilities on each slice. Accordingly, for $P(t)\neq 0$, we define the expectation value at time $t$ of the $j$ component of the electric field by
\begin{equation}
\langle E_j\rangle|_{\Sigma_t} := P(t)^{-1}\int_{\Sigma_t} N(p)\,\epsilon|_{\Sigma_t}(p) \int_{O(3,1)} d\mu_L(\Lambda)\; F_{ab}{\bf e}^a_\alpha {\bf e}^b_\beta\, \Lambda^\alpha_j\Lambda^\beta_0 \left|\Psi({\bf P}_{\Lambda{\bf e}})\right|^2 ;
\label{EvalueQRFtime}
\end{equation}
and similarly for any classical quantity $A$. 

Note that: $P(t)$ is finite for almost every $t$ (by Fubini's theorem); no normalisation ``at a time'' needs to be imposed separately; all states of motion of the apparatus are still allowed; and the dependence on the choice of time function $t$ (including through the lapse $N$) is explicit---different foliations of $\cal O$ define different conditional probabilities, as they should (cf.\ the end of this item). Eqs.\ \ref{normpjfiducialfnoinfy}, \ref{marginaltime} and \ref{norm20Oct} thus exhibit a uniform pattern: conditioning on a point, conditioning on a time, and no conditioning (a region).

Alternatively, one may wish to model an apparatus that is designed to exist, and fire, {\it only} on a given slice. Given our idea of a r-QRF, the obvious suggestion is to make this a special case of our treatment in item (3) above, of expectation values for a spacetime region $\cal O$: namely the special case where $\cal O$ is squashed to a spacelike slice $\Sigma$. (One might also require that the slice be of some preferred kind.)  Here  we need to be careful to distinguish 4-dimensional and (3+1)-dimensional thinking. The suggestion is {\it not}  just that the r-QRF has support confined to a single  section of ${\cal F}_{\rm o}[g]$, i.e.\ a tetrad field on the 4-dimensional spacetime region $\cal O$. Rather, the suggestion is ``(3+1)-dimensional". It amounts to imposing two conditions, as follows. The first is the core of the suggestion; the second is not mandatory, but it is natural.\\
\indent \indent \indent \indent (i):  The r-QRF $\Psi$ is to have  support confined to the various possible tetrads at all the points $p$ of a single spacelike slice $\Sigma$, taken as a subset of $\cal O$. That is, $\Psi$ is to be a function on the ``column" $\pi_{\rm o}^{-1}[\Sigma] \subset {\cal F}_{\rm o}[g]$ that stands above $\Sigma \subset {\cal O}$---more precisely, an element of $L^2(\pi_{\rm o}^{-1}[\Sigma],\epsilon|_\Sigma\otimes\mu_L)$, rather than of $L^2(\pi_{\rm o}^{-1}[{\cal O}],\epsilon\otimes\mu_L)$, since $\pi_{\rm o}^{-1}[\Sigma]$ has zero measure in the latter. In this sense the r-QRF $\Psi$ is required to be instantaneous. It is assumed to be located, and to fire i.e.\ register a result, only on $\Sigma$; though there can still be quantum uncertainty about its spatial location across $\Sigma$, and its spatial configuration (orientation of axes) or state of motion.  (This uncertainty is  of course encoded in the amplitudes $\Psi(p;\{\Lambda {\bf e}\})$ for $p \in \Sigma$.)  \\
\indent \indent \indent \indent (ii): It is also natural to think of the three spacelike vector fields ${\bf e}_i^a$ of tetrads in the support ${\rm supp}(\Psi)$  as tangent to the spacelike slice $\Sigma$. 
 But note that since for any tetrad at $p$, each of the three spacelike vectors  ${\bf e}_i^a$ are orthogonal to the tetrad's unit timelike vector, the requirement that the spacelike vector {\it fields} ${\bf e}_i^a$ are tangent to $\Sigma$ is tantamount to requiring that there is, at all $p \in \Sigma$, no quantum uncertainty in the r-QRF's state of motion (its unit timelike vector; apart from time orientation). In other words: it amounts to requiring that at each $p$, $\Psi$'s only quantum uncertainty is about its spatial configuration, i.e.\ about the spatial directions of its spatial vectors (related, generically, by an orthogonal transformation $R \in O(3)$).  Note that when (ii) is imposed, the support of $\Psi$ in each fibre is (a copy of) $O(3)$---together with its time-reversed copy, if one does not fix the time orientation---which has zero Haar measure in $O(3,1)$. So in that case, the fibre integrals over $O(3,1)$ with the measure $\mu_L$ in Eqs.\ \ref{norm23Feb} and \ref{EvalueQRFnewsptSigma} below are to be understood as integrals over $O(3)$ with its own Haar measure; just as the integral over $\cal O$ with $\epsilon$ is replaced by an integral over $\Sigma$ with $\epsilon|_\Sigma$. The relevant bundle is then that of frames adapted to $\Sigma$ (with ${\bf e}_0^a = \pm n^a$, the unit normal to $\Sigma$), which is essentially the orthonormal frame bundle of $\Sigma$ with its induced metric; and since $O(3)$ is compact, the integrands of expectation values such as Eq.\ \ref{EvalueQRFnewsptSigma} below are bounded along each fibre.

We can readily adapt the formulas above to this suggestion. We write the slice that is (maybe a superset of) the canonical projection of  $\Psi$’s support, i.e.\ $\pi_{\rm o}[{\rm supp}(\Psi)]$, as $\Sigma$. And since we wrote $\epsilon$ for the positive 4-volume element defined by the metric $g$: let us also  write $\epsilon|_{\Sigma}$ for the positive 3-volume element induced by $\epsilon$ on $\Sigma$. Then our basic normalisation equation, Eq.\ \ref{norm20Oct}, becomes an integral over $\Sigma$. Using again the notation $\|\Psi(p)\|^2$ defined by Eq.\ \ref{normpjfiducialfnoinfy} for the fibre-integral  at  $p \in \Sigma$, we have:  
\begin{eqnarray}
 \|\Psi\|^2:=\int_{\Sigma}   \epsilon|_{\Sigma}(p)  \;
\int_{O(3,1)}d\mu_L(\Lambda) \;  \left|\Psi({\bf P}_{\Lambda{\bf e}})\right|^2 \, 
\equiv \int_{\Sigma}  \epsilon|_{\Sigma}(p)  \; \|\Psi(p)\|^2 \; =  \, 1 \, .
\label{norm23Feb}
\end{eqnarray}

Similarly, we adapt  item (3)'s  treatment of expectation values of a classical field for a region $\cal O$, by integrating over $\Sigma$ rather than over  $\cal O$. In particular, Eq.\ \ref{EvalueQRFnewspt} for the expectation value of the $j$ component of the electric field, becomes
\begin{eqnarray}
\langle E_j\rangle|_{\Sigma} := \int_{\Sigma}  \epsilon|_{\Sigma}(p)   \;
\int_{O(3,1)}d\mu_L(\Lambda) \; 
F_{ab}{\bf e}_\alpha^a{\bf e}_\beta^b
\Lambda_j^\alpha \Lambda_0^\beta
\left|\Psi({\bf P}_{\Lambda{\bf e}})\right|^2 ,
    \label{EvalueQRFnewsptSigma}
\end{eqnarray}
where, again, $\{{\bf e}_\mu^a\}$ is any (fixed)  tetrad  field. Similarly of course for the expectation value at time $t$, i.e.\ on $\Sigma$, of any classical quantity $A$ whose measurement results can be written by coupling the system and the tetrad characterizing the apparatus. But note that such an instantaneous r-QRF is not obtained by restricting a r-QRF on $\cal O$ to $\pi_{\rm o}^{-1}[\Sigma]$ (a set of zero measure): it is a different object, and instantaneous r-QRFs for different slices belong to different Hilbert spaces, whose relation would require the dynamics that we set aside. It is the conditioning of Eqs.\ \ref{marginaltime} and \ref{EvalueQRFtime} that connects a r-QRF on $\cal O$ with expectation values at a time. \\

We close this Section by returning to the  conceptual point we made just before (1) above. Thus we stress that  if one considers a r-QRF defined on the whole of $\cal O$ (i.e.\ not confined to $\pi_{\rm o}^{-1}[\Sigma]$) and conditions on the time of firing, as in Eqs.\ \ref{marginaltime} and \ref{EvalueQRFtime}, the results will of course not be invariant under a change of the spacelike slice across $\cal O$, say from $\Sigma$ to $\Sigma'$; (nor, indeed, under a change of the foliation to which a given slice belongs, since the lapse $N$ enters the conditional measure).
 And it is not just that the values of the measured system's quantity, say $F_{ab}$, can be very different on the two slices (and even differ between two points on the slices that in some way naturally correspond). Also, the weight determined by the (state of the) r-QRF, i.e.\ $\left|\Psi({\bf P}_{\Lambda{\bf e}})\right|^2$ in the integrand of an expectation value such as Eq.\ \ref{EvalueQRFnewsptSigma}, can be very different on the two slices (and again: even differ between two points on the slices that in some way naturally correspond).

 %%%%%%%%%%%%%%%%%%%%%

\section{The general framework}\label{sec:basicsing}

For the general  framework our overall plan is, as in Section \ref{sec:perspsing}, to first define a quantum reference frame, and then discuss normalisation and expectation values. But unlike Section \ref{sec:perspsing}, we need first to clarify some conceptual issues about diffeomorphism invariance that arise from the key ideas of the framework.  This is the task of Section \ref{ssec:conceptual}. After that, we will follow Section \ref{sec:perspsing}’s pattern. That is, we will first formally define a general quantum reference frame, and make some comments about it (Section \ref{ssec:bs:definition}). Then we will discuss normalisation, and the ensuing expectation values of measurement results  (Section \ref{ssec:bs:normnexpn}).
\bigskip

\subsection{On diffeomorphism invariance}\label{ssec:conceptual}
The topic of diffeomorphism invariance has a large literature, wholly independent of quantum reference frames. So without of course attempting a survey of it, we need to explain how our general framework relates to it---hence this Section. But the discussion is conceptual, and its details will not affect the technicalities of normalisation and expectation values. So readers less interested in it are invited to skip to Section \ref{ssec:bs:definition}.

This Section's discussion will proceed in three  stages, (I) (II) and  (III) below. But to prepare for them we first recall, and add to,  key ideas of the framework as summarized at the end of Section \ref{ssec:bundle}. This preparation consists of three comments. The third is much longer, and will introduce (I), (II) and  (III)  in more detail. 

\bigskip

\noindent First, about the formalism: the general framework works with the {\it frame bundle} ${\cal F}({\cal M})$,  whose fibre above a point $p$ is the set of all {\it bases} of the tangent space $T_p{\cal M}$.  So this bundle makes no assumption that there is a metric (despite its name!); and the structure group of ${\cal F}({\cal M})$ is the 16-dimensional general linear group $GL(4)$. 
There are two main points to make: the first about the structure of a fibre, the second about sections. Combining them will then yield the main point we will need about normalisation of our general quantum reference frames, g-QRFs. \\
\indent \indent \indent \indent (1): 
We recall that $O(3,1)$ is not normal in $GL(4)$. Instead, its conjugate sub-groups $g^{-1} O(3,1)g$ (writing  just in this paragraph, but as usual in group theory: the generic element of $GL(4)$ as $g$)
 preserve the {\it transformed} symmetric bilinear form. That is: $g^{-1} O(3,1)g$ preserves, not the given Lorentzian form $\eta$,  but the form: $g^T \eta g$. Thus one verifies that if $h \in O(3,1)$ [equivalently: $h^T \eta h = \eta$], and if $g$ is any element of $GL(4)$, then: $g^{-1}hg$ preserves the transformed symmetric bilinear form $g^T \eta g$. That is: $(g^{-1}hg)^T  (g^T \eta g ) (g^{-1}hg) =  g^T \eta g$. Thus the set $Sym_L$ of non-degenerate symmetric bilinear forms on ${\mathbb R}^4$ with Lorentzian signature (an open subset of the set $Sym$ of all symmetric bilinear forms) is in bijection with the quotient set $GL(4)/O(3,1)$.  Thus in the frame bundle ${\cal F}({\cal M})$, the 16-dimensional fibre $\pi^{-1}(p)$ is foliated by 6-dimensional leaves. (In particular: $Sym$ is 10(=(4+3+2+1))-dimensional, and this corresponds to the cosets of  $O(3,1)$ having codimension 16 - 6 = 10.)\\
\indent \indent \indent \indent (2): We recall from  the end of  Section \ref{ssec:bundle} that we can specify a spacetime $({\cal M}, g)$ by specifying a  section $\sigma$ of ${\cal F}({\cal M})$. But a section ``over-specifies" a spacetime, i.e.\ contains redundant information; since it includes a specific basis-field. But we can, using this bundle, specify a spacetime non-redundantly by taking  the union of a section $\sigma$  (taken as the embedded sub-manifold) and all its pointwise Lorentz transforms. This union, which is a subset of ${\cal F}({\cal M})$ (more precisely: of the ``column" $\pi^{-1}[{\cal M}]$), is
 the orthonormal frame bundle ${\cal F}_{\rm o}[g]$ for the metric $g_{ab}$ defined by the section (embedded sub-manifold)  $\sigma$---and of course, defined equally well by any of $\sigma$'s pointwise Lorentz transforms. And these remarks also of course apply locally. That is: the union of  a local section $\sigma$ on a region ${\cal O} \subset {\cal M}$ with all its pointwise Lorentz transforms is a subset of ${\cal F}({\cal O})$; and it {\it is} the orthonormal frame bundle ${\cal F}_{\rm o}[g]$ for the metric $g$ on $\cal O$ defined by $\sigma$ (and by any of its transforms).\footnote{As also discussed at the end of  Section \ref{ssec:bundle}: the infinite-dimensional set $Sec({\cal F}({\cal M}))$ of sections is partitioned by the equivalence relation `is a pointwise Lorentz transform of' on sections; so these equivalence classes are in 1-1 correspondence with metrics $g$ on $\cal M$.  Similarly locally, i.e.\ for a region ${\cal O} \subset {\cal M}$.}

Combining (1) and (2): we will see in Section \ref{ssec:bs:normnexpn}, how the volume-element $\epsilon(p)$, as  defined by a metric,  varies across the fibre $\pi^{-1}(p)$, so that one needs to be careful when it occurs in the integrand of normalisation integrals. In effect, this varying volume-element will be the infinitesimal, i.e.\ at-a-point, ``shadow" of the way that the frame bundle ${\cal F}({\cal M})$ accommodates varying metrics.

\bigskip

\noindent Second, about interpretation. We stress that the general framework involves a {\it twofold interpretation} of tetrads. First,  the tetrads in a section of ${\cal F}({\cal M})$ collectively specify a geometry, i.e.\ a metric: as we just recalled. In this role, the tetrads do not represent a system, or degrees of freedom, separate from spacetime itself. But as discussed in Section \ref{sec:intro}'s (C) `Without hats': we do not in this paper (nor in $\bf {[I]}$ and $\bf {[II]}$) address the ``real physics" of these spacetime i.e.\ metrical degrees of freedom.  But second: we do also interpret a tetrad (i.e.\ a basis at a point of ${\cal F}({\cal M})$, that ``becomes" a tetrad by our specifying a section of ${\cal F}({\cal M})$)  as representing (a possible configuration of) an infinitesimal apparatus with its own material degrees of freedom, separate from (but subject to) the spacetime  geometry.

\bigskip

\noindent Third, this Section's topic, diffeomorphism invariance, relates closely to the last issue raised in Section \ref{ssec:bundle}: how we treat a spacetime point existing in two  (possibly non-isometric) spacetimes.   Recall that we said:---\\ 
\indent \indent (a) Our formalism seems {\it haecceitistic} since it treats a spacetime point as existing in two spacetimes, while having very different metrical properties and relations in them. And this seems wrong; (in particular, because it seems to lead to a spurious indeterminism in view of Einstein's hole argument).\\
\indent \indent (b) But we then claimed  that this haecceitism is only apparent. It  is a formal convenience, to simplify our formalism. And we added that it has the advantage of being  flexible about how one resolves questions, when comparing two spacetimes, about ``which spacetime point in this spacetime is (or corresponds to) this point in that other spacetime?".\\
Therefore, we now need to justify these claims---which is this Subsection's task. We will proceed in three  stages, (I),  (II) and  (III).

 It will be clearest to first summarize the themes of (I), (II) and (III) {\it seriatim}. It will be clear that much of their content is independent of our proposals about quantum reference frames. So after the {\it seriatim} summaries,  we will make two points, [A] and [B], about how our proposals will relate to (I), (II) and (III).

Note also two minor points, to make the discussion clearer, and notation simpler, throughout this Section. (i): We will consider just two spacetimes (and thus comparisons between, and superpositions of, them); from which the general case can be inferred---provided the cross-identification maps between more than two spacetimes are mutually consistent, i.e.\ $d_{13} = d_{23}\circ d_{12}$ for the maps $d_{ij}:{\cal M}_i \to {\cal M}_j$.
(ii): We will mostly talk of an entire manifold; but as in previous discussions, all
that we say will carry over to discussion of an appropriate region $\cal O$ of the manifold.

 (I) focusses on what `diffeomorphism invariance'  means. The phrase is used in various ways, and we need to say what we will mean by it, and how our formalism obeys it, i.e.\ is diffeomorphism invariant. But rest assured that what we mean is straightforward and is also a widespread usage. It also makes precise the sense in which our formalism attributes no intrinsic physical meaning to individual spacetime points, thus rebutting the accusation of haecceitism. Besides, this explanation and rebuttal will be independent of our use of the frame bundle to model configurations of an infinitesimal apparatus.

(II) is about a question that will remain after the discussion in (I): how {\it should} one ``cross-identify'' spacetime points between different spacetimes? More precisely: given two spacetimes---that of course, are in general {\it not} isometric---and a point $p$ in one of them, which point $q$ (if any) in the other spacetime ``is'',  or corresponds to, $p$? Clearly, this is a question that remains even after one has rejected  haecceitism. For haecceitism is the doctrine that there is a factually right answer: indeed, a very simple one---that the point $q$ in the other spacetime must be the very same point as $p$, i.e.\ $q=p$! To almost all physicists and philosophers, this has seemed over-simple, even wrong-headed---and we agree. But after rejecting this wrong answer, the question  remains. 

And it is  a question addressed by philosophers as well as physicists. As we will see in (II), the philosophical literature assesses  the various answers mostly in terms of their implications for interpreting classical general relativity. But the  physics literature focusses on  how we should make sense of {\it superpositions of spacetimes}. For such a superposition will surely involve a spacetime point $p$ being somehow equipped with both the metric tensor (and-or associated quantities like curvature) of a point, say $q$, in the manifold of one ``branch" (one of the superposed spacetimes), and the metric tensor etc.\ of a point, say $r$, in the manifold of the other ``branch".  So the physical question is: in defining such superpositions, what are the rules or constraints about which points $q$ in one spacetime's manifold can be ``coalesced" with which points $r$ in the other  spacetime's manifold?  We believe---and we believe the physics literature agrees---that this is a hard open question:  new physical insight is needed.

But fortunately, our task is much easier. In (II), we will very briefly review these literatures (including their rejection of haecceitism), but we will also argue for a more general and positive point: that one should be {\it flexible}. That is: we expect that whichever rules or constraints about cross-identifying points (or perhaps better, in the quantum context: coalescing points) eventually get agreed on (whether by philosophy or by physics) will vary from one context to another. 

Then in (III), we argue that our formalism can accommodate almost any such rules or constraints. More precisely: our formalism can  accommodate almost  any cross-identification of points expressed with a {\it diffeomorphism}. The reason will be, in short, that our use of the frame bundle, with  ``the selfsame point $p$ under the fibre", amounts merely  to ``unwinding" the prescribed diffeomorphism, say $d$, that cross-identifies points between the spacetimes. That is: our use of the frame bundle means that we implicitly apply the inverse diffeomorphism $d^{-1}$ so as to re-label the point $q:=d(p)$ as $p$---a merely apparent haecceitism. 

Here we should admit a limitation; (it will also be evident from our always using the frame bundle $\cal {F(M)}$ of a single manifold $\cal M$, and similarly for regions $\cal O$). Namely, our formalism cannot treat superpositions of spacetimes (or of their regions) that are not diffeomorphic; or, {\it a fortiori}, not homeomorphic. In short: superpositions of topology are beyond our ken at this point. 
But we will see that within this limitation, we {\it can} describe superposed spacetimes (without being haecceitistic)---and using our tetrads,  superposed perspectives thereon.  

\bigskip

So much by way of an introduction {\it seriatim} to the tasks undertaken by (I), (II) and (III). But we should add here two points, [A] and [B], that arise from this paper’s basic proposal---also in the restricted framework---to model a quantum reference frame as an assignment of complex amplitudes to the possible configurations of an infinitesimal apparatus; (where, recall, a configuration is a point in the bundle, i.e.\ a spacetime location, directions of three spatial axes and a state of motion). [B] will make precise what we mean in the preceding paragraph by `almost any cross-identification', and so explain why our formalism’s limitation is acceptably mild. Besides, both these points [A] and [B] apply to the restricted framework as well as the general framework.\footnote{For both the issues, diffeomorphism invariance and the cross-identification of points, also apply to the restricted framework. But there, the use of one fixed metric makes the issues much less tricky (as will be clear in (I) and (II)); so that the literature hardly discusses them there. Hence our first raising them within the general framework, i.e.\ in this Section.}   

Both points [A] and [B] are prompted by a question that arises by recalling that discussions, both of diffeomorphism invariance and of cross-identification of points, within general relativity, tend to focus on the vacuum case, i.e.\ spacetimes $({\cal M},g)$, without matter or radiation fields. So for example, it is usual to consider whether only isometries can give legitimate cross-identifications of points. But of course, if matter or radiation fields are included, then discussion of the status of isometries gets replaced by corresponding discussion of isomorphisms. In (I) and (II), we shall for the most part follow this common pattern, since it simplifies exposition; and so we only occasionally mention matter or radiation fields. But one naturally asks: what about the quantum reference frame itself? Is it to be treated, as regards diffeomorphism invariance and the cross-identification of points, as ``just another''  matter or radiation field---or does it get a different treatment? 

It is this question that prompts our two points, [A] and [B]. 

\medskip

[A]: Our concern to take reference frames ``seriously''  in the sense of attributing degrees of freedom to them {\it does} affect the question how to cross-identify points. For almost all discussions take it that one cross-identifies points by requiring the values of metrical and-or material fields at the two points to be equal, or suitably close. Hence our mention, just now, of isometries and-or isomorphisms. (For details, cf.\ (II).) So if we attribute degrees of freedom to a reference frame and treat it as a material field, then in assessing whether two points should be cross-identified, the equality or near-equality of the values of this field should surely be given some weight, along with the values of the other material fields. Besides, this point clearly applies classically as well as in our quantum context.  For recall from Section \ref{sec:intro}'s (B) the classical treatment of a reference frame: one interprets a tetrad field on a region $\cal O$ as a possible configuration across $\cal O$ (so: over time as well as space) of a continuous field of infinitesimal apparatuses (i.e.\ three rulers and a clock). Thinking of the apparatuses as material bodies, the equality or near-equality of the values of the tetrad field at two points should surely count in assessing whether to cross-identify the points. And similarly in our quantum context, with the single (not field) conception of a tetrad field which we adopted in Section \ref{sec:intro}'s (B). Besides, as we mentioned above: this point applies equally to the restricted framework; although there, the issues of diffeomorphism invariance and of cross-identification of points are  much less tricky. 

\medskip

[B]: This second point is longer and more substantial. It returns us to the general framework’s {\it twofold interpretation} of tetrads; (cf.\ the second comment at the start of this Section). (i): The tetrads in a section of ${\cal F}({\cal M})$ collectively specify a metric. But as discussed in Section \ref{sec:intro}'s (C) `Without hats': we do not in this paper (nor in ${\bf {[I]}}$ and ${\bf {[II]}}$) address the ``real physics" of these metrical degrees of freedom. But (ii): We also interpret a tetrad i.e.\ a basis at a point of ${\cal F}({\cal M})$---that ``becomes" a tetrad by our specifying a section of ${\cal F}({\cal M})$---as representing (a possible classical configuration of) an infinitesimal apparatus with its own material degrees of freedom, separate from (but subject to) the spacetime geometry. 

The point now is that this twofold interpretation implies a constraint on how one defines the cross-identification map between spacetime points. It is an entirely acceptable constraint, but worth noting.

The constraint follows from mathematical points which, though very elementary, are helpful to bear in mind. (Indeed, they are helpful for (I) and (II)'s discussion of the literature, not just for our concern with whether cross-identification should give some weight to matching, or near-matching, of the quantum reference frame’s degrees of freedom.) Namely: A diffeomorphism, say $d$, on a manifold or from one manifold to another, lifts to give a smooth map $d^\ast$ on vectors, tensors and bases. But if the manifold or manifolds are spacetimes, i.e.\ have a metric, then $d$ of course need not be an isometry. That is: for a single spacetime $({\cal M},g)$, a diffeomorphism $d$ on it need not push forward $g$ to itself. We can have $d^*g \neq g$. And similarly for two spacetimes: if $d$ is from $({\cal M}_1,g_1)$ to $({\cal M}_2,g_2)$, we can have: $d^*g_1 \neq g_2$.\footnote{And even if $d$ {\it is} an isometry, it need not be an isomorphism of tetrad fields. That is: for a single spacetime $({\cal M},g)$, an isometry $d$ need not push forward a tetrad field ${\bf e}$ to itself. We can have $d^*({\bf e}) \neq {\bf e}$ as fields.} Similarly of course for a local diffeomorphism defined on a region $\cal O$ of $\cal M$ or of ${\cal M}_1$.

Now let us consider a cross-identification map for spacetime points that: (i) is a (perhaps local) diffeomorphism $d$, and (ii) is between non-isometric spacetimes (where if $d$ is only local: its domain and range are not isometric). And we must consider such maps. For as we will see in (I) and (II): the literature almost always accepts feature (i). And feature (ii) is undoubtedly needed: both in order to make sense of superpositions of non-isometric spacetimes, as both we (more details in (II) and later) and the physics literature wish to do; and in order to make sense of counterfactual statements in classical general relativity (`If this point were to have a different scalar curvature, \dots'), as the philosophy literature wishes to do.\footnote{\label{ReJim}As we will see in (II): unfortunately, the philosophy literature is dominated by the topic of how to reply to the hole argument, and thus by the question whether two spacetimes being isometric is wholly sufficient for their physical equivalence. For if so, a given spacetime $({\cal {M}},g)$ is physically equivalent to its hole-diffeomorph, which is by construction isometric to it. So one would escape the hole argument’s conclusion of indeterminism.  Thus \cite{W18} argues for such sufficiency. But whatever one's answer to that question, one anyway needs to accommodate cross-identification between non-isometric spacetimes. This point is stressed in e.g.\ \cite{GB23}, \cite{GB23a} and \cite{L23}.} 
In particular, and as we will see in more detail in (III): the general framework needs maps with feature (ii) in its treatment of superposed spacetimes. 

The constraint on cross-identification maps is now clear.  For consider how such a cross-identification diffeomorphism $d$ between non-isometric spacetimes, i.e.\ from $({\cal M}_1,g_1)$ to $({\cal M}_2,g_2)$,  can fit in with our twofold interpretation of tetrads. That is: with our interpreting a tetrad field, i.e.\ a section of ${\cal F}({\cal M})$,  as both (i) specifying a metric and (ii) specifying some possible (classical) configurations (through time as well as space) of our infinitesimal apparatus. Namely: since $d$ is not an isometry, it cannot push forward a tetrad field on its domain ${\cal M}_1$  (with `tetrad' of course understood according to the metric $g_1$) to a tetrad field on ${\cal M}_2$  (now with `tetrad' understood according to the metric $g_2$). And similarly if $d$ is a local diffeomorphism, with some region ${\cal O} \subset {\cal M}_1$ as its domain. For if $d$ {\it did} push forward one tetrad field into equality with the other, it would be an isometry: we would have $d^*g_1 = g_2$ throughout the range of $d$. 

This veto on $d$'s pushing forward one tetrad field into equality with the other is, in our view, entirely acceptable. But it does imply a constraint, albeit a mild one, when one goes about defining $d$ in terms of equality or near-equality of the values of vectors, tensors etc.\ (as will be detailed in (I) and (II)). Namely, even though one may want---following part (ii) of our twofold interpretation of tetrads---to give weight to ``matching of tetrad fields'', now understood as representing the apparatus’ degrees of freedom: one cannot give {\it over-riding} weight to such matching, on pain of forcing $d$ to be an isometry. 

But we submit that this veto on the apparatus having over-riding weight is entirely acceptable. We say this for two reasons. First: after all, there are {\it other} matter and-or radiation fields whose values should surely also be compared in assessing whether two points should be cross-identified; i.e.\ whose equality or near-equality of values should be given weight in defining the cross-identification map. 

The second reason applies even if there are no matter and-or radiation fields to consider, i.e.\ in the vacuum sector of the theory. This reason follows from the fact that our apparatus is {\it quantum}. So its state is---unlike that of a classical reference frame---{\it not} a tetrad field on a region ${\cal O}  \subset {\cal M}_1$; but an assignment of complex amplitudes (possibly zero) to all of the column of the frame bundle above ${\cal O}$, i.e.\ $\pi^{-1}[{\cal {O}}]$. So the desideratum that a cross-identification map $d$ should   ``match across the spacetimes the states of the apparatus'',  though legitimate in itself, does {\it not} imply that $d$ should (once lifted to bases) push forward a particular tetrad field, viz.\ the field by which we specify $g_1$, to another particular tetrad field, which specifies $g_2$.\footnote{\label{NeitherLrtzTrsfm}Nor does the desideratum imply the weaker claim that $d$ should (once lifted) push forward some pointwise Lorentz transform of the field by which we specify $g_1$ to some pointwise Lorentz transform of the field by which we specify $g_2$. Fortunately for us: since this weaker claim would of course also make $d$ an isometry. For recall from (B) of Section \ref{ssec:bundle} that a metric $g_1$ corresponds, not to a single tetrad field (section of the frame bundle), but to such a field (section) together with all its pointwise Lorentz transforms. Indeed, the orthonormal frame bundle ${\cal F}_{\rm o}[g_1]$ for the metric $g_1$ can be identified with the union of all these sections, taken as embedded sub-manifolds.}

\bigskip

Here ends our long introduction to (I), (II) and (III). As we mentioned at the start of this Section \ref{ssec:conceptual}, the details to follow will not affect the technicalities of normalisation and expectation values. So readers less interested in conceptual discussion are invited to skip to Section \ref{ssec:bs:definition}.\\

\bigskip    

(I): {\it Diffeomorphism invariance secured}:---\\
Our first task is to show that working with a single manifold $\cal M$ does not prevent a formalism, in particular ours, being diffeomorphism invariant in a precise sense. So we begin by considering two spacetimes that are given as being on different manifolds; and then show how to map one’s discussion of, and constructions on, them on to discussion and constructions about a single manifold. 

So disregarding for a moment our formalism's use of  the frame bundle $\cal {F(M)}$: consider two spacetimes, $({\cal M}_1,g_1)$ and $({\cal M}_2,g_2)$ (as usual: suppressing the subscript $ab$), built from distinct manifolds ${\cal M}_1$ and ${\cal M}_2$ that are however assumed to be diffeomorphic.  In order to keep the notation symmetric between these two spacetimes---and for no deeper reason---let us consider a fiducial manifold ${\cal M}$ diffeomorphic to ${\cal M}_1$ and ${\cal M}_2$ (obviously,  either one of ${\cal M}_1$ or ${\cal M}_2$ could play the role of ${\cal M}$). Let ${\cal D}_j:={\text{\it Diff}({\cal M}_j,{\cal M})}$, $j=1,2$,  be the set of diffeomorphisms between ${\cal M}_j$ and ${\cal M}$.

Then for {\it each} $j=1,2$, given any $\varphi_j\in{\cal D}_j$,  the spacetimes $({\cal M}_j,g_j)$ and $({\cal M},\bar{g}_j)$,  with 
$\bar{g}_j:=\varphi_j^\ast g_j$ (the push-forward metric induced on ${\cal M}$ by  $\varphi_j$) are agreed by all parties/by all physicists and philosophers to be physically indistinguishable. And similarly if we equip the spacetimes with matter and-or radiation (particles or fields),  and-or (as in  our formalism) tetrad fields representing infinitesimal apparatuses.  That is: once {\it all} physical quantities on $({\cal M}_j,g_j)$---characterizing the metric, the matter and-or radiation, and the apparatuses living there---are correspondingly mapped to quantities on $({\cal M},\bar{g}_j)$, the ``equipped spacetimes" are physically indistinguishable. That is: {\it no} measurement by the apparatuses characterized by the tetrads $\{{\bf e}_\mu^a\}$ at $p\in {\cal M}_j$ and $\{\varphi_j^\ast{\bf e}_\mu^a\}$ at $\varphi_j(p)\in {\cal M}$ can distinguish  between them.

Now we turn to the comparison of the two spacetimes $({\cal M}_1,g_1)$ and $({\cal M}_2,g_2)$. The first point to make is the obvious one: which applies whether or not they are also equipped with matter and-or radiation and-or tetrad fields. Namely: once  both spacetimes (perhaps thus equipped) are represented on the {\it same} manifold ${\cal M}$---i.e.\ {\it given $\varphi_1$ and $\varphi_2$}---comparing 
physical quantities becomes simply a matter of comparing pointwise the various tensorial quantities that are induced on ${\cal M}$. In particular: the metrics $\bar{g}_1$ and $\bar{g}_2$---being symmetric, nondegenerate rank-(0,2) tensors---are no exceptions.

All this discussion is of course valid for any choice of $(\varphi_1,\varphi_2)\in {\cal D}_1\times{\cal D}_2$. And this means that now, the accusation of a suspicious haecceitism can be made precise, as follows. Namely:  different choices, say $(\varphi_1,\varphi_2)$ and $(\varphi'_1,\varphi'_2)$,  will in general lead to  distinct {\it physical}  configurations---understood as in the last paragraph in terms of pointwise comparisons. Restricting ourselves (for simplicity, and as usual) to the comparison of metric tensors, the accusation is thus that: (i) we can have  $\bar{g}_1\neq\bar{g}'_1:=\varphi^{\prime\ast}_1 g_1$ and-or $\bar{g}_2\neq \bar{g}'_2:=\varphi^{\prime\ast}_2 g_2$ as tensor fields on ${\cal M}$; and (ii) these could represent distinct {\it physical}  geometries. To this, we reply: (i) is of course true.\footnote{For the inequations in (i) are of course generic. For, taking the case $j = 1$ as an example: if the equation $\bar{g}_1 = \bar{g}'_1:=\varphi^{\prime\ast}_1 g_1$ were true throughout some small region  of $\cal M$, then the map $\varphi^{\prime {-1}}_1 \circ \varphi_1$, as restricted to a suitable small region (call it $\varphi^{-1}_1({\cal O})$, with ${\cal O} \subset {\cal M}$), would be a local isometry on  $({\cal M}_1,g_1)$. And there being such a local isometry is not generic. 

Here we should add the more general point that our entire discussion, in both  (I) and (II), is formulated, so as to accommodate cases where one or both spacetimes (or the pertinent region of them being discussed) each have an isometry to themselves. That is:  cases where there is a map $\varphi: {\cal M}_1 \rightarrow {\cal M}_1$ other than the identity map that is an isometry, at least on the pertinent region.
%{\jtod Is this footnote OK? the second paragraph is to set aside (avoid endless footnotes about!) the fuss about spacetimes with non-trivial isometries, which has beset the philosophical literature...} 
} 
 But (ii) is not true---for the following reason.  
 
Let ${\cal D}:=\text{\it Diff}({\cal M})$ be  the group of diffeomorphisms of ${\cal M}$.
Then  consider the quotient $({\cal D}_1\times{\cal D}_2)/{\cal D}$---i.e.\ the set of equivalence classes
$[(\varphi_1,\varphi_2)]:=\{(\phi\circ\varphi_1,\phi\circ\varphi_2); \;\phi\in {\cal D}\}$,  $\varphi_j\in{\cal D}_j$.
If $(\varphi_1,\varphi_2)$ and $(\varphi'_1,\varphi'_2)$ belong to the same equivalence class, then
the corresponding constructions on ${\cal M}$ which we have just presented differ by the {\it same} $\phi^\ast$ applied to {\it all} quantities
induced on ${\cal M}$. In other words: although, for instance,  $\bar{g}_1$ (resp.,  $\bar{g}_2$) may differ from 
$\bar{g}'_1$ (resp., $\bar{g}'_2$),  the {\it relation} between $\bar{g}_1$ and  $\bar{g}_2$ is the same as the
one between $\bar{g}'_1$ and  $\bar{g}'_2$. More generally,  the relation between {\it any} two  quantities
induced on ${\cal M}$ (from ${\cal M}_1$ or from ${\cal M}_2$) is the same for any representative  of the
same equivalence class $[(\varphi_1,\varphi_2)]$. And this, we submit, refutes (ii). For {\it physical} configurations of fields, in particular of metrics $g$, should be identified only up to a choice of ``which point of $\cal M$ is which", as encoded in a choice of $\phi \in {\cal D}$---meaning a single choice for all the quantities to be compared. 

To sum up: although we have introduced the background manifold ${\cal M}$, and used its points as ``locations" at which to compare quantities induced from different spacetimes $({\cal M}_1,g_1)$ and $({\cal M}_2,g_2)$, our discussion and constructions are invariant under diffeomorphisms $\phi$ of ${\cal M}$. In that sense, we have attributed no intrinsic physical meaning or ``identity'' to such points---refuting  the accusation of haecceitism. 

Accordingly, we maintain that no harm is done by adopting our formalism, with the frame bundle $\cal {F(M)}$. For this amounts, if one is given two spacetimes ${\cal M}_j$ as above,  to:\\
\indent \indent \indent \indent  (i) adopting fixed ``reference'' diffeomorphisms $^{(R)}\varphi_j$ to ``identify'' ${\cal M}_j$ and ${\cal M}$; and then \\
\indent \indent \indent \indent (ii) varying over $\phi\in {\cal D}$ to obtain all the
 {\it different} ways of ``co-locating"  a set of field  values possessed by a given point in one spacetime, say $({\cal M}_2,g_2)$, at one or another point in  $({\cal M}_1,g_1)$.\\
 Of course, in the quantum context, these co-locations of field values are considered, not just so as to compare values pointwise (as explained above), but also so as to   obtain  all the different possible superpositions of $({\cal M}_1,g_1)$ and $({\cal M}_2,g_2)$---up to the weighting amplitudes, of course---relative to our choice of $[(^{(R)}\varphi_1,\,\phi\circ ^{(R)}\!\varphi_2)]$. Note that we have here specified the equivalence class with one occurrence of $\phi$, not two. Thus we think of $\phi$ as ``stirring" the points of ${\cal M}$, so as to drag the ($\cal M$-images of) points of ${\cal M}_2$ into comparison with  the ($\cal M$-images of) points of ${\cal M}_1$.\footnote{Note that any element of $({\cal D}_1\times{\cal D}_2)/{\cal D}$ can be written in this form with some $\phi\in{\cal D}$, for any fixed $^{(R)}\varphi_1,  \,^{(R)}\varphi_2$.} \\
 
 \bigskip

(II): {\it How should one ``cross-identify”  points between different spacetimes?}:---\\
So far, so good. But our phrases at the end of (I), `all the different ways to co-locate' and `all the different possible superpositions of $({\cal M}_1,g_1)$ and $({\cal M}_2,g_2)$' (up to the weighting amplitudes) signals a question that (I) has not addressed. Namely the question: How {\it should} one ``cross-identify''  points between different spacetimes? 

Recall that when we introduced this question (before the start of (I)) we contrasted how philosophers  have addressed it, and how physicists have done so. We said that philosophers tended to address this question as a probe of the nature of spacetime points in classical general relativity. On the other hand, physicists have addressed it with a view to properly defining a quantum state-space containing superpositions of spacetimes. 

Of course we cannot here review these literatures thoroughly. But we will review some aspects that bear on our formalism. This will include developing the contrast above, but also exhibiting some common themes. The main common theme will be that one needs to be flexible about how to cross-identify points; and for this, the philosophical literature makes a proposal that could be helpful for physics. So although we of course see the main merit of our general framework as its offering  insight into superpositions of spacetimes---i.e.\ the concern of the physics literature---we will begin with the philosophers.

The philosophy literature is dominated by exploring the ramifications of Einstein's 1913 hole argument for general relativity (cf.\ \cite{S14}).  One main endeavour has been formulating criteria for ``which point is, or corresponds to, which'' that: (i) render general relativity deterministic (i.e.\ escape the conclusion of Einstein's argument), once the criteria are combined with a suitable definition of determinism; and (ii) make good sense of counterfactual statements like `If this point (region) had had a different (mean) scalar curvature, then \dots'.\footnote{This literature was set going by \cite{EN87}. In addition to the references in footnotes \ref{ReJim}, and \ref {philcpartpoints} below, cf.\ also e.g.\ \cite{D25}.}  For this endeavour, it is natural to think that the criteria must be along the lines that, in order to be cross-identified, two points' metrical properties (e.g.\ a scalar curvature), and-or metrical relations to yet other points (e.g.\ their mutual distance), and-or material properties and relations (e.g.\ an energy density) must be suitably similar. And it is also natural to think that this similarity should be expressed in terms of quantities, such as curvature or distance, having close values. But exactly which quantities, and how close in value? Indeed, to seek utterly precise criteria seems likely to be a wild goose chase. 

In this connection, the philosopher Lewis has advocated an attractively flexible approach. It applies not just to cross-identifying points, but to cross-identifying objects, in general, between distinct possibilities; and it explains why one should {\it not} demand utterly precise criteria. It is called {\it counterpart theory} (\cite{L73} pp.\ 39-43; \cite{L86} Chapter 4). According to this approach, a pair of objects in distinct possibilities are never literally {\it identical}, but just more or less similar.\footnote{\label{saycross}But we will continue to sometimes say `cross-identify'  and `cross-identification', even though Lewis---and we ourselves---mean a correspondence, not literal identity. Some authors, including in the physics literature, say `comparison map', e.g.\ \cite{KHACGBB24}.} And which respects of similarity are relevant is a flexible, context-dependent and even vague matter, and therefore ``up to us".  Thus counterfactual statements are not about how the properties of this selfsame object might have been different, but instead are about another object that in the possibility considered is most relevantly similar to the given object (so that Lewis dubs it a {\it counterpart}). This approach's allowance that what counts as relevant is context-dependent and even vague makes good sense of the countless examples of counterfactuals about objects in everyday thought and talk whose truth or falsity is evidently hard to assess. But everyday thought and talk shades into technical science; and there is a strong case that also counterfactuals about e.g.\ spacetime points can be context-dependent and even vague.\footnote{\label{philcpartpoints}For philosophers applying this approach to spacetime points, cf.\ \cite{B89}, \cite{GB23}, \cite{GB23a}, \cite{F20} and \cite{C24}. This comment is further supported by the physics literature's flexible treatment of cross-identification; cf.\ below.}

We turn to the physics literature. As we explained (already before the start of (I)), physicists' main interest in cross-identifying points is---not of course to give a semantics for counterfactuals, but---to better understand superpositions of spacetimes. As we put it, their question is: what are the rules or constraints about which points $q$ in one spacetime's manifold (one branch) can be ``coalesced'' with which points $r$ in the other spacetime's manifold (the other branch), so as to define a manifold that ``hosts'' the superposition? {It is natural to think that these rules will concern closeness, or even equality, of the values of quantities defined at the points concerned (or maybe involving regions around them); though the rules will of course have to allow for the fact that for tensorial quantities other than scalars, even to {\it define} such equality requires a diffeomorphism between neighbourhoods of the points concerned. Thus the question is closely connected with the famously difficult question, within {\it classical} general relativity, of how to define observables: where ‘observable’ means, for the simpler case of vacuum general relativity, a function on spacetimes (Lorentzian manifolds) that is constant on isometric manifolds.\footnote{The use of `observable’, to indicate gauge-invariance in a classical theory, was introduced by \cite{B56} p.\ 1178, and \cite{B61} p.\ 511, as a deliberate echo of the quantum terminology.} Of course, we cannot enter into details here. For discussions highlighting this connection, cf.\ e.g.\ \cite{R91a}, \cite{R91b}. Discussions focussing on our question about how to cross-identify points include: \cite{Penrose96}, Sections 4 and 6; \cite{KdHCB24}; and \cite{KHACGBB24} especially pp.\ 2-8. Here,} it must suffice to say that (unsurprisingly, given the difficulty of the questions; and as we also announced before the start of (I)): it seems to be a consensus in the physics literature that to settle what these rules or constraints are, we must all await new physical insight.

Our own views are, in short, an admission of ignorance and an advocacy of flexibility! For first, we endorse the apparent physics  consensus that we  await new physical insight; (for more details, cf.\ \cite{VB24}, Sections V, VI). And we also advocate flexibility about ``which point is, or corresponds to, which'' questions about classical general relativity. Not only do we reject haecceitism. Also and more positively: we endorse  Lewis'  counterpart theory approach: cross-identifying objects between distinct possibilities is indeed a context-dependent, even vague matter, that turns on which respects of similarity count as relevant. And we see this context-dependence, even in physics; (cf.\ the reference in footnote \ref{philcpartpoints}). 

Let us support and illustrate these views by considering some envisaged criteria about how  to cross-identify points between two spacetimes  $({\cal M}_1,g_1)$ and $({\cal M}_2,g_2)$.\footnote{As we said in footnote \ref{saycross}:  now that counterpart theory is being considered, the word `cross-identify' need not imply literal identity.} As we have discussed, both philosophers and physicists expect any such criteria to be based on the two points in the two spacetimes having sufficiently similar properties and relations (geometrical and-or about matter and radiation). But we urge flexibility:  one should not commit to there being a set of factually correct criteria that  associate to each point in $({\cal M}_1,g_1)$ the uniquely correct corresponding point in $({\cal M}_2,g_2)$. Indeed,  one should be flexible in two ways, (A) and (B): of which (B) will be more important.\\ 

\indent \indent \indent \indent (A): For the given criteria, there might well be more than one point in $({\cal M}_2,g_2)$, all of them ``tied for first-equal" for corresponding to (being the best match to) the given point in $({\cal M}_1,g_1)$. Think for example of criteria that require exactly matching geometrical properties and relations, while $({\cal M}_2,g_2)$ has a symmetry, i.e.\ an isometry $d: {\cal M}_2 \rightarrow {\cal M}_2$. Then if  $q \in {\cal M}_2$ corresponds to (exactly matches, in the required sense)  $p \in {\cal M}_1$, so will $d(q) \in {\cal M}_2$. (Similarly of course if the spacetimes also have matter and-or radiation fields: one just replaces isometry by isomorphism.) \\ 

\indent \indent \indent \indent (B): More generally and importantly, one should not require (and we do not believe) that there are some criteria that are factually correct (though of course yet to be discovered by physics and-or philosophy), independent of the context. For even though physics is a quantitative, and mostly precise, science, it is important {\it in physics} to be flexible about ``which point is which" between two spacetimes. Most relevant for us here (and for the companion paper  [II]), this flexibility is important for the topic with which (II) above began: viz.\ precisely defining superpositions of spacetimes.  \\
 \indent \indent  \indent \indent \indent \indent Thus we agree that given two isometric spacetimes, it is often natural to cross-identify points between them by the isometry.\footnote{This is often said, also for classical general relativity: in both the physics literature, eg \cite{HE73}, p.\ 56, and the philosophy literature e.g.\ \cite{W18}.  Of course, if each (and thus both!) of the spacetimes is symmetric, i.e.\ has an isometry {\it to itself}, then this proposed cross-identification suffers the ambiguity described in (A) above.}  But note that this is {\it not} always the right way to cross-identify points, even between two isometric spacetimes.\footnote{This is  sometimes said, also for classical general relativity: e.g.\ \cite{L23}, pp 4-6, 10-11, and \cite{G24}. But we concentrate on the quantum context; cf.\ immediately below.}  For consider how to model the gravitational field due to a superposition of a massive source being localized in two distinct spatial regions; (call them `left’ and `right’, lying along some spatial direction). Here, it is natural: (i) to  take two {\it isometric} spacetimes, each containing a single localized mass-density ``hump”, and (ii) to define the superposition by cross-identifying points {\it non-isometrically}, even though the spacetimes are isometric. To state (ii) more precisely: we cross-identify the point lying below the peak of the first spacetime’s hump (which we take as supplying the `left’ branch of the superposition) with a point in the second spacetime lying to the left of (not under!) the peak of the second spacetime’s hump. In short: given two ``duplicate copies”, i.e.\ isometric spacetimes, each with a single hump, we define a superposition of spacetimes by ``putting” both humps on a single manifold, but with one copy’s hump to the left of the other copy’s hump.\\
 \indent \indent  \indent \indent \indent \indent  Note that for the sake of clarity, our exposition has set aside the topic of how far apart the two peaks in the superposition are to be. But one should of course allow for different inter-peak distances, and so different superpositions will be defined by different cross-identification maps that vary about how far ``along to the left” (in the given spatial direction) the point under the peak of the first spacetime’s hump is to be ``sent” in the manifold that is to support the superposed massive source. (For more discussion of how to model the gravitational field due to a spatially superposed massive source, including our awaiting new physical insight,  cf.\ \cite{VB24}, Section V and VI.  See also Ref.~\cite{KHACGBB24}.)  \\
 
Evidently, both types of flexibility (A) and (B) go well with Lewis' counterpart-theoretic approach to cross-identifying objects, including spacetime points.  Indeed, in the philosophical literature (cf.\ references in footnote \ref{philcpartpoints}) both (A)'s allowance of ``tied for first-equal" points, and (B)'s allowance that the ``correct" cross-identification can depend on the context (especially, on what one wants to model), are widely agreed merits of the approach. 

As we have stressed, we urge such flexibility. For our purposes, the only limitation to note is as we explained in [B] before the start of stage (I) above. Namely: for two {\it non}-isometric spacetimes, a cross-identification diffeomorphism cannot push forward one tetrad field into another, though (as  [A]  suggested) the criteria for cross-identification can give some weight to matching the state of the quantum reference frame.

\bigskip

(III): {\it ``Unwinding” the cross-identification map}:---\\
At last, we can gather all the threads of this Section. We will combine the ideas of (I) and (II). More precisely, we use the ideas of (II) to augment and clarify (I)’s justification of our using the frame bundle as {\it not} committing us to haecceitism. As we announced in the introductory summary before the start of (I):  the idea is that our formalism ``unwinds” the diffeomorphism that expresses the  cross-identification map, so as to re-label points as a single point $p$. The ``extra ingredient" supplied by the discussion in (II) is that we can be very flexible about what exactly is the cross-identification diffeomorphism between two non-isometric spacetimes.\footnote{To repeat: we only need to veto its pushing forward a tetrad field in its domain spacetime to a tetrad field in the range: cf.\ [B] before the start of (I).}

So  we are given two spacetimes $({\cal M}_1,g_1)$ and $({\cal M}_2,g_2)$: in general, not isometric,  though we require them to be diffeomorphic. (Or similarly for regions of spacetimes; but as often, we will for clarity and simplicity talk only of entire spacetimes.) Also we are given  a cross-identification diffeomorphism $d: ({\cal M}_1,g_1) \rightarrow ({\cal M}_2,g_2)$. (Again: Following (II), $d$ need not be uniquely correct or context-independent. It can depend on what we are trying to model or describe. But this context-dependence does not affect the details of what follows.)  

Let us map these ideas onto (I)’s idea of a fiducial manifold $\cal M$ and two given ``reference” diffeomorphisms $^{(R)}\varphi_j$, i.e.\ $^{(R)}\varphi_j: {\cal M}_j \rightarrow {\cal M}$. Since the cross-identification map $d$ is a diffeomorphism and so are the $^{(R)}\varphi_j$, the composite map induced on $\cal M$, i.e.\ $^{(R)}\varphi_2 \circ d \circ (^{(R)}\varphi_1)^{-1}: {\cal M} \rightarrow {\cal M}$ is also a diffeomorphism. 

We have thus transferred the discussion of cross-identification between ${\cal M}_1$ and ${\cal M}_2$ onto $\cal M$. Using as we did in (I), the letter $\phi$ as notation for a diffeomorphism of $\cal M$, we will now write this induced composite map as $\phi_{((R),d)}$, so as to show its dependence on the choices of $^{(R)}\varphi_j$, and on $d$. (We write $(R)$ as a subscript in $\phi_{((R),d)}$, not as a superscript, so as to signal $\phi_{((R),d)}$'s dependence on a previous choice of the  $^{(R)}\varphi_j$. There is not a ``free" choice of a ``reference" $\phi$, like there was a ``free" choice of the  $^{(R)}\varphi_j$.)

Now observe that it is perfectly possible for $\phi_{((R),d)}$ to fix a point $p \in {\cal M}$, i.e.\ map $p$ to itself, $\phi_{((R),d)}(p) = p$. Whatever $d$ is---indeed, no matter how disparate the metrics (or other tensorial quantities) are at a point $q \in  {\cal M}_1$ and its $d$-image $d(q) \in {\cal M}_2$---the $^{(R)}\varphi_j$ might happen to map $q$ and $d(q)$ respectively to the same point $p \in {\cal M}$. That is: $^{(R)}\varphi_1(q) = p = \,^{(R)}\!\varphi_2(d(q))$.

Furthermore, it is clear that by appropriately choosing, for our given  $^{(R)}\varphi_1$ and $d$, a $^{(R)}\varphi_2$ (that is probably different from the one we envisaged choosing initially!), we can make $\phi_{((R),d)} \; := \; ^{(R)}\varphi_2 \circ d \circ (^{(R)}\varphi_1)^{-1}$ be the identity map on $\cal M$. All we need to do is define it as the inverse of $d \circ (^{(R)}\varphi_1)^{-1}$. That is, we define: $^{(R)}\varphi_2 := (d \circ (^{(R)}\varphi_1)^{-1})^{-1} \equiv \; ^{(R)}\varphi_1 \circ d^{-1}$.  Then $\phi_{((R),d)}$ is trivially the identity map on $\cal M$.

 This is of course the main idea of what we announced as our frame bundle formalism ``unwinding” a cross-identification diffeomorphism.

In a moment, we will put this in terms of the frame bundle. But let us first adapt the exposition so far, to allow for $d$ being only a local diffeomorphism i.e.\ defined on a region ${\cal O}_1 \subset {\cal M}_1$.

So let $d: {\cal O}_1 \rightarrow d[{\cal O}_1] \subset {\cal M}_2$. (As usual, we use square brackets for the image of a non-singleton subset of a function’s domain. In this example, $d[{\cal O}_1]$ is of course $d$’s entire range.)  Since the restriction of the diffeomorphisms $^{(R)}\varphi_j$ are also diffeomorphisms, the induced composite map from $^{(R)}\varphi_1[{\cal O}_1]$ to $^{(R)}\varphi_2[d[{\cal O}_1]]$ is also a diffeomorphism. That is:
\begin{equation}
\phi_{((R),d)} \;  := \;  ^{(R)}\varphi_2|_{ d[{\cal O}_1] } \circ d \circ (^{(R)}\varphi_1|_{{\cal O}_1})^{-1}
 \end{equation}
is a diffeomorphism. Note that we again write the map as $\phi_{((R),d)}$. So we leave the fact that it is local, i.e.\ defined only on $^{(R)}\varphi_1[{\cal O}_1]$ as implicit, thanks to our understanding that $d$ was defined only on ${\cal O}_1$.

As in the case of a global  diffeomorphism $d$, it is perfectly possible for $\phi_{((R),d)}$ to fix a point of its domain, i.e.\ to map a point $p \in \,^{(R)}\!\varphi_1[{\cal O}_1]$ to itself: $\phi_{((R),d)}(p) = p$. For again the $^{(R)}\varphi_j$ might happen to map $q \in {\cal O}_1$ and $d(q) \in  d[{\cal O}_1]$ respectively to the same point $p \in {\cal M}$. That is, we can have: $^{(R)}\varphi_1(q) \; = \; p \; = \; ^{(R)}\varphi_2(d(q))$.

And on the other hand (and unlike the case of $d$ being a global  diffeomorphism): the ranges of the restricted reference diffeomorphisms, i.e.\ the sets $^{(R)}\varphi_1[{\cal O}_1]$ and $^{(R)}\varphi_2[d[{\cal O}_1]]$ might happen to be disjoint: in which case $\phi_{((R),d)}$ cannot fix any points.

But the main idea remains as in the case of global $d$. That is: by appropriately choosing, for our given  $^{(R)}\varphi_1$ and $d$, a $^{(R)}\varphi_2$, we can make $\phi_{((R),d)} := \,^{(R)}\varphi_2 \circ d \circ (^{(R)}\varphi_1|_{{\cal O}_1})^{-1}$ be the identity map on $^{(R)}\varphi_1[{\cal O}_1] \subset {\cal M}$. Again, all we need to do is define $^{(R)}\varphi_2$ as the inverse of $d \circ (^{(R)}\varphi_1|_{{\cal O}_1})^{-1}$. 
This trivially makes  $\phi_{((R),d)}$ the identity map on $^{(R)}\varphi_1[{\cal O}_1] $. So to sum up: we see that the main idea of  ``unwinding” a cross-identification diffeomorphism can be implemented for a local diffeomorphism $d$.

Finally, let us put all this in terms of the frame bundle. It is really just a matter of combining the above with the comments at end of  Section \ref{ssec:bundle}. Recall from there\footnote{Or from the first comment, about formalism, at the very start of this Section. The main idea we need is in  its second point, (2). Its first point (1) added detail about the structure of each 16-dimensional fibre $\pi^{-1}(p)$ of ${\cal F}({\cal M})$. Namely: it is foliated by the 6-dimensional cosets of  $O(3,1)$, with $GL(4)/O(3,1)$ being in bijection with the set $Sym_L$ of Lorentzian-signature symmetric bilinear forms on ${\mathbb R}^4$.} that although a section $\sigma$ of ${\cal F}({\cal M})$  ``over-specifies" a spacetime,  since it includes a specific basis-field,  the union of a section $\sigma$  (taken as the embedded sub-manifold) and all its pointwise Lorentz transforms---which is thus a subset of ${\cal F}({\cal M})$ (more precisely: of the ``column" $\pi^{-1}[{\cal M}]$)---is the orthonormal frame bundle ${\cal F}_{\rm o}[g]$ for the metric $g$ defined by the section $\sigma$. (Here, ${\cal F}_{\rm o}[g]$ is of course defined equally well by any of $\sigma$'s pointwise Lorentz transforms.) And these remarks also of course apply locally. That is: the union of  a local section $\sigma$ on a region ${\cal O} \subset {\cal M}$ with all its pointwise Lorentz transforms is a subset of ${\cal F}({\cal O})$; and it {\it is} the orthonormal frame bundle ${\cal F}_{\rm o}[g]$ for the metric $g$ on $\cal O$ defined by $\sigma$ (and by any of its transforms).

So suppose again that we are given two spacetimes $({\cal M}_1,g_1)$ and $({\cal M}_2,g_2)$ (in general, not isometric) and a cross-identification diffeomorphism between them $d: ({\cal M}_1,g_1) \rightarrow ({\cal M}_2,g_2)$. Then whatever constructions we  wish to make about  $({\cal M}_1,g_1)$ and-or $({\cal M}_2,g_2)$,  in particular constructions that advert to the diffeomorphism  $d$ (for example,  so as to define a superposition of the spacetimes, with  associated  complex amplitudes),  can of course be done in the setting of their orthonormal frame bundle(s) ${\cal F}_{\rm o}[g_i], i = 1,2$. For since a spacetime  $({\cal M},g)$ determines its orthonormal frame bundle ${\cal F}_{\rm o}[g]$, any constructions about $({\cal M},g)$ that ``go beyond" ${\cal F}_{\rm o}[g]$ must also ``go beyond" $({\cal M},g)$. So whatever one does for one's desired construction about $({\cal M},g)$ can be transferred  to the ``richer'' setting of ${\cal F}_{\rm o}[g]$. 

We  then apply to the construction the  discussion above. That is: we first transfer the construction by the reference diffeomorphisms $^{(R)}\varphi_i$ ($i =1$ and-or 2) to our fiducial manifold $\cal M$, given as diffeomorphic to both ${\cal M}_1$ and ${\cal M}_2$. Then we note that whatever the cross-identification diffeomorphism $d$, and our choice of  $^{(R)}\varphi_1$, we can appropriately choose  $^{(R)}\varphi_2$ so as to make the induced map $\phi_{((R),d)}$ the identity map. This point is of course independent of what  the construction might be.  

So to conclude:--- We have shown how for any such construction, we can:\\
\indent \indent \indent \indent  (i) ``unwind" the cross-identification diffeomorphism $d$, whatever it might be---a choice about which we are flexible, cf.\ (II)---and;\\
\indent \indent  \indent \indent  (ii)  represent the construction on the frame bundle $\cal {F(M)}$ of our fiducial manifold $\cal M$.\\
Besides, as noted above: This result---that  we can make the induced map $\phi_{((R),d)}$ be the identity map---holds true both when the diffeomorphism $d$ is global, and when it is local i.e.\ defined on a region ${\cal O}_1 \subset {\cal M}_1$. It is just that in the latter case, $\phi_{((R),d)}$ is the identity map, not on all of $\cal M$, but only on  $^{(R)}\varphi_1[{\cal O}_1] \subset {\cal M}$.

Now, we will again proceed in two stages. First, we will formally define a general quantum reference frame, and make some comments about it (Section \ref{ssec:bs:definition}). Then we will discuss normalisation, and the  ensuing expectation values of measurement results  (Section \ref{ssec:bs:normnexpn}).\\

\subsection{The definition}\label{ssec:bs:definition}
We now {\bf define}: 
\begin{quote} 
A {\bf g-quantum reference frame} (a g-QRF) $\Psi$ on $\cal O$ is an assignment of a complex amplitude (possibly zero) to every element of the column $\pi^{-1}[{\cal O}] \subset {\cal {F(M)}}$. We will write  ${\bf {P_b}} := (p ; \{{\bf b}_\mu^a(p)\})$ for a point of this column (`b' for basis); for short, we will also write ${\bf {P_b}}$ as  $(p, {\bf b}_p)$. Then the complex amplitude will be: 
$\Psi({\bf {P_b}}) \equiv \Psi(p ; \{{\bf b}_\mu^a(p)\}) \equiv \Psi(p, {\bf b}_p)$.  This assignment is also required to obey the normalisation condition Eq.\ \ref{normgeneral} in Section \ref{ssec:bs:normnexpn} below; and its fibre integrals are required to be finite, Eq.\ \ref{normpgeneralinfy}.\footnote{Again: the comments in footnotes \ref{config}, \ref{config2} and \ref{avoidstate1}, about our using the phrase `configuration space' and our not using the phrase `state of quantum reference frame', apply  here in the general framework.} 
\end{quote}

Now we can see how  what we said at the end of Section \ref{ssec:bundle} makes more specific the second comment in this Section's Preamble, about the twofold interpretation of tetrads. Namely,  the amplitude $\Psi({\bf {P_b}}) \equiv \Psi(p ; \{{\bf b}_\mu^a(p)\})$ is of course the amplitude for the apparatus to be (i) at the point $p$ and (ii) in the configuration $\{{\bf b}_\mu^a(p)\}$. But note that  by (ii) we mean  `the configuration $\{{\bf b}_\mu^a(p)\}$ understood as orthonormal in some or other metric/geometry on $\cal O$'.

We said in Section \ref{ssec:bundle} that this assignment will be subject to constraints. But in this paper, we will impose only the mathematical constraint that the g-QRF be suitably well-behaved so as to allow integration, i.e.\ so as to be in an $L^2$-space (and, as stated in the definition, normalised). For we of course want, on analogy with Section \ref{sec:perspsing}, to consider complex  linear combinations of g-QRFs in the obvious way, i.e.\ point-wise; so that for each  region ${\cal O}$, the g-QRFs on $\cal O$ are the normalised elements of a complex vector space of wave-functions: (cf.\ footnote \ref{footdelta}, after the restricted framework's normalisation condition Eq.\ \ref{norm20Oct}). There are two reasons for this simplicity. 

(1): The first reason is a warning. Since we are thinking of bases in the bundle as (collectively) specifying a metric on $\cal O$, one might think we should require that for each spacetime point $p \in {\cal O}$, there is some $(p, {\bf b}_p)$ in $\pi^{-1}[\{p\}]$ whose amplitude $\Psi(p ; \{{\bf b}_\mu^a(p)\})$ is non-zero.  For one might think: ``surely we need this, so as to avoid ``gaps" in the specification of ``one or another" geometry?" But this proposed constraint confuses: (a) our using, as theorists, bases in the bundle to specify a metric throughout $\cal O$; and (b) the g-QRF as a physical system having, for any point in $\cal O$, a non-zero amplitude to be located there. We of course uphold (a). But (a) does not imply (b).\footnote{Agreed, if we had adopted the more ambitious conception of g-QRFs as {\it defining} the spacetime geometry---i.e.\ what footnote \ref{setasidebasic} called `the basic conception',  and then set aside---then the proposed constraint (b), i.e.\ $\pi[{\rm supp}(\Psi)] = {\cal O}$, would be justified.}

(2): The second reason is more substantive. Throughout this paper (and in {\bf [I]} and {\bf [II]}), we are setting aside dynamics. In the general framework, this means ignoring the dynamics of both: the (single) infinitesimal apparatus, and of the spacetime geometry (whether treated as the temporal evolution of a 3-geometry on an appropriate spatial slice, or as a constraint equation for the 4-geometry). Agreed: in a full treatment, we would of course have to incorporate dynamics. But once we set dynamics aside, there is in the general framework no {\it a priori} geometry to be respected, and no further constraint is needed for our purposes here.

We also wish to interpret a g-QRF on $\cal O$ on analogy with Section \ref{sec:perspsing}'s  restricted framework, and specifically Section \ref{ssec:ps:normn}'s opening assumption that the apparatus, although  pointlike,  is coherently spread over the  region ${\cal O}$: which made Eq.\ \ref{norm20Oct} the restricted framework's correct normalisation. Thus for the general framework: we intuitively expect that $\left|\Psi(p,{\bf b}_p)\right|^2$ will be a Born-rule probability density for:\\
\indent \indent \indent \indent (i)  the metric tensor {\it at $p$}, i.e.\ at the apparatus' location, to be one according to which ${\bf b}_p$ is an orthonormal basis---i.e., to be
\begin{eqnarray}
g_{ab} = \eta_{\mu\nu} {\bf B}^\mu_a{\bf B}^\nu_b,
\label{eq:metricb}
\end{eqnarray}
with $\{{\bf B}_a^\mu\}$ being the dual basis (of $T_p^\ast {\cal O}$) to $\{{\bf b}_\mu^a\}$ and $\eta_{\mu\nu} = {\rm diag}(-1,+1,+1,+1)$ uniform {\it scalar} fields; and for :\\
\indent \indent \indent \indent (ii) the apparatus, assumed {\it ab initio} to  be in a region contained in $\cal O$, to be at  the location and (orthonormal!)~orientation $(p, {\bf b}_p)$.\\
 We stress that $\Psi$ assigns no probability to a metric {\it field} on $\cal O$: the values of the metric at distinct points are not correlated by $\Psi$, which describes a single apparatus at a single, though uncertain, event. This is in line with our setting aside, in (A) of Section \ref{sec:intro}, amplitudes for sections, i.e.\ for geometries.\footnote{\label{noprobgeom}One might think that we can  extract probabilities for geometries by integrating $|\Psi|^2$ over the region of ${\cal F}({\cal O})$ swept out by the orthonormal frame bundles ${\cal F}_{\rm o}[g]$ of some set of metrics $g$ on $\cal O$. But this does not define a probability measure on geometries. For each point $(p,{\bf b}_p)$ of ${\cal F}({\cal O})$ lies in ${\cal F}_{\rm o}[g]$ for {\it every} metric $g$ on $\cal O$ whose value at $p$ is given by Eq.\ \ref{eq:metricb}; so disjoint sets of geometries can sweep out overlapping regions. (Besides, each ${\cal F}_{\rm o}[g]$ is a 10-dimensional submanifold of the 20-dimensional ${\cal F}({\cal O})$, and so has zero measure for the measure we will adopt in Eq.\ \ref{eq:muO}.) Probabilities for geometries would require a measure on the set of sections $Sec({\cal F}({\cal O}))$---i.e.\ the basic conception set aside in footnote \ref{setasidebasic}.}
 
This expectation will indeed be upheld in what follows, {\it modulo} the discussion in Section \ref{ssec:bs:normnexpn} below about our needing care over normalisation. In particular, we note that:\\
\indent \indent \indent (i): The comments (3)  and (4) at the end of Section \ref{ssec:ps:normn}, about this paper not being concerned with non-commuting quantities, nor with the dynamics of our quantum reference frames, apply again in the general framework---word for word. \\
\indent \indent \indent (ii): When we turn in Section \ref{ssec:bs:normnexpnregion} to defining expectation values of measurement results {\it on a region $\cal O$}, there will again be no difficulties compared with the restricted framework, thanks to our assuming, as  in  (3) and (4) of  Section \ref{ssec:ps:expnvals}, that the apparatus is somehow designed so as to measure the field in question (the electric field, or the energy density) {\it just once}, somewhere sometime in the region $\cal O$ and with some or other internal configuration of the apparatus.\footnote{\label{severeideal}{Admittedly, this assumption involves an idealisation here, additional to what it involved for the restricted framework. For it is no doubt more difficult to engineer that an apparatus register (``fire") just once in a given spacetime region, whatever the geometry of that region might be, than to engineer that it register just once in the region as endowed with a single geometry. This difficulty is even larger when one remembers that our phrase `in a given spacetime region' leans on our haecceitistic-looking---but we say: harmless---formalism.}}

\bigskip

\subsection{Normalisation and expectation values}\label{ssec:bs:normnexpn}

\noindent We will introduce and justify our normalisation condition, given in Eq.\ \ref{normgeneral} below, by pursuing the analogy with the restricted framework. The main novelty will be that now, thanks to our defining fibres by the action of $GL(4)$ not $O(3,1)$, the volume-element {\it form} (or volume element, for short) varies along a fibre $\pi^{-1}(p)$, in order to ensure that every basis at $p$ be orthonormal in the corresponding metric.\footnote{\label{epsilon} But note that  the pseudo-scalar obtained from the volume form {\it is} $\pm1$ {\it throughout the fibre}, since every element of the fibre is a tetrad in the corresponding metric.  This is why the volume {\it form} varies along the fibre. For if it were constant, the pseudo-scalar obtained when applied to different elements of the fibre could not be $\pm 1$ throughout, but would satisfy the tensor transformation rule for the {\it single} (volume) tensor.} So  the volume element  cannot be taken out of the fibre integral and appear only in an integration over $\cal O$, as it did in the restricted framework's  normalisation condition Eq.\ \ref{norm20Oct} and in its ensuing expectation values like  Eq.\ \ref{EvalueQRFnewspt}. 

So we will,  first:\\
\indent  \indent   \indent  (A) calculate how the volume element varies along a fibre; and then \\
\indent  \indent  \indent   (B) discuss how this affects  the formulation of our normalisation condition, especially the attempt to proceed on analogy with Eq.\ \ref{norm20Oct}. \\
In both (A) and (B), we will as usual  ``localize" the discussion to a region $ {\cal O} \subset {\cal M}$, and so we consider the bundle ${\cal {F(O)}}$. 

Then in (C), we return to how the general framework addresses the interpretative issue of  {\it  haecceitism}. For we need to note a subtlety about what, in this framework, the phrase `expectation value at a point' (or `at a region') really means. 

Then with that point in hand, we can, from (D) onwards, turn to the details of defining expectation values. \\

\noindent (A):  {\it How the volume element varies}:---\\
In the general framework, any element of $\cal {F({\cal O})}$ above a point $p \in {\cal O}$ can be interpreted as a tetrad. So assuming ${\bf {P_b}} := (p ; \{{\bf b}_\mu^a(p)\}) \in {\cal {F({\cal O})}}$ is so interpreted (i.e.\ forms part of the basis field we wish to stipulate to be orthonormal), the 4-volume element  form thereby defined satisfies
\begin{equation}
\epsilon_{abcd} {\bf b}_{\alpha}^a {\bf b}_{\beta}^b {\bf b}_{\gamma}^c {\bf b}_{\delta}^d =  \epsilon_{\alpha \beta \gamma \delta},
\label{setepsilon}
\end{equation}
with the r.h.s.\ defined by $\epsilon_{0123} = +1$ and complete antisymmetry among the indices.\footnote{Again, we write `${\bf b}$' to signal that this is an element of an arbitrary basis, as well as writing $b$ as superscript or subscript as part of the abstract-index notation. But no confusion will arise.} It is important to stress that Eq.\ \ref{setepsilon} defines $\epsilon_{abcd}$, not $ \epsilon_{\alpha \beta \gamma \delta}$. In fact, since any element $\{{\bf b}_\mu^a(p)\}$ of $\cal {F({\cal O})}$ can be interpreted as a tetrad, $ \epsilon_{\alpha \beta \gamma \delta}$ are {\it uniform scalar fields} throughout  $\cal {F({\cal O})}$; what varies is the tensor $\epsilon_{abcd}$. Inverting this relation, we have
\begin{equation}
\epsilon_{abcd} = \epsilon_{\alpha \beta \gamma \delta}  {\bf B}^{\alpha}_a {\bf B}^{\beta}_b {\bf B}^{\gamma}_c {\bf B}^{\delta}_d \; ,
\label{invertepsilon}
\end{equation}
where again $\{{\bf B}^\mu_a(p)\}$ is the dual basis (of $T^*_p {\cal O}$) to $\{{\bf b}_\mu^a(p)\}$.  This equation plays the same role for the volume element as Eq.~\ref{eq:metricb}  for the metric tensor.

So changing $\{{\bf b}_\mu^a(p)\}$ along the fibre $\pi^{-1}(p)$, by $\{{\bf b}_\mu^a\} \mapsto \{M^\nu_\mu {\bf b}_\nu^a \} =: \{ {\tilde {\bf b}}_\mu^a\} $ where $M \in GL(4)$,  changes  $\epsilon_{abcd} $ through
\begin{eqnarray}
{\tilde \epsilon}_{abcd} &=& \epsilon_{\alpha \beta \gamma \delta}  {\tilde {\bf B}}^{\alpha}_a  {\tilde {\bf B}}^{\beta}_b  {\tilde {\bf B}}^{\gamma}_c  {\tilde {\bf B}}^{\delta}_d \nonumber \\ 
&=& (M^{-1})^{\alpha}_{\mu}(M^{-1})^{\beta}_{\nu}(M^{-1})^{\gamma}_{\rho}(M^{-1})^{\delta}_{\sigma}  \epsilon_{\alpha \beta \gamma \delta} 
{\bf B}^{\mu}_a {\bf B}^{\nu}_b {\bf B}^{\rho}_c {\bf B}^{\sigma}_d  \nonumber\\
&=& {\rm det}(M^{-1})  \epsilon_{\mu \nu \rho \sigma} 
{\bf B}^{\mu}_a {\bf B}^{\nu}_b {\bf B}^{\rho}_c {\bf B}^{\sigma}_d  \nonumber\\ 
&=& \frac{1}{{\rm det}(M)} \epsilon_{abcd}  \, .
\label{trsfmepsilon1}
\end{eqnarray}
Thus in the general framework the 4-volume element varies along the fibres $\pi^{-1}(p)$, in the way shown in Eq.\ \ref{trsfmepsilon1}. \\

\noindent (B): {\it The normalisation condition}:---\\
 So if we plan to formulate a normalisation condition on analogy with Eq.\ \ref{norm20Oct}, the volume element (i.e.\ its absolute value) will have to be (at least partially) included in the integrand of the fibre integral (together with $\left|\Psi(p;\{M{\bf b}(p)\})\right|^2$---cf.\ below). We write this absolute value as $\epsilon$ i.e.\ without the modulus bars.\footnote{For in the differential geometry literature, omitting the modulus bars is usual, since $n$-volumes {\it are} taken as positive. We stress that our topic here is $\epsilon$, not the tensor (4-form) $\epsilon_{abcd}$ defined by Eq.\ \ref{setepsilon}; cf.\ \ footnote \ref{epsilon}.}

Thus we see that we {\it cannot} take  as our normalisation condition for $\Psi$ on $\cal O$, in ``naive" analogy with Eq.\ \ref{norm20Oct}, the formula
\begin{equation}
\int_{{\cal O}}   \, \epsilon(p)  \int_{GL(4)} d\mu_G(M)\,    \left|\Psi(p;\{M{\bf b}(p)\})\right|^2 = 1 , 
\label{normgeneraltrial}
\end{equation} 
where:\\
\indent \indent \indent \indent (i) $\{{\bf b}(p)\}$ is a {\it fiducial}  basis field defined on all points $p \in  {\cal O}$ (which we are thinking of as  stipulated to be a tetrad field); \\
\indent \indent \indent \indent (ii) 
$\epsilon$ the volume element associated to the fiducial basis ${\bf b}$; (and on analogy with comment (1) just after Eq.\ \ref{norm20Oct}, we secure a positive volume element by taking the absolute value); and \\
\indent \indent \indent \indent (iii) $\mu_G$ is the (unique up to a multiplicative factor) invariant Haar measure defined on $GL(4)$. 

For there are two problems about Eq.\ \ref{normgeneraltrial}. The first is about  $\mu_G$'s multiplicative factors mentioned in (iii). This problem is familiar from the restricted framework: that is, from comment (2) just after Eq.\ \ref{norm20Oct}---and fortunately, it can be solved in the same way as we adopted there. Namely,  the fixed fiducial basis field $\{{\bf b}_\mu^a\}$---which is a 
section of $\pi^{-1}[{\cal O}]$---used to make sense of the fibre integration,  provides a corresponding identification among measures in different fibres. This is so because
the $GL(4)$ acts freely and transitively on 
each fiber $\pi^{-1}(p)$, which means that each fibre is a ``copy'' of  $GL(4)$ with the identity ``rubbed out''.
The role of the fiducial basis field is exactly selecting, at each $p\in {\cal O}$, the fibre element
$(p;\{{\bf b}_\mu^a(p)\})$ which is to be identified with the group identity.  Once this is done,  all fibers become
uniquely identified among themselves and to $GL(4)$: $(p;\{M^\nu_\mu{\bf b}_\nu^a(p)\})$ is identified to $(q;\{M^\nu_\mu{\bf b}_\nu^a(q)\})$ and to $M$,  for all $p,q\in{\cal O}$, $M \in GL(4)$.
Hence,  fixing one invariant Haar measure on $GL(4)$ fixes the integration measure on all fibres of $\pi^{-1}[{\cal O}]$,  and the remaining uniform multiplicative-constant indeterminacy is eventually absorbed
by the normalization of $\Psi$. 

The second problem is new, i.e.\ specific to the general framework. It is the one announced above, at the start of this item (B). Namely, the  volume element $\tilde{\epsilon}(p)\equiv \epsilon(p;M)$ that is determined by stipulating   all $\tilde{\bf b}= M{\bf b}$, with $M\in GL(4)$,  to be a tetrad field on $\cal O$, varies along the fibres $\pi^{-1}(p)$,  according to Eq.\ \ref{trsfmepsilon1}. So in our normalisation condition,  the absolute value of the volume element has to be included in the fibre integral's integrand. 

Fortunately,  the dependence of $\tilde{\epsilon}$ on $M\in GL(4)$ and on $p\in{\cal O}$  is {\it separable  once the fiducial basis field {\bf b} on ${\cal O}$}---i.e.,  a reference foliation
of $\pi^{-1}[{\cal O}]\subset {\cal F}({\cal M})$ (whose leaves are the sections $\{M{\bf b}\}$, with $M\in GL(4)$ the same at all points of ${\cal O}$)---{\it is given}. This allows the  definition of 
a {\it natural}  integration measure on 
$\pi^{-1}[{\cal O}]$---hence, on ${\cal F}({\cal M})$---such that the frame-bundle integration
of a function $f:{\cal F}({\cal O})\to {\mathbb C}$
is given by
\begin{eqnarray}
\int_{\pi^{-1}[{\cal O}]}d\mu_{\cal F}({\bf P}_{\tilde{\bf b}}) \;f({\bf P}_{\tilde{\bf b}})&:=& 
\int_{\cal O}\int_{GL(4)} d\mu_G(M)\;\tilde{\epsilon} ({p})\; f(p;\{M{\bf b}(p)\})\nonumber \\
&=&\int_{\cal O}\epsilon(p) \int_{GL(4)} \frac{d\mu_G(M)}{|{\rm det}(M)|} \;f(p;\{M{\bf b}(p)\}),
\label{eq:muO}
\end{eqnarray}
where $\epsilon$ is the {\it fixed} volume element on ${\cal O}$ associated to the metric that makes 
the fiducial basis field ${\bf b}(p)$ a tetrad field, and hence is independent of $M$. $\epsilon$ being independent of $M$  is of course a necessary condition
for this definition Eq.\ \ref{eq:muO} to make sense. 
The other conditions are, of course, that the result should not depend
on the choice of fiducial basis field ${\bf b}(p)$, nor on the choice of coordinates to cover ${\cal O}$. We leave the
proof of these conditions to the Appendix. The Appendix (part (C)) also shows that the construction fixes the overall multiplicative constant of this measure---which is anyway immaterial, since it would be absorbed in the normalisation of $\Psi$.

So in analogy with Eq.\ \ref{norm20Oct},  our normalisation condition for $\Psi:\pi^{-1}[{\cal O}]\to {\mathbb C}$ 
is\footnote{Again, for the sake of rigour, cf.\ footnote \ref{footdelta}.}
\begin{equation}
\fbox{ $
\int_{\pi^{-1}[{\cal O}]}d\mu_{\cal F}({\bf P}_{\tilde{\bf b}}) \;\left|\Psi({\bf P}_{\tilde{\bf b}})\right|^2   \equiv
\int_{\cal O}\epsilon(p) \int_{GL(4)} \frac{d\mu_G(M)}{|{\rm det}(M)|} \;\left|\Psi(p;\{M{\bf b}(p)\})\right|^2 
= 1  \; .
$}
\label{normgeneral}
\end{equation}

Besides, just as in the restricted framework the fibre integral was a marginal probability density function (for location at the point $p$), so that we  therefore required it to be finite (cf.\ Eq.\ \ref{normpjfiducialfnoinfy}): so also here we interpret the fibre integral as the marginal probability density function for location at $p$. So we require that 
\begin{equation}
\|\Psi(p) \|^2 :=  \int_{GL(4)} \frac{d\mu_G(M)}{|{\rm det}(M)|} \; \left|\Psi(p;\{M{\bf b}(p)\})\right|^2  < \infty  \; .
\label{normpgeneralinfy}
\end{equation} 
One important comment should be made here.  Note that $\| \Psi(p)\|^2$ is a (spatio-temporal) {\it density}. Comparing Eq.~\ref{normpgeneralinfy}  with Eq.~\ref{eq:muO}, we immediately see that it is a density w.r.t.\ the volume
element $\epsilon$, which is determined by the metric according to which the fiducial basis ${\bf b}$ is orthonormal. Changing the fiducial basis  changes the density $\|\Psi(p)\|^2$ in just the right way to make its spacetime integral,
Eq.~\ref{normgeneral},  fiducial-basis independent. Although this dependence of $\|\Psi(p)\|^2$ on an arbitrary choice may at first seem undesirable, we will see below (Secs.~\ref{ssec:bs:normnexpnpoint} and \ref{ssec:bs:normnexpnregion}) that expectation values of observables are well-defined
basis-independent quantities.\\

\noindent (C): {\it Addressing haecceitism when defining expectation values}:---\\
  In defining expectation values within the general framework, we must be clear what exactly the phrases `expectation value {\it at a point}', and `expectation value {\it in a region}'  mean.  For recall Section \ref{ssec:conceptual}'s rebuttal of the accusation that our formalism's use of $\cal {F(O)}$, with a fibre above each point $p$, is {\it  haecceitistic}. We agreed that haecceitism is wrong. But we explained that our use of $\cal {F(O)}$ amounts to just ``unwinding" whatever is one's adopted cross-identification map between spacetimes. Thus this rebuttal prompts us to clarify that we intend phrases like `marginal probability density for location at $p$' and `expectation value {\it at a point}/{\it in a region}' to reflect this ``unwinding".

To put this more precisely is quite a mouthful, i.e.\ cumbersome to state. But we spell it out as follows: as usual,  for a g-QRF $\Psi$ defined on a region $\cal O$. We discuss only the case of `expectation value {\it at a point}'. For that case will make equally clear the meanings of `marginal probability density for location at $p$', and `expectation value {\it in a region}'.

Thus: `Expectation value at the point $p \in {\cal O}$' will mean a weighted average of  expectation values of the quantity concerned, where these expectation values: \\
\indent \indent \indent (i) are calculated much as they were in the {\it restricted} framework i.e.\ using a given metric; and \\
\indent \indent \indent  (ii) are evaluated at the sequence of {\it physical} spacetime points that are the images of $p$ under the adopted cross-identification maps between spacetimes; and \\
\indent \indent \indent  (iii) are assigned, in the weighted average, weights  that are fixed by the Born-rule probabilities  
 for the apparatus, given that it is at $p$, to be in a basis that is orthonormal for the corresponding value of the metric {\it at $p$} (with some or other spatial configuration and time-axis); that is, for each such value of the metric at $p$, the weight is fixed by the integral of $| \Psi |^2$ (with the measure in Eq.\ \ref{normpgeneralinfy}, and divided by $\|\Psi(p)\|^2$) over the corresponding leaf, i.e.\ coset of $O(3,1)$, in the fibre $\pi^{-1}(p)$. Note that these weights are {\it not} probabilities for entire spacetimes, i.e.\ for metric fields on $\cal O$; cf.\ footnote \ref{noprobgeom}.

Let us illustrate this general statement with a simple example. Consider a g-QRF $\Psi$  on $\cal O$ whose support  in the fibre above $p$ is concentrated near the two leaves of bases that are respectively orthonormal for the values at $p$ of two metrics $g_1$ and $g_2$ on $\cal O$: where the Lorentzian manifolds $({\cal O}, g_1)$ and $({\cal O}, g_2)$ are {\it not} isometric, and $g_2$ is understood as ``unwound'' by a cross-identification map $d$ mapping a point  $p$ in $({\cal O}, g_1)$ to  $d(p)$ in $({\cal O}, g_2)$---cf.\ (III) of Section \ref{ssec:conceptual}. 
 
 Applying the general statement above to this example, we infer: the expectation value according to $\Psi$ of a quantity $A$ at the point $p$ in $({\cal O}, g_1)$ is
a weighted average of two  expectation values of $A$, that are:\\
\indent \indent \indent  (i) both calculated much as  in the {\it restricted} frameworks ${\cal F}_{\rm o}[g_1]$ and ${\cal F}_{\rm o}[g_2]$ respectively, but \\
\indent \indent \indent (ii) with the first evaluated at  $p$ in $({\cal O}, g_1)$, and the second evaluated at  $d(p)$ in $({\cal O}, g_2)$; and \\
\indent \indent \indent (iii) with each given as its weight: the integral of $| \Psi |^2$ over its respective 
 part of the fibre $\pi^{-1}(p)$, i.e.\ the part near the leaf corresponding to the value at $p$ of the metric $g_i$ (with the measure in Eq.\ \ref{normpgeneralinfy}, and divided by $\|\Psi(p)\|^2$).

Note that only the values at $p$ of $g_1$ and $g_2$ enter here: $\Psi$ does not distinguish $g_1$ (or $g_2$) from any other metric on $\cal O$ with the same value at $p$. The metric fields $g_1$ and $g_2$ matter only through the classical quantity $A$ whose expectation value is taken. This point will be further clarified by a {\it simplifying assumption} that we will introduce in (1) of Section \ref{ssec:bs:normnexpnpoint} below.

We can sum up this discussion, as follows. We will give the phrase `expectation value at the point $p$' a meaning that involves contributions from different spacetimes (geometries). But doing this while mentioning just one point $p$ does not really commit us to haecceitism as a doctrine. For  $p$ functions merely as our label for the  identificatory ``thread" through various spacetimes. Namely: the thread that ``identifies" the point $p$  with those points in other spacetimes that, according to the adopted cross-identification diffeomorphisms, each correspond to $p$.

\bigskip

\noindent (D): {\it Defining expectation values}:---\\
 So much by way of: justifying our normalisation condition, Eq.\ \ref{normgeneral} in (B); and  explaining in (C)  how our expectation values will not ``re-introduce"  haecceitism. We can now, at last, define the expectation values.
 
  For clarity of exposition, we will divide our discussion in two Subsubsections. In Section \ref{ssec:bs:normnexpnpoint}, we will  treat  expectation values at a point $p$. Then in Section \ref{ssec:bs:normnexpnregion}, we treat expectation values in a region $\cal O$. So this division will echo Section \ref{ssec:ps:expnvals}'s division of its four examples, (1) to (4), into two pairs: the first pair, viz.\ (1) and (2), at a point; the second pair for a region and at a time.  But note that the echo is not exact. For in the restricted framework, with a single metric, the notion of a spacelike slice is unproblematic; so that example (4) took a time as a spacelike slice $\Sigma$, and treated this as a special case of a region. But in the general framework, with more than one metric in play, defining an appropriate  spacelike slice to take as the time in question is problematic. So there will be no analogue of Section \ref{ssec:ps:expnvals}'s example (4). In any case, we make the parallel with Section \ref{ssec:ps:expnvals}'s ordering of its examples, (1) to (3), as close as possible, by discussing in order, with the same labels:\\
\indent \indent  \indent \indent  (1):  The expectation value, at a point $p$, of time and spatial separations; \\
\indent \indent  \indent \indent  (2):  The expectation value, at a point $p$,  of the classical electric field;\\
\indent \indent  \indent \indent  (3):  The expectation value, in region $\cal O$, of the classical electric field;\\

But we should note at the outset a limitation of the definitions that follow, in (1) to (3) below. It arises from the fact that, when there {\it is} a definite metric (as in our restricted framework), the value of a classical quantity on a system in general depends, of course, on the metric: and not just on the metric at the point $p$ in question, but also on the metric elsewhere---most obviously because the metric elsewhere, in particular in the past of $p$, contributes to the system's dynamics over time. Besides, the metric defines a canonical, i.e.\ basis-independent, isomorphism between $T_p$ and $T_p^*$, so that we could define an expectation value equally well in terms of: (i) coupling a tetrad to a tensor for the measured system e.g.\ the electromagnetic tensor $F_{ab}$; or in terms of: (ii) coupling a co-tetrad to a corresponding tensor e.g.\ $F^{ab}$. So when we turn to the general framework, the definition of expectation values needs to address questions about how quantities' values (and so their expectation values) are determined when more than one metric is ``in play''. We will address these questions by making a simplifying assumption: details in (1) below. In effect, the assumption will be a natural, indeed mild, restriction on what we will mean by a `classical quantity'.

But apart from this restriction,  and (as explained above) there being no analogue of Section \ref{ssec:ps:expnvals}'s example (4):---  Our endeavour to define expectation values will involve no further difficulties or surprises, compared with the restricted framework. (That is indeed a relief, given this Section's long preliminary stage-setting.) In particular:---\\
\indent \indent \indent \indent  (1): The remarks at the end of Section \ref{ssec:ps:normn}, and the start of Section \ref{ssec:ps:expnvals}---about  (i)  this paper not being concerned with non-commuting quantities, nor with the dynamics of our quantum reference frames, and about (ii) expectation values obviously being sensitive to where in the bundle the support of $\Psi$ happens to be---apply again. Indeed, these remarks  apply again, word for word. \\
\indent \indent \indent \indent  (2): In Section \ref{ssec:bs:normnexpnregion}'s  definition of expectation values on a region $\cal O$, we will assume, just as we did in  (3) of  Section \ref{ssec:ps:expnvals} (and as we announced in (ii) at the end of Section \ref{ssec:bs:definition}), that the apparatus, though infinitesimal in temporal and spatial extension, measures the field in question just once, somewhere in the region $\cal O$ and with some or other internal configuration.\footnote{A familiar qualification, about this assumption involving a more severe idealisation than in the restricted framework: cf.\ again footnote \ref{severeideal}.}

\subsubsection{Expectation values at a point}\label{ssec:bs:normnexpnpoint}

\noindent (1):  {\it The expectation value, at a point, of time and spatial separations}:--- \\
We first recall the ideas and notation from Section \ref{ssec:ps:expnvals}; then we discuss the subtlety about integrating across different geometries/metrics; and then we give our proposed definition of the expectation values. 

In (1) of Section \ref{ssec:ps:expnvals}, we were given two infinitesimally close points  $p$ and $q$, joined by a curve with tangent vector $\xi^a$. We then  proposed that the expectation value of the time separation and spatial separations (along each direction) be given by Eqs.~\ref{ps:expntime} and \ref{ps:expnlength}: which we now repeat, for convenience. (Recall that: (i) the expectation values are conditional on the measurement occurring at the point, so  that we divide by the squared norm of the fibre integral,  which  Eq.\ \ref{normpjfiducialfnoinfy} required to be finite; and (ii) we write the dual basis of the tetrad  $\{{\bf e}_\mu^a(p)\}$ at $p$, as $\{{\bf E}^\mu_a(p)\}$, so that $ g_{ab} {\bf e}^b_\mu = \eta_{\mu \nu} {\bf E}^\nu_a$.) 
\begin{eqnarray}
\langle d\tau \rangle &=& - \|\Psi(p)\|^{-2}
\int_{O(3,1)}d\mu_L(\Lambda) 
g_{ab}(p){\bf e}_\alpha^a \Lambda_0^\alpha d\lambda \,{\xi}^b
\left|\Psi(p;\{\Lambda{\bf e}\})\right|^2 	\nonumber \\
&
=& 
-\|\Psi(p)\|^{-2}
\int_{O(3,1)}d\mu_L(\Lambda) 
\eta_{\alpha\beta}{\bf E}^\alpha_a \Lambda_0^\beta d\lambda\, {\xi}^a
\left|\Psi(p;\{\Lambda{\bf e}\})\right|^2\nonumber\\
 & =&
  \|\Psi(p)\|^{-2}\int_{O(3,1)}d\mu_L(\Lambda) 
(\Lambda^{-1}{\bf E})^0_a d\lambda \,{\xi}^a
\left|\Psi(p;\{\Lambda{\bf e}\})\right|^2 \,
\label{ps:expntimerepeat}
\end{eqnarray}
and
\begin{eqnarray}
\langle dl_j \rangle &=&  \|\Psi(p)\|^{-2}
\int_{O(3,1)}d\mu_L(\Lambda) 
g_{ab}(p){\bf e}_\alpha^a \Lambda_j^\alpha d\lambda \,{\xi}^b
\left|\Psi(p;\{\Lambda{\bf e}\})\right|^2 	\nonumber \\
&=& \|\Psi(p)\|^{-2}
\int_{O(3,1)}d\mu_L(\Lambda) 
\eta_{\alpha\beta}{\bf E}^\alpha_a \Lambda_j^\beta d\lambda\, {\xi}^a
\left|\Psi(p;\{\Lambda{\bf e}\})\right|^2 \, \nonumber\\
 & =&
  \|\Psi(p)\|^{-2}\int_{O(3,1)}d\mu_L(\Lambda) 
(\Lambda^{-1}{\bf E})^j_a d\lambda\, {\xi}^a
\left|\Psi(p;\{\Lambda{\bf e}\})\right|^2 \, ;
\label{ps:expnlengthrepeat}
\end{eqnarray}
where  $\|\Psi(p)\|^{2}$ is defined by Eq.\ \ref{normpjfiducialfnoinfy}.  Recall (see below Eq.~\ref{ps:expnlength}) that 
in passing from the second to the third lines of the expressions above we used the identity
$\eta_{\alpha\beta} \Lambda^\beta_\gamma= \eta_{\mu\beta}\Lambda^\beta_\gamma \Lambda^\mu_\nu (\Lambda^{-1})^\nu_\alpha =\eta_{\gamma\nu} (\Lambda^{-1})^\nu_\alpha $,  which implies 
 $\eta_{\alpha\beta}{\bf E}^\alpha_a \Lambda^\beta_0 \xi^a=-(\Lambda^{-1})^0_\alpha{\bf E}^\alpha_a \xi^a \equiv
-(\Lambda^{-1}{\bf E})^0_a\xi^a$ and
$\eta_{\alpha\beta}{\bf E}^\alpha_a \Lambda^\beta_j \xi^a=(\Lambda^{-1})^j_\alpha{\bf E}^\alpha_a \xi^a \equiv (\Lambda^{-1}{\bf E})^j_a \xi^a$. 
As pointed out in Section \ref{ssec:ps:expnvals},
this  confirms that one can express the scalar integrands equally well in terms of the tetrad $\{{\bf e}_\mu^a\}$ or of its dual basis, $\{{\bf E}^\mu_a\}$,  as long as one recalls that if 
${\bf e}^a_\mu \mapsto (\Lambda {\bf e})^a_\mu$,  then the duality relation demands 
${\bf E}_a^\mu \mapsto (\Lambda^{-1}{\bf E})^\mu_a$.

Now we turn to trying to adapt these equations to the general framework. The point to address is of course  that in the bundle $\cal {F({\cal O})}$, the metric $g_{ab}$ varies across the fibre (a copy of $GL(4)$)  above $p$ over which one wishes to integrate, so as to sum contributions to the expectation value. In other words:  we need to sum contributions that vary, not just due to different spatial orientations and motions of the apparatus as in the restricted framework, but also due instead to different  metrics on $\cal O$---while in the above equations, $g_{ab}$ is fixed.  But how may expectation values depend on the metric?

Going back to the basics, recall that observables can always be expressed as  {\it scalar} 
quantities constructed out of (i) tensors that characterize the system being measured   {\it and} (ii) the ``configuration'' (i.e.\ state of motion and spatial orientation) of the apparatus doing the measurements. This consideration alone
was enough to suggest how to calculate expectation values in the restricted framework  (see Section \ref{ssec:ps:expnvals}). The additional piece of information that we must  consider now, in the general framework, is that although the tensors characterizing the system being measured cannot depend on the  apparatus doing the measurements, they {\it can} (and, in general,  {\it do}) depend on the geometry of the spacetime---usually through the system's dynamical equations. 
And this dependence is, in general, non-local---in the sense that the tensors that characterize the system at $p\in {\cal M}$  will in general vary with the metric field $g_{ab}$ even if $g_{ab}$ is kept fixed
on a neighborhood of $p$  (thanks to the dynamical equations).

Let $A[^{({\bf b})}\!g; {\bf P}_{\bf b})$ denote the {\it scalar} which represents, at $p\in {\cal M}$,  a given classical observable constructed out of (i)~the tensors which characterize the system being measured, 
when the spacetime metric field is $^{({\bf b})}\!g_{ab}$ (i.e.\ the metric according to which
the basis  {\it field} $\{{\bf b}_\mu^a\}$ is a tetrad field) and (ii) the apparatus ``configuration''
characterized by the {\it single} (at $p$) tetrad $\{{\bf b}_\mu^a\}$. 
The reader may object to our notation: saying that it seems redundant, since the basis field ${\bf b}$ already fixes the metric $^{({\bf b})}\!g_{ab}$.
However,  our notation, particularly the use of different brackets to enclose the arguments of $A$,  is meant to
emphasize that the dependence of $A$ on the metric $^{({\bf b})}\!g_{ab}$ is of a different nature than its dependence on the apparatus's configuration characterized by ${\bf b}$. For the dependence is {\it local} on the latter (${\bf b}$ at each $p$; hence we write ${\bf P}_{\bf b})$); while it is non-local on the former, as discussed above,  so that we write $[^{({\bf b})}\!g$.

At this point, we make a {\it simplifying assumption}.
It amounts to clarifying what we will mean by a `{\it classical} observable'. Namely: We assume that {\it classical} observables assume well-defined, single values on a region  ${\cal O}\subset{\cal M}$ 
on which a geometry is given. 
Here, `given' just means `for a given first argument $^{({\bf b})}\!g$ of $A[^{({\bf b})}\!g; {\bf P}_{\bf b})$'.\footnote{It does not mean that $\Psi$'s support within the column $\pi^{-1}[{\cal O}]$ is a subset of the union of a section of ${\cal {F(O)}}$  with its pointwise Lorentz-transforms. Recall the discussion in (B) of Section \ref{ssec:bundle} about the set $Sec({\cal {F(M)}})$ of sections of ${\cal {F(M)}}$ being partitioned into equivalence classes by the relation `...is a pointwise Lorentz-transform of...'. }
   This  is not in conflict with the discussion above that observables can depend non-locally on the spacetime geometry. It only means that a {\it classical} (in our sense)  quantity cannot have superposed values on a region ${\cal O}$  that has a  given geometry.  That is consistent with  different {\it global} (i.e.\ across $\cal O$) spacetime geometries  leading to different values for the classical observables on ${\cal O}$.
The assumption is only that on ${\cal O}$ each  classical observable takes
{\it one} value at each point---either given by one realization of the different possible values or, say, by an average over those values.  

With this simplifying assumption, we obtain the following chain of dependence. The value of an observable $A$ at $p\in{\cal O}$ can  depend on (i) the given  metric at $p$, $g_{ab}(p)$,  {\it and} (ii), for each possible $g_{ab}(p)$, on the apparatus's configuration $\{{\bf b}_\mu^a\}$ such that ${\bf P}_{\bf b}\equiv (p; \{{\bf b}_\mu^a\}) \in \pi^{-1}_{\rm o}(p)\subset {\cal F}_{\rm o}[g]$.\footnote{Cf.\ again the discussion in (B) of Section \ref{ssec:bundle} about $Sec({\cal {F(M)}})$.}  So, thanks to this simplifying assumption, we can, indeed,
represent the observable simply as the {\it local} quantity $A(^{({\bf b})}\!g(p);{\bf P}_{\bf b})$. Notice that we chose to keep the explicit dependence on the metric---even though ${\bf P}_{\bf b}$ determines
$^{({\bf b})}\!g_{ab}(p)$---so as to emphasize, once more, that $A$ can only depend on the specific choice of ${\bf b}$ (among all possible bases that are tetrads according to $^{({\bf b})}\!g_{ab}(p)$) through the {\it coupling}  
of ${\bf b}$ with the tensor which characterizes the system being measured (which characterizes the system-apparatus coupling). The only ``external'' information that the tensor which characterizes ``the system'' can depend on is $^{({\bf b})}\!g_{ab}$.

To illustrate this,  suppose we are interested in the electric field along the spatial direction labelled $j\in \{1,2,3\}$---as in Eq.~\ref{EvalueQRFnew}. Then,  the relevant scalar quantity would be given by 
 $E_j=F_{ab}(^{({\bf b})}\!g){\bf b}_j^a {\bf b}_0^b$.  The Faraday tensor depends only on the system being measured {\it and} on the given   metric. It does {\it not} depend on different choices of
${\bf b}$ that are related by Lorentz transformations.  Besides,  extra metric dependence can  be  accommodated   by using tensors of different rank to characterize the system, where we use the given metric to raise and lower indices. For example, the relevant scalar quantity can be constructed using 
$F^{ab}(^{({\bf b})}\!g) = F_{cd}(^{({\bf b})}\!g) ^{({\bf b})}\!g^{ac}\,^{({\bf b})}\!g^{bd}$, instead of $F_{ab}(^{({\bf b})}\!g)$.  For using Eq.~\ref{eq:metricb}, we have  $E_j = -F^{ab}(^{({\bf b})}\!g){\bf B}^j_a {\bf B}^0_b$.
And using manipulations analogous to the ones below Eq.~\ref{ps:expnlengthrepeat},  namely
 $^{({\bf b})}\!g_{ab}(\Lambda{\bf b})_\mu^b= \,^{({\bf b})}\!g_{ab}\Lambda^\nu_\mu {\bf b}_\nu^b= 
\eta_{\alpha\nu}\Lambda^\nu_\mu{\bf B}^\alpha_a=\eta_{\mu\nu}(\Lambda^{-1})^\nu_\alpha{\bf B}^\alpha_a \equiv
\eta_{\mu\nu}(\Lambda^{-1}{\bf B})^\nu_a$---i.e.,  $^{({\bf b})}\!g_{ab}(\Lambda{\bf b})_0^b=
-(\Lambda^{-1}{\bf B})^0_a$ and $^{({\bf b})}\!g_{ab}(\Lambda{\bf b})_j^b=
(\Lambda^{-1}{\bf B})^j_a$---we see that: for each {\it fixed}  metric $^{({\bf b})}\!g_{ab}$,  
it does not matter which
tensor,  $F_{ab}(^{({\bf b})}\!g)$ or $F^{ab}(^{({\bf b})}\!g)$,  we use to construct our observable: 
 $F_{ab}(^{({\bf b})}\!g)(\Lambda{\bf b})_j^a(\Lambda{\bf b})_0^b = -
F^{ab}(^{({\bf b})}\!g)(\Lambda^{-1}{\bf B})^j_a(\Lambda^{-1}{\bf B})^0_b $.

This last equality {\it seems} to rely strongly on the defining property of $\Lambda \in O(3,1)$,  viz.\
$\Lambda_\mu^\alpha \Lambda^\beta_\nu\eta_{\alpha\beta} = \eta_{\mu\nu}$.   So it may  seem hopeless
to expect that this same freedom to define the (scalar) observables using $\Lambda{\bf b}$ and-or
$\Lambda^{-1}{\bf B}$
could be extended to
$M{\bf b}$ and
$M^{-1}{\bf B}$
when integrating over $GL(4)\ni M$.  But on the other hand,  the manipulations which led to  this equality 
are valid {\it whatever} the fiducial basis ${\bf b}$ is.  Hence,  using again the electric-field example and 
considering two bases at $p$,  ${\bf b}$ and $\tilde{\bf b}$, that are {\it not} related by a Lorentz transformation:  the equations 
 $F_{ab}(^{({\bf b})}\!g)(\Lambda{\bf b})_j^a(\Lambda{\bf b})_0^b = -
F^{ab}(^{({\bf b})}\!g)(\Lambda^{-1}{\bf B})^j_a(\Lambda^{-1}{\bf B})^0_b $
and
 $F_{ab}(^{(\tilde{\bf b})}\!g)(\Lambda\tilde{\bf b})_j^a(\Lambda \tilde{\bf b})_0^b = -
F^{ab}(^{(\tilde{\bf b})}\!g)(\Lambda^{-1}\tilde{\bf B})^j_a(\Lambda^{-1}\tilde{\bf B})^0_b $
must be equally valid. So if $l\in GL(4)$ gives the transformation between ${\bf b}$ and $\tilde{\bf b}$
(so, $l\notin O(3,1)$),  the latter equation becomes:
 $F_{ab}(^{(l{\bf b})}\!g)[(l\Lambda){\bf b}]_j^a[(l\Lambda){\bf b}]_0^b = -
F^{ab}(^{(l{\bf b})}\!g)[(l\Lambda)^{-1}{\bf B}]^j_a[(l\Lambda)^{-1}{\bf B}]^0_b $.\footnote{Keeping track of the indices:  $\tilde{b}_\mu^a = l^\alpha_\mu {\bf b}_\alpha^a$, which implies $\tilde{B}^\mu_a = (l^{-1})^\mu_\alpha {\bf B}^\alpha_a$,  $(\Lambda \tilde{\bf b})^a_\mu = 
\Lambda^\alpha_\mu (l^\beta_\alpha {\bf b}^a_\beta)= (l\Lambda)^\beta_\mu {\bf b}^a_\beta = [(l\Lambda){\bf b}]^a_\mu$, and $(\Lambda^{-1} \tilde{\bf B})_a^\mu = 
(\Lambda^{-1})_\alpha^\mu [(l^{-1})_\beta^\alpha {\bf B}_a^\beta]= (\Lambda^{-1}l^{-1})_\beta^\mu {\bf B}_a^\beta = [(l\Lambda)^{-1}{\bf B}]_a^\mu$.} Since ${\bf b}$ and $\tilde{\bf b}$ are arbitrary,  $l\Lambda =:M$
is a generic element of $GL(4)$.

The subtle point here, which justifies our substituting $M = l\Lambda \in GL(4)$ for $\Lambda\in O(3,1)$, while keeping
the equality in the electric-field example,  is that $F_{ab}(^{(l{\bf b})}\!g)$ and $F^{ab}(^{(l{\bf b})}\!g)$, despite being insensitive to $\Lambda$ (idea: ``the system i.e.\ the  electromagnetic field is independent of the apparatus''), {\it do} depend on $l\in GL(4)$
(idea: ``the system depends on the background geometry'').  Thus each $^{(l{\bf b})}\!g_{ab}$ defines its 
own canonical isomorphism between vectors and covectors---i.e., it raises and lowers indices---in such a way as to guarantee that the scalar observables are insensitive to the tensor representation used in their construction (i.e.\ $F_{ab}$ or $F^{ab}$ in the example above).

Motivated by the discussion above, we can break the group $GL(4)$ into equivalence classes (viz.\ $O(3,1)$ and its cosets) using the following equivalence relation: for any $M,M'\in GL(4)$,  $M\sim M' $ iff there exists $\Lambda\in O(3,1)$
such that $M'= M\Lambda$.   Adapting the notation introduced above,  we will use $l$ (`l' for `leaf') to denote these equivalence classes, i.e. $l\in GL(4)/\!\sim$; as well as using $l$ for a representative member of these classes. 
Since $O(3,1)$ is not a normal subgroup of $GL(4)$,  $GL(4)/\!\sim$ has no natural group structure.  This, however,  does not prevent us from
writing
\begin{eqnarray}
d\mu_G (M) =: d\mu_Q(l)\, d\mu_L(\Lambda)
\label{factorizemeasure}
\end{eqnarray}
($Q$ for quotient) at a formal level so as to make explicit the dependences on,  respectively,  different metrics and on different   
apparatus configurations.

 In summary, the discussion above suggests that it can be convenient to interpret the integral over $GL(4)$ (a 16-dimensional integral), at each $p$,  as two integrals:\\
\indent \indent \indent \indent  (i) a 6-dimensional integral over the Lorentz group $O(3,1)$: so this will be similar to the integrals throughout Section \ref{ssec:ps:expnvals}; and for which we again write $\Lambda \in O(3,1)$ as the  integration variable, rather than $M$ as in our normalisation condition Eq.\ \ref{normpgeneralinfy}; and \\
\indent \indent \indent \indent (ii) a 10-dimensional integral over the ``leaves", i.e.\ the co-sets of  $O(3,1)$ as a subgroup of $GL(4)$, $GL(4)/\!\sim$---but we recall  that $O(3,1)$ is not a normal subgroup of $GL(4)$, so that this is only a set-theoretic decomposition, not a quotient group. 

This also means that,  as in Eq.\ \ref{factorizemeasure}, we take the Haar measure $\mu_G$ on $GL(4)$, as in our normalisation condition Eq.\ \ref{normpgeneralinfy}, to be a product of:\\
\indent \indent \indent \indent   (i) the Haar measure on  $O(3,1)$ which we have written (above, and throughout Section \ref{ssec:ps:expnvals}) as $\mu_L$; and \\
\indent \indent \indent \indent (ii) the ``quotient Haar measure" on the 10-dimensional set of leaves/co-sets of the Lorentz group; which we  write as  $\mu_Q$ ($Q$ for `quotient'); and for which we use as integration variable `$l$' ($l$ for `leaf'). 

We also need to bear in mind  that as one moves (i.e.\ as one considers different integrand contributions) in the integral over $p$'s fibre, from one leaf to another leaf, say from ${\bf {P_b}} \equiv (p; \{ {\bf b}^a_{\mu}(p) \})$ to ${\bf {P_{b'}}} \equiv (p; \{ {\bf b'}^a_{\mu}(p) \})$: the apparatus also ``moves", in general,  from being in an orientation orthonormal with respect to one geometry (viz.\ $\{ {\bf b}^a_{\mu}(p) \}$ taken as a tetrad), to  being in an orientation that is orthonormal with respect to {\it another} geometry (viz.\ $\{ {\bf b'}^a_{\mu}(p) \}$ taken as a tetrad---for this other geometry).\footnote{\label{subtleleaf} Note that this is {\it not} to say that a leaf of the fibre is somehow equivalent to a local geometry or metric. For the leaves form a merely 10-dimensional space, while the set of geometries/metrics is of course infinite-dimensional, even for a region $\cal O$ as small as you please. Thus the apparatus `being in $\{ {\bf b}^a_{\mu}(p) \}$ taken as a tetrad' means that the apparatus' orientation is orthonormal for one or other of continuously many geometries.  Cf.\ again the discussion in (B) of Section \ref{ssec:bundle} about $Sec({\cal {F(M)}})$.}

To signal this fact, we  keep  the leaf-label $(l)$ as a superscript to the metric $g$ when it occurs in the integrand. So we will write  $^{(l)}g_{ab}$. But we stress that this does {\it not} mean that our 10-dimensional  integration over leaves is an integration over  geometries; cf.\ footnote \ref{subtleleaf}. \\ 
 
  So with this break-down (at each point $p$) of the integral over $GL(4)$, and these notational conventions about measures and integration variables, and with  $\|\Psi(p)\|^2$ defined by our normalisation condition Eq.\ \ref{normpgeneralinfy},  we return (at last!) to this item (1)'s announced topic, viz.\ the expectation value, at a point, of time and spatial separations. Thus we now define  the expectation value of the time separation and the spatial separations (along each direction) at the point $p$,  according to the g-QRF $\Psi$,  by
  \begin{eqnarray}\label{bs:expntime}
\langle d\tau \rangle &=&  - \|\Psi(p)\|^{-2} \;
\int_{GL(4)/{\!\sim}}  \frac{d\mu_Q(l)}{| {\rm{det}} \, l |}  \; 
\int_{O(3,1)}d\mu_L(\Lambda) \,
^{(l)}g_{ab}(p){\bf b}_\alpha^a M_0^\alpha d\lambda {\xi}^b
\left|\Psi(p;\{ M{\bf b}\})\right|^2 	\nonumber \\
&=& - \|\Psi(p)\|^{-2} \;
\int_{GL(4)}  \frac{d\mu_G(M)}{| {\rm{det}} \, M |}  \; 
 ^{(M{\bf b})}g_{ab}(p){\bf b}_\alpha^a M_0^\alpha d\lambda {\xi}^b
\left|\Psi(p;\{ M{\bf b}\})\right|^2 
\end{eqnarray}
and
\begin{eqnarray}\label{bs:expnlength}
\langle dl_j \rangle &=&  \|\Psi(p)\|^{-2}  \;
\int_{GL(4)/{\!\sim}}  \frac{d\mu_Q(l)}{| {\rm{det}} \, l |}  \; 
\int_{O(3,1)}d\mu_L(\Lambda)  \,
^{(l)}g_{ab}(p){\bf b}_\alpha^a M_j^\alpha d\lambda {\xi}^b
\left|\Psi(p;\{ M{\bf b}\})\right|^2 	\nonumber \\
&=& \|\Psi(p)\|^{-2} \;
\int_{GL(4)}  \frac{d\mu_G(M)}{| {\rm{det}} \, M |}  \; 
 ^{(M{\bf b})}g_{ab}(p){\bf b}_\alpha^a M_j^\alpha d\lambda {\xi}^b
\left|\Psi(p;\{ M{\bf b}\})\right|^2   \; .
\end{eqnarray}
Note that, in the end, the integration is really performed explicitly in terms of $M\in GL(4)$;  though we can,  for notational purposes,  represent $M$ as $M = l\Lambda$ for a particular (fiducial) choice of representative
$l$ of each equivalence class of $GL(4)/\!\sim$.\footnote{There is no need to show independence w.r.t.\ the choice of the representative $l$ since this is just a notation.  The integration is really performed in the variable $M$.}\\

\bigskip

\noindent (2):  {\it The expectation value, at a point, of the classical electric field}:---\\
We again recall the ideas and notation from  Section \ref{ssec:ps:expnvals}.  The electromagnetic field is described, at $p$, by the Faraday tensor $F_{ab}$. For a {\it classical} apparatus passing through $p$ with 4-velocity $u^a$, the electric field measured along the spatial direction given by the unit vector $n^a$ (with $g_{ab}u^a n^b = 0$) is $E_n=F_{ab}n^au^b$. So in the restricted framework, we defined the expectation value at $p$ for the $j$ component of the electric field by Eq.\ \ref{EvalueQRFnew}. To repeat: it was \begin{eqnarray}
\langle E_j\rangle|_{p} :=\|\Psi(p)\|^{-2} 
\int_{O(3,1)}d\mu_L(\Lambda) F_{ab}{\bf e}_\alpha^a{\bf e}_\beta^b
\Lambda_j^\alpha \Lambda_0^\beta
\left|\Psi(p;\{\Lambda{\bf e}\})\right|^2 \, ,
    \label{EvalueQRFnewrepeat}
\end{eqnarray}
where  $\|\Psi(p)\|^{2}$ is of course defined by Eq.\ \ref{normpjfiducialfnoinfy}.

To adapt this to the general  framework (while adopting our normalisation condition, Eq.\ \ref{normgeneral}), we use the discussion in item (1) above.  We ``replace" (i)  the orthonormal bundle  ${\cal F}_{\rm o}[g]$ by the frame bundle ${\cal {F({\cal O})}}$; and (ii) an integration over $O(3,1)$ with integration variable $\Lambda$,  by one over $GL(4)$ with integration variable $M$.  So given a g-QRF $\Psi$, we define the expectation value at $p$ for the $j$ component of the electric field by:
\begin{equation}
\langle E_j\rangle|_{p} :=  
\|\Psi(p)\|^{-2} 
\int_{GL(4)}  \frac{d\mu_{G}(M)}{| {\rm det} M | } \; 
F_{ab}{(^{(M{\bf b})}\!g)}{\bf b}_\alpha^a{\bf b}_\beta^b
M_j^\alpha M_0^\beta \, 
\left|\Psi(p;\{M{\bf b}\})\right|^2 ,
    \label{EvalueQRFnewgeneral}
\end{equation}
where we again take $\|\Psi(p)\|^2$ as defined by  Eq.\ \ref{normpgeneralinfy}.\footnote{We stress again that in the general  framework, the apparatus `being in $\{ {\bf b}^a_{\mu}(p) \}$ taken as a tetrad' means that the apparatus' orientation is orthonormal for one or other of continuously many geometries. Cf.\ again footnote \ref{subtleleaf}.} 

As we announced after Eq.\ \ref{normpgeneralinfy}, these expectation values at a point, Eqs.\ \ref{bs:expntime}, \ref{bs:expnlength} and \ref{EvalueQRFnewgeneral}, are independent of the fiducial basis field. For if ${\bf b}$ is replaced by $\bar{\bf b} = \bar{M}_p{\bf b}$ (as in the Appendix, part (A)), then $M\bar{\bf b} = (\bar{M}_pM){\bf b}$, so that the integrands depend on $M$ only through $M' := \bar{M}_pM$; and the change of variable $M\mapsto M'$, together with the invariance of $\mu_G$ and $|{\rm det}M|^{-1} = |{\rm det}\bar{M}_p|\,|{\rm det}M'|^{-1}$, shows that both the fibre integral in the numerator and $\|\Psi(p)\|^2$ acquire the same factor $|{\rm det}\bar{M}_p|$, which cancels. As in the restricted framework (cf.\ the remark after Eq.\ \ref{ps:expnlength}), these expectation values are finite only for g-QRFs decaying sufficiently fast along the (here, $GL(4)$) fibres.

And just as in Section \ref{ssec:ps:expnvals}, we say: Similarly of course for any classical quantity, $A$ say, whose measurement results can be written in terms of  the coupling between the system and the tetrad characterising the apparatus.  But here, `classical quantity' is to be understood, not just as in Section \ref{ssec:ps:expnvals}, but also as obeying our simplifying assumption which we adopted in item (1) above.  That is: we would similarly define the expectation value $\langle A\rangle|_{p}$ at $p$ for a g-QRF $\Psi$.

%%%%%%%%%%%%%%%%%%%%

\subsubsection{Expectation values in a region}\label{ssec:bs:normnexpnregion}

As announced in (D) above,  we assume, just as we did in  (3) and (4) of  Section \ref{ssec:ps:expnvals}, that the apparatus is somehow designed so as to measure the field in question {\it just once}, somewhere in the region $\cal O$, while having some or other internal configuration (i.e.\ spatial configuration and state of motion). So it is to have a definite lifetime, being ``turned off" outside ${\cal O}$. \\

\noindent (3):  {\it The expectation value, in a region $\cal O$, of a classical  field}:---\\
We again take the example of the electric field.
 Again, we include the  volume element  $ \epsilon(p;M)$ in the fibre integral's integrand; and overall,  we ``replace"  the orthonormal bundle  ${\cal F}_{\rm o}[g_{ab}]$ with the frame bundle ${\cal {F({\cal O})}}$. So we define:
\begin{eqnarray}
\langle E_j\rangle|_{\cal O} := \int_{\cal O}  \; 
\int_{GL(4)} \, d\mu_G(M) \;   \epsilon(p;M)  \; 
F_{ab}{(^{(M{\bf b})}\!g)}{\bf b}_\alpha^a{\bf b}_\beta^b
M_j^\alpha M_0^\beta
\left|\Psi(p;\{ M{\bf b} \})\right|^2 ,
    \label{EvalueQRFnewsptgeneral}
\end{eqnarray}
where: $\{{\bf b}_\mu^a\}$ is the  fiducial basis field; and  $\epsilon(p;M)$ is the  volume element for the metric  $^{(M{\bf b})}\!g$, 
  i.e.\ $\epsilon(p;M) = \epsilon(p)/|{\rm det}M|$, with $\epsilon$ the volume element associated to ${\bf b}$ (cf.\ Eqs.\ \ref{trsfmepsilon1} and \ref{eq:muO}).

 Similarly of course for any classical quantity whose measurement results can be written in terms of  the coupling between the system and the tetrad characterizing the apparatus. For example, the quantity could be the energy density $\rho$,  as in Section \ref{ssec:ps:expnvals}'s item (2) and specifically  Eq.\ \ref{EvalueQRFnewclassicalrho}. \\

   \bigskip

\section{Conclusion}\label{sec:conclusion}

This paper has developed the concrete, albeit idealized, conception of a quantum reference frame as an infinitesimal apparatus (``three rulers and a clock”), with quantum uncertainty as to its spatial configuration and state of motion. But this is the first of three papers about this conception. The other two papers will give more details about two frameworks for developing it, which we call respectively `the restricted framework'  and `the general framework'.

The core idea is that the apparatus' quantum uncertainty means that, instead of classical relativity theory's representation of such an apparatus as a tetrad, we represent it as an assignment of complex amplitudes to tetrads (i.e.\ to possible classical configurations of the apparatus): an idea which we express in terms of fibre bundles (Section \ref{ssec:bundle}). 

We  first developed this conception for an apparatus in a classical spacetime i.e.\ with a given Lorentzian metric. This is the setting which we call `the restricted framework' (Section \ref{sec:perspsing}). In this framework, the main physical idea is that a quantum reference frame provides a superposition of perspectives on a classical spacetime. And the main mathematical notion for expressing this is the bundle over the spacetime manifold of orthonormal frames (i.e.\ bases that are orthonormal for the given metric).  

Then we developed the conception for an apparatus in a spacetime without a given metric: a setting which we call `the general framework' (Section \ref{sec:basicsing}). Here, the key idea is to think of specifying a metric by stipulating that a field of bases (of the manifold's points' tangent spaces) is to be orthonormal in the metric that one is thereby defining. This leads us to use the bundle over the spacetime manifold of {\it all} bases; (ordinarily called `the frame bundle', although the bases are generic and {\it not} given as being orthonormal). In this way, we could make precise the idea of quantum reference frame as providing a superposition of perspectives on a superposition of spacetimes---in the pointwise sense that the metric at the apparatus' (uncertain) location is itself uncertain; the quantum reference frame assigns no amplitudes, or probabilities, to metric fields (cf.\ (A) in Section \ref{sec:intro}, and footnote \ref{noprobgeom}).

For both frameworks, our main effort in this paper has been to write down expectation values for measurements of {\it classical} quantities associated with the spacetime, such as the electromagnetic field tensor. The main idea has been to adapt the classical formulas stating how a tetrad couples to a tensorial quantity, giving the observer-dependent (but coordinate-independent)
 value of the corresponding observable. This adaptation involved weighting a specific tetrad's coupling to the quantity concerned, with the Born-rule probability for the quantum reference frame to be in the configuration which is represented by that tetrad.  In this endeavour, the main topic needing care was the normalisation of the state of the quantum reference frame. This care was especially needed in the general framework, where one needs the normalisation to be independent of the chosen basis field. \\

It is clear from this summary that there remains much to do in developing and assessing this conception of a quantum reference frame. Three broad topics are obviously timely, not least because they have been addressed in detail in other formulations of  the idea of a quantum reference frame.  These topics are: how to treat the measurement of {\it quantum} quantities; how to treat transformations between quantum reference frames; how to treat the gravitational field of a massive source in a superposition of spatial locations (i.e.\ a quantum reference frame ``immersed in a superposition of spacetimes''). These topics, and also the relation between our conception and some other proposals in the quantum reference frame literature, will be addressed in the two companion papers.

%%%%%%%%%%%%%%%%%%%%%%%%%

 \section{Appendix}\label{sec:appdx}
 \noindent We consider the general framework, and so the bundles  ${\cal {F(M)}}$, and locally ${\cal {F(O)}}$. We proceed in three Parts. 
 
 In (A),  we show that the measure $d\mu_{\cal F}$ defined in Eq.~\ref{eq:muO}, and therefore the normalisation condition we adopted in  Eq.\ \ref{normgeneral}, is independent of the choice of  fiducial basis field ${\bf b}$ used to make the identification $\pi^{-1}(p) \leftrightarrow
GL(4)$ at each $p\in{\cal O}$. As we stressed in Section \ref{ssec:bs:normnexpn}, this is needed for our proposals to make physical sense. This is not just because the normalisation should of course not change under a pointwise Lorentz transform $\Lambda_p$ of a given basis field. Also, we are proposing that a g-QRF is a physical system ``immersed" in a superposition of geometries on $\cal O$. So  its ``wave function" $\Psi$ should be normalized once and for all, independently of which geometry on $\cal O$ is being considered---i.e., invariant under a pointwise $GL(4)$ transformation. And this implies that if we state the normalisation condition in terms of one fiducial basis field (taken as a tetrad field), the condition should be invariant under a change of basis field.

    In (B), we show that  the measure $d\mu_{\cal F}$ defined on $\pi^{-1}[{\cal O}]$ thanks to (A) can be used to define a measure on the whole ${\cal F}({\cal M})$ by applying (A) to an open covering of ${\cal M}$. 
    
    Finally, in (C), we show the explicit form of the measure $d\mu_{\cal F}$.

\bigskip

{\bf (A) The normalisation condition is independent of the  fiducial  basis field}\\
We  show that the normalisation we have proposed, Eq.\ \ref{normgeneral},  is independent of the choice of  the basis field {\bf b} appearing in Eq.~\ref{eq:muO}.

 Suppose we choose another fiducial basis field $\bar{\bf b}$ to play the role that ${\bf b}$ plays in Eq.~\ref{eq:muO}.  Let $\bar{M}_p\in GL(4)$ be the transformation relating these two basis at $p\in{\cal O}$: ${\bar{\bf b}}^a_\mu = (\bar{M}_p)^\nu_\mu {\bf b}^a_\nu$; and let $\bar{\epsilon}$ be the corresponding volume element associated to the metric which makes $\bar{\bf b}$ a tetrad.  Thus,  the concrete integration on $\pi^{-1}[{\cal O}]$ would be
\begin{eqnarray}
\int_{{\cal O}} {\bar \epsilon}(p)\int_{GL(4)} 
\frac{d\mu_G(M)}{|{\rm det}M |}\,f(p;M{\bar{\bf b}(p)})&=&
\int_{\cal O}\int_{GL(4)} 
\frac{
d\mu_G(M)\,\epsilon(p)}{|{\rm det}M \,{\rm det}\bar{M}_p|}\,f(p;(\bar{M}_pM){{\bf b}(p)})\nonumber \\
&=&
\int_{\cal O}\int_{GL(4)} 
\frac{
d\mu_G(\bar{M}_p^{-1}M'_p)\,\epsilon(p)}{|{\rm det}M'_p|}\,f(p;M'_p{{\bf b}(p)})\nonumber \\
&=&
\int_{\cal O}\int_{GL(4)} 
\frac{
d\mu_G(M'_p)\,\epsilon(p)}{|{\rm det}M'_p|}\,f(p;M'_p{{\bf b}(p)})\nonumber \\
&=&
\int_{\cal O}\epsilon(p)\int_{GL(4)} 
\frac{
d\mu_G(M)}{|{\rm det}M|}\,f(p;M{{\bf b}(p)}),
\label{eq:dFtransf}
\end{eqnarray}
where in the first equality we  used Eq.~\ref{trsfmepsilon1} for changing between bases $\bar{\bf b}$ and
${\bf b}$ (pointwise related by $\bar{M}_p$, so that $M^\nu_\mu\bar{\bf b}_\nu^a = M^\nu_\mu (\bar{M}_p)_\nu^\sigma{\bf b}_\sigma^a = (\bar{M}_p M)^\sigma_\mu {\bf b}_\sigma^a$),  in the second we made the 
change of matrix variable $M'_p :=\bar{M}_p M$,  in the third we made use of the fact that the Haar measure is invariant under the change $M\mapsto M_p'$,  and finally in the fourth 
we just renamed the integration variable at each $p$, pulling out $\epsilon(p)$ from the fibre integration since
it depends only on $p$. These manipulations prove that the integration 
defined in Eq.~\ref{eq:muO} is independent of the choice of fiducial basis.

\bigskip

{\bf (B) The measure on ${\cal {F(M)}}$}\\
Now, we show that the well-defined integration on $\pi^{-1}[{\cal O}]$ defined in Eq.~\ref{eq:muO}  can be consistently extended to the whole
frame bundle ${\cal F}({\cal M})$. This is necessary to ensure that the construction above works even when
${\cal M}$ is not parallelizable; i.e.,  when there is no continuous basis field defined throughout ${\cal M}$. 
The strategy is simple: cover ${\cal M}$ with open sets ${\cal O}_{\cal I}$,  each being the domain of
a chart (i.e., a coordinate system) ${\cal O}_{\cal I}\ni p \mapsto \,^{({\cal I})}\!x^\mu(p)\in{U}_{\cal I}\subset {\mathbb R}^4$---which
ensures each ${\cal O}_{\cal I}$ is parallelizable.  Then,  on each ${\cal O}_{\cal I}$ define 
a  basis field $^{({\cal I})}{\bf b}$.  These fiducial basis fields play the same role for ${\cal F}({\cal M})$ as
the coordinate systems 
${\cal O}_{\cal I}\ni p\mapsto \,^{({\cal I})}\!x^\mu(p)\in{U}_{\cal I}$ do for the manifold ${\cal M}$.  More specifically,  on each
${\cal O}_{\cal I}$, use  $^{({\cal I})}{\bf b}$ to define integration on $\pi^{-1}[{\cal O}_{\cal I}]$ via Eq.~\ref{eq:muO}. Then, all we need to show is that these definitions are consistent; i.e., that wherever two (or more) elements of the
covering of ${\cal M}$,  say ${\cal O}_{\cal I}$ and ${\cal O}_{\cal J}$,  intersect, the integration defined on 
$\pi^{-1}[{\cal O}_{\cal I}]\cap \pi^{-1}[{\cal O}_{\cal J}]$ using $^{({\cal I})}{\bf b}$ or $^{({\cal J})}{\bf b}$ 
leads to the same result. But this is exactly what we proved in (A),  now 
applied to $\pi^{-1}[{\cal O}_{\cal I}\cap{\cal O}_{\cal J}]=\pi^{-1}[{\cal O}_{\cal I}]\cap \pi^{-1}[{\cal O}_{\cal J}]$. This completes our proof.

\bigskip

{\bf (C) The concrete form of the measure $d\mu_{\cal F}$}\\
Although not strictly necessary for our arguments,  we now show, for the benefit of the reader
interested in concrete calculations,  the explicit form of the measure
$d\mu_{\cal F}$ defined in Eq.~\ref{eq:muO}; in other words, how to express Eq.~\ref{eq:muO} as a
20-dimensional real integral.

Given a coordinate system $p \mapsto x^{\mu}(p)$ defined on the open set  $ {\cal O} \subset {\cal M}$, we induce a coordinate system on $\pi^{-1}[{\cal O}]$ as follows.   To a point  ${\bf {P_b}} := (p ; \{{\bf b}_\mu^a(p)\}) \in \pi^{-1}[{\cal O}]$ is assigned the coordinates $(x^{\mu}, y^{\alpha}_{\beta}) \in {\mathbb {R}}^{20}$ by:
\begin{equation}
x^{\mu} := x^{\mu}(p) \;\;  ; \;\; y^{\alpha}_{\beta} := {\bf b}_\beta^a\,d{\bf x}^{\alpha}_a,
\label{inducedcoordsF(M)}
\end{equation}  
where $\{d{\bf x}^{\alpha}_a\}$ is the dual basis to  the coordinate basis $\{{\bf \partial}^a_\mu\}$, both evaluated at $p$.  The second equation implies that $y^{\alpha}_{\beta}$ is the component of ${\bf b}_\beta^a$ in the direction given by ${\bf \partial}^a_\alpha$: 
\begin{equation}
 {\bf b}_\beta^a =  y^{\alpha}_{\beta}  {\bf \partial}^a_\alpha \, .
 \label{yascomponents}
\end{equation}  
Since all $ x^{\mu}$ and $y^{\alpha}_{\beta}$ ($\mu, {\alpha}, {\beta} = 0,1,2,3$) are real numbers,  consider the  {\it coordinate  volume element}  (or ``coordinate measure") on $\pi^{-1}[{\cal O}]$ given by the mere product of the differentials:
\begin{equation}
d{\cal {F}}_x  \equiv d^4x \,d^{16}y:= \Pi_{\mu = 0}^3 dx^{\mu} \Pi_{{\alpha}, {\beta}  = 0}^3 dy^{\alpha}_{\beta}.
\label{coordvol}
\end{equation}  
 Thus far, the coordinate system $x^\mu$ is completely general.

Our plan is to  write the invariant volume element $d\mu_{\cal F}$ in terms of  $d{\cal {F}}_x$. To do this, we will proceed in four steps, (1) to (4), as follows. Then in (5) we will give a summary. \\
\indent \indent (1): First, we will analyse how  $d{\cal {F}}_x$ transforms when we change coordinates, $x^{\mu} \mapsto x'^{\mu}(x)$, on ${\cal O}$, and thereby induce, by Eq.\ \ref{inducedcoordsF(M)}, a coordinate transformation on $\pi^{-1}[{\cal O}]$.      \\
\indent \indent (2): We then relate the 4-dimensional physical i.e.\ geometric volume element on  ${\cal M}$, that is defined by interpreting  ${\bf b}_\beta^a$ as a tetrad, to coordinate systems of the form given by  Eq.\ \ref{inducedcoordsF(M)}.\\
\indent \indent (3): With (1) and (2) in hand, we can then define  a  volume element on $\pi^{-1}[{\cal O}]$ and express it in any  coordinate system  of Eq.\ \ref{inducedcoordsF(M)}'s form (and thus relate it to any such coordinate system's coordinate volume element, $d{\cal {F}}_x$, defined by Eq.\ \ref{coordvol}).\\  
\indent \indent (4): The definition of this volume element on $\pi^{-1}[{\cal O}]$ immediately implies that it factorizes into a product of (a) the 4-dimensional volume element on  ${\cal M}$, and (b) the Haar measure on  $GL(4)$ (with its multiplicative constant also fixed).

\bigskip

\noindent (1): {\it How the coordinate volume element transforms}\\
 It is easy to find the induced transformation on $y^{\alpha}_{\beta}$, as follows:
\begin{equation}
 {\bf b}_\beta^a =  y^{\alpha}_{\beta}  {\bf \partial}^a_\alpha =  y^{\alpha}_{\beta} \frac{\partial x'^\gamma}{\partial x^\alpha} {\partial}'^a_\gamma
\, =: \, y'^{\gamma}_{\beta}  {\partial}'^a_\gamma \; \, \Rightarrow \; \, y'^{\alpha}_{\beta} = \frac{\partial x'^\alpha}{\partial x^\gamma}y^{\gamma}_{\beta} \, .
\label{inducedy}
\end{equation} 
Then, calculating the Jacobian ${\bf J}_{\cal F} \equiv {\bf J}_{\pi^{-1}[{\cal O}]}$ of the transformation $(x^{\mu},  y^{\alpha}_{\beta}) \mapsto (x'^{\mu},  y'^{\alpha}_{\beta})$, we have\\

\begin{equation} 
{\bf J}_{\cal F} := 
\left( \begin{array}{ll}
\left( \frac{\partial x'^\alpha}{\partial x^\gamma}\right)_{4 \times 4} &

\left( \frac{\partial x'^\alpha}{\partial y^\gamma_\delta}\right)_{4 \times 16} \\

\left( \frac{\partial y'^\alpha_\beta}{\partial x^\gamma}\right)_{16 \times 4}  &

\left( \frac{\partial y'^\alpha_\beta}{\partial y^\gamma_\delta}\right)_{16 \times 16} \\

\end{array} \right )  
=
\left( \begin{array}{ll}
 {\bf J}_{\cal O} &

{\bf 0}_{4 \times 16} \\

\left( \frac{\partial y'^\alpha_\beta}{\partial x^\gamma}\right)_{16 \times 4}  &

\left( \frac{\partial y'^\alpha_\beta}{\partial y^\gamma_\delta}\right)_{16 \times 16} \\

\end{array} \right )  \; ;
\end{equation} 
where in the last equation we have written $ {\bf J}_{\cal O}$ for the Jacobian of the coordinate transformation $x^{\mu} \mapsto x'^{\mu}(x)$ on ${\cal O}$, and we have used the fact that $ x'^{\mu}$ does not depend explicitly on $y^\alpha_\beta$.  

Using Eq.\ \ref{inducedy}, we have
\begin{equation}
\frac{\partial y'^{\alpha}_{\beta}}{\partial x^\gamma} =   y^{\delta}_{\beta}\frac{\partial^2 x'^\alpha}{\partial x^\delta \partial x^\gamma} \;  \; ; \; {\rm and} \; \;
\frac{\partial y'^{\alpha}_{\beta}}{\partial y^\gamma_\delta} = \delta^\delta_\beta \frac{\partial x'^{\alpha}}{\partial x^\gamma} \; .
\end{equation} 
This last result implies that we can conveniently re-order the indices so as to write
\begin{equation} 
\left( \frac{\partial y'^{\alpha}_{\beta}}{\partial y^\gamma_\delta} \right)_{16 \times 16} \; = \;
\left( \begin{array}{cccc}
 {\bf J}_{\cal O} & {\bf 0} & {\bf 0}  & {\bf 0}  \\
 
{\bf 0}  &  {\bf J}_{\cal O} & {\bf 0}  & {\bf 0}  \\

{\bf 0} & {\bf 0} &  {\bf J}_{\cal O}   & {\bf 0}  \\

{\bf 0} & {\bf 0}    & {\bf 0} &  {\bf J}_{\cal O}

\end{array} \right )  \; ;
\end{equation} 
As a consequence, ${\rm det} {\bf J}_{\cal F} = ({\rm det} {\bf J}_{\cal O})^5$. Hence
\begin{equation} 
d{\cal {F}}_{x'} = | {\rm det} {\bf J}_{\cal O} |^5 d{\cal {F}}_x \; .
\label{induceddF}
\end{equation} 
It is important to stress what Eq.\ \ref{induceddF} means. It gives the change in the {\it expression} of the coordinate volume element at each (fixed) point of $\pi^{-1}[{\cal O}]$. It does not refer to the dependence of a physical (geometric) volume element along the fibre, or across ${\cal O}$ (let alone, ``further afield" in  $ {\cal M}$). It is ``just'' a $\pi^{-1}[{\cal O}]$-pointwise statement of how the {\it expression} of the coordinate volume element changes, under (i) a coordinate transformation on ${\cal O}$ and (ii) the corresponding induced coordinate transformation on $\pi^{-1}[{\cal O}]$, as given by Eq.\ \ref{inducedcoordsF(M)}.    

\bigskip

\noindent  (2): {\it The volume element on $\cal M$ in terms of the coordinate systems}\\
Our task now is: Given a coordinate system $p \mapsto x^{\mu}(p)$ on ${\cal O}$, and a point  ${\bf {P_b}} := (p ; \{{\bf b}_\mu^a(p)\}) \in \pi^{-1}[{\cal O}] \subset {\cal {F({\cal M})}}$, how is the physical i.e.\ geometric volume element on $\cal M$ that is defined by interpreting  ${\bf b}_\mu^a$ as a tetrad, expressed in terms of the coordinate system? 

Since the basis $\{ {\bf b}_\alpha^a \}$ is to be interpreted as a tetrad, it defines a volume element form on $\cal M$ through
\begin{equation}
^{({\bf b})} \epsilon_{abcd} := \epsilon_{\alpha \beta \gamma \delta}  {\bf B}^{\alpha}_a {\bf B}^{\beta}_b {\bf B}^{\gamma}_c {\bf B}^{\delta}_d \; ,
\end{equation} 
where $\epsilon_{0123} = +1$, $\epsilon_{\alpha \beta \gamma \delta}$ is totally antisymmetric, and $\{{\bf B}^\alpha_a \}$ is the dual basis (of $T^*_p{\cal M}$) to $\{{\bf b}_\alpha^a\}$, i.e.\ ${\bf B}^\alpha_a {\bf b}_\beta^a = \delta^\alpha_\beta$.  Notice that $^{({\bf b})} \epsilon_{abcd}$ varies along the fibre $\pi^{-1}(p)$, whereas the $ \epsilon_{\alpha \beta \gamma \delta}$ are {\it uniform scalar fields} on $\pi^{-1}[{\cal O}]$. (Indeed, repeating our discussion at each open set ${\cal O}$ in an open cover of ${\cal M}$, the $ \epsilon_{\alpha \beta \gamma \delta}$ are uniform scalar fields over all of  $\cal {F({\cal M})}$.)

Now we recall that for the coordinate system $x^\mu$, where $y^{\alpha}_{\beta}$ are the components of ${\bf b}_\beta^a$ in the basis ${\bf \partial}^a_\alpha$, the components of an element of the dual basis  ${\bf B}^\alpha_a$ are given by the inverse of the matrix $y$. That is, we have:  ${\bf B}^\alpha_a = (y^{-1})^\alpha_\beta d{\bf x}^\beta_a$, where $(y^{-1})^\alpha_\beta y^\beta_\gamma = \delta^\alpha_\gamma$. Therefore
\begin{equation}
^{({\bf b})} \epsilon_{abcd} = \epsilon_{\alpha \beta \gamma \delta}  (y^{-1})^\alpha_\mu (y^{-1})^\beta_\nu (y^{-1})^\gamma_\rho (y^{-1})^\delta_\sigma  \, d{\bf x}^\mu_a d{\bf x}^\nu_b d{\bf x}^\rho_c d{\bf x}^\sigma_d  \; = ({\rm det} y^{-1}) \epsilon_{\mu \nu \rho \sigma}   \, d{\bf x}^\mu_a d{\bf x}^\nu_b d{\bf x}^\rho_c d{\bf x}^\sigma_d,
\label{4-element}
\end{equation} 
This last expression shows how the 4-dimensional geometric volume element varies along the fibre $\pi^{-1}(p)$: namely, through a dependence on the components $y^{\alpha}_{\beta}$ of ${\bf b}_\beta^a$.

\bigskip

\noindent  (3): {\it Definition of the  volume element on $\pi^{-1}[{\cal O}]$}\\
We are now in a position to define the volume element on $\pi^{-1}[{\cal O}]$ determined by interpreting
 all bases as tetrads  in some metric.  It will turn out to be uniquely defined, once we make the natural assumption that it has the form of the coordinate volume element, i.e.\ Eq.\ \ref{coordvol}, in  particular coordinate systems,  of Eq.\ \ref{inducedcoordsF(M)}'s form, in some way privileged at each given point ${\bf P}_{\bf b}\in \pi^{-1}[{\cal O}]$. 

Given    ${\bf {P_b}} = (p ; \{{\bf b}_\alpha^a(p)\}) \in \pi^{-1}[{\cal O}]$, let  $p \mapsto {\bar x}^{\mu}(p)$ be a coordinate system such that,  at the given ${\bf {P_b}}$,  ${\bf b}_\alpha^a(p) = {\bar {\bf \partial}}_\alpha^a(p)$, i.e.\ ${\bar y}^\alpha_\beta = \delta^\alpha_\beta$ at ${\bf {P_b}}$.  Coordinate systems satisfying this always exist, for any ${\bf {P_b}}$. In fact, the Riemann normal coordinates adapted to ${\bf {P_b}}$ is just an example of such a coordinate system. Since we are interpreting  ${\bf b}_\alpha^a$ as a tetrad, we now {\it define} the  volume element $d\mu_{\cal F}$ by requiring that, in the coordinate system on ${\cal F}({\cal O})$ around ${\bf {P_b}}$ that is induced by $p \mapsto {\bar x}^{\mu}(p)$ via Eq.\ \ref{inducedcoordsF(M)}, $d\mu_{\cal F}$ is given by the coordinate volume element, i.e.\ Eq.\ \ref{coordvol}.  That is: we define
\begin{equation}
d\mu_{\cal F} := d{\cal F}_{{\bar x}} = \Pi_{\mu = 0}^3 d{\bar x}^{\mu} \Pi_{{\alpha}, {\beta}  = 0}^3 d{\bar y}^{\alpha}_{\beta} \; .
\end{equation} 

So for a  coordinate system around  ${\bf {P_b}}$ that is arbitrary---though required to be of Eq.\ \ref{inducedcoordsF(M)}'s form,  so that ${\bf b}_\alpha^a(p) = y_\alpha^\beta{\bf \partial}_\beta^a$, i.e.\ ${\bf J}_{\cal O} = (y^\alpha_\beta)$---we infer from Eq.\ \ref{induceddF} that
 \begin{equation}
d{\cal F}_x = |{\rm det}(y)|^5 \, d{\cal F}_{{\bar x}} =  |{\rm det}(y)|^5 \, d\mu_{\cal F}  \; .
\end{equation} 
Using Eq.\ \ref{coordvol} and rearranging, we get our desired formula
 \begin{equation}
 d\mu_{\cal F} =  |{\rm det}(y)|^{-5}  \, \Pi_{\mu = 0}^3 dx^{\mu} \Pi_{{\alpha}, {\beta}  = 0}^3 dy^{\alpha}_{\beta} \; .
 \label{desired}
\end{equation} 
Eq.\ \ref{desired} is our desired formula, because it expresses  the volume element on $\pi^{-1}[{\cal O}]$, that is determined  by interpreting a basis-field as a tetrad-field,  in any coordinate system $(x^\mu, y^\alpha_\beta)$ of Eq.\ \ref{inducedcoordsF(M)}'s  form  (i.e.\ with $x$ i.e.\ $p \mapsto  x^{\mu}(p)$ on ${\cal O}$ being arbitrary, and the $y^\alpha_\beta$ being induced by $x$ via Eq.\ \ref{inducedcoordsF(M)}). 

\bigskip

\noindent  (4): {\it Factorization of the volume element into the 4-dimensional volume element  and the Haar measure}\\
Eq.\ \ref{desired}  is very convenient. For it shows that in $d\mu_{\cal F}$, one factor viz.\ $|{\rm det}(y)|^{-1} \equiv |{\rm det}(y^{-1} )|$ times another factor viz.\ $\Pi_{\mu = 0}^3 dx^{\mu}$ is the coordinate expression of the positive volume element on $\cal M$, i.e.\ is 
 $^{({\bf b})} \epsilon$, the modulus of the 4-form $^{({\bf b})} \epsilon_{abcd}$ given by Eq.\ref{4-element}.

Moreover, the remaining factor in Eq.\ \ref{desired},  i.e.\ $|{\rm det}(y)|^{-4}  \, \Pi_{{\alpha}, {\beta}  = 0}^3 dy^{\alpha}_{\beta}$, is precisely one of the invariant Haar measures of the group $GL(4)$, with its multiplicative factor determined by Eq.\ \ref{desired}.  This follows from the fact that $ |{\rm det}(y)|^{-4}  \, \Pi_{{\alpha}, {\beta}  = 0}^3 dy^{\alpha}_{\beta}$ is (left and right) invariant under $y \mapsto My$ and $y \mapsto yM$ with $M \in GL(4)$, as shown by:
 \begin{eqnarray}
  |{\rm det}(My)|^{-4}  \, \Pi_{{\alpha}, {\beta}  = 0}^3 d(My)^{\alpha}_{\beta} & = &  
  |{\rm det} M|^{-4} |{\rm det} y|^{-4}  \, 
  \Pi_{{\beta}  = 0}^3 [ \Pi_{{\alpha}  = 0}^3 d(M^\alpha_\gamma y^\gamma_\beta ) ]   \nonumber \\
&=&
  |{\rm det} M|^{-4} |{\rm det} y|^{-4}  \,  \Pi_{{\beta}  = 0}^3 [ |{\rm det} \, M| \Pi_{{\alpha}  = 0}^3 d y^\alpha_\beta  ]  \nonumber \\
  & = & 
   |{\rm det} M|^{-4} |{\rm det} y|^{-4}   |{\rm det} M|^4 \Pi_{{\alpha}, \beta  = 0}^3 d y^\alpha_\beta \nonumber \\
 & \equiv &
    |{\rm det} y|^{-4}   \Pi_{{\alpha}, \beta  = 0}^3 d y^\alpha_\beta  \; .
     \label{MHaar}
 \end{eqnarray}
A similar argument shows the invariance under right multiplication by $M$, i.e.\ under  $y \mapsto yM$.\footnote{The only non-obvious transition in Eq.\ \ref{MHaar} is from the first to the second line. To see this, consider 
at the end of the first line of \ref{MHaar},  $\Pi^3_{\alpha = 0}d(M^\alpha_\gamma y^\gamma_\beta)$ for each {\it fixed} 
$\beta$. To be concrete, consider $\beta = 0$: $\Pi^3_{\alpha = 0}d(M^\alpha_\gamma y^\gamma_0)$. We can define new variables $y'^\alpha :=M^\alpha_\gamma y^\gamma_0$, in terms of which
\begin{eqnarray}
\Pi^3_{\alpha = 0}d(M^\alpha_\gamma y^\gamma_0) = dy'^0\,dy'^1\,dy'^2\,dy'^3
=|{\rm det}(\partial y'/\partial y_0)| dy_0^0\,dy_0^1\,dy_0^2\,dy_0^3,
\nonumber 
\end{eqnarray}
where $(\partial y'/\partial y_0)$ is the Jacobian of the transformation $y^\mu_0 \mapsto y'^\mu = M^\mu_\alpha y^\alpha_0$. But the Jacobian is exactly $M$. Hence, 
\begin{eqnarray}
\Pi^3_{\alpha = 0}d(M^\alpha_\gamma y^\gamma_0) 
=|{\rm det}M| \Pi^3_{\alpha = 0}dy_0^\alpha.
\nonumber 
\end{eqnarray}
Since this is equally true for $\beta = 1,2,3$,  by taking the product of all of 
them we finally get the desired result, viz.\
$\Pi^3_{\alpha,\beta = 0}d(M^\alpha_\gamma y^\gamma_\beta) 
=|{\rm det}M|^4 \Pi^3_{\alpha,\beta = 0}dy_\beta^\alpha$, that is used in passing from the first to the second line of 
\ref{MHaar}.}

\bigskip

\noindent  (5): {\it Summary}:---
Finally, let us {\it summarize} what we have done in (1) to (4) above. We have shown that if we define a metric on ${\cal O}$ by  stipulating that a basis field on ${\cal O}$ is to be a tetrad field,
then there is a natural volume element on $\pi^{-1}[{\cal O}]$---and hence,  by (B),  on the bundle $\cal {F({\cal M})}$. This  volume element  can be written as
\begin{equation}
d\mu_{\cal F}  = \, ^{({\bf b})}\epsilon(p) \; d\mu_G \, ,
\end{equation}
where $^{({\bf b})} \epsilon$ is the $\{ {\bf b}^a_\alpha\}$-dependent 4-dimensional volume element  on ${\cal O}$, and $d\mu_G$ is the invariant Haar measure on $GL(4)$ which is given in  coordinate form by 
 \begin{equation}
 d\mu_G  = |{\rm det} y|^{-4} \Pi_{{\alpha}, {\beta}  = 0}^3 dy^{\alpha}_{\beta}  \, .
\end{equation}
We also note that although in general invariant Haar measures on noncompact Lie groups are defined only up to a multiplicative constant,  our construction, based on defining a metric on the base-manifold by stipulating a basis field  to be a tetrad field, determines the multiplicative constant, and so privileges one such measure.
The intuitive reason why the  constant gets determined is of course that in stipulating $\{ {\bf b}^a_\alpha\}$ to be a tetrad, we are stipulating orthonormality, not just orthogonality, i.e.\ we are stipulating standards of spatial and temporal length. This holds along the fibre too: in the adapted coordinates, $\bar{\partial}^a_\alpha = {\bf b}^a_\alpha$ at $p$, so that the fibre coordinates $\bar{y}^\alpha_\beta$ measure the components of the basis vectors in units given by the basis itself; this is what normalises the Haar factor as $|{\rm det}\,y|^{-4}\,\Pi_{\alpha,\beta=0}^3 dy^\alpha_\beta$. \\
 
\medskip

\acknowledgments
This work was partially developed while D.V.\ was on a sabbatical leave
at the Institute for Quantum Optics and Quantum Information
(IQOQI-Vienna), of the Austrian Academy of Sciences. D.V.\ 
acknowledges
partial financial support from
the S\~ao Paulo State Research Foundation
(FAPESP) under grant no.\ 2023/04827-9. 
D.V.\ thanks IQOQI, the University 
of Vienna, and
\v Caslav Brukner and his group for the
stimulating environment and their extended hospitality. D.V.\ also thanks
Henrique Gomes 
for an 
interesting exchange on the subject, and Viktoria Kabel and 
Anne-Catherine de la Hamette for their kind attention and enlightening discussions about
Ref.~\cite{KHACGBB24}. J.B.\ also thanks IQOQI, \v{C}aslav Brukner and  Viktoria Kabel and 
Anne-Catherine de la Hamette, for their hospitality, kind attention and enlightening discussions, over the years. 

 For corrections and comments on a previous version of this paper, we are very grateful to 
two referees and to: Anne-Catherine de la Hamette,  Henrique Gomes, Philipp H\"ohn, Viktoria Kabel, Markus M\"uller, Eduardo Am\^ancio Oliveira and James Read. We are especially grateful to Sam Fedida and Jan G\l owacki. We are also grateful to  audiences at seminar and conference talks in: Logar, Slovenia (New Directions conference 2024);  Castiglioncello, Italy (DICE conference 2024); Bonn, Germany; Oxford, U.K.; OIST, Okinawa, Japan; IQOQI Vienna, Austria; and York, U.K. (Semi-Local Quantum Physics conference 2026).

\end{document}